\documentclass{biometrika}
\usepackage{natbib,lscape,longtable}
\usepackage[pdflatex]{graphicx}

\usepackage{amsmath,amssymb,epsfig,color,enumitem,blkarray}
\usepackage{url}
\usepackage{comment}
\usepackage{algorithm}

\DeclareGraphicsExtensions{.pdf,.eps}
\bibpunct{(}{)}{;}{a}{,}{,}

\newcommand{\red}{\color{red}}

\newcommand{\ignore}[1]{}

\DeclareMathOperator{\diag}{diag}

\DeclareMathOperator\D{D}

\def\rank{{\rm rank}}
\newcommand{\argmin}{\mathop{\rm arg\min}}

\allowdisplaybreaks[1]

\begin{document}

\jname{Biometrika}
\jyear{2019}
\jvol{}
\jnum{}

\accessdate{}
\copyrightinfo{}

\markboth{Y. Cao, A. Zhang \and H. Li}{Composition Estimation}

\title{Multi-sample Estimation of Bacterial Composition Matrix in Metagenomics Data}

\author{Yuanpei Cao}\affil{Department of Biostatistics and Epidemiology, Perelman School of Medicine, University of Pennsylvania, Philadelphia, Pennsylvania 19104, U.S.A. \email{yuanpeic@sas.upenn.edu}}

\author{Anru Zhang}\affil{Department of Statistics, University of Wisconsin-Madison, Madison, Wisconsin 53706, U.S.A. \email{anruzhang@stat.wisc.edu}}

\author{\and Hongzhe Li}\affil{Department of Biostatistics and Epidemiology, Perelman School of Medicine, University of Pennsylvania, Philadelphia, Pennsylvania 19104, U.S.A. \email{hongzhe@upenn.edu}}

\maketitle
\begin{abstract}
	Metagenomics sequencing is routinely applied to quantify  bacterial abundances in microbiome studies, where the bacterial composition is estimated based on the sequencing  read counts. Due to limited sequencing depth and DNA dropouts, many rare bacterial taxa  might not be captured  in the final  sequencing reads, which results in many zero counts. Naive composition estimation using count normalization  leads  to many zero proportions, which tend to result in inaccurate estimates of bacterial abundance and diversity.  This paper takes a multi-sample approach to the estimation of bacterial abundances in order to borrow information across samples and across species.  Empirical results from real data sets suggest  that the composition matrix over multiple samples is approximately low rank, which  motivates a regularized maximum likelihood estimation with a nuclear norm penalty. An efficient optimization algorithm using the generalized accelerated proximal gradient and Euclidean projection onto  simplex space is developed. The theoretical upper bounds and the minimax lower bounds of the estimation errors, measured by the Kullback-Leibler divergence and the Frobenius norm, are established. Simulation studies demonstrate that the proposed  estimator outperforms the naive estimators. The method is applied  to  an analysis of a human gut microbiome dataset.
\bigskip
\end{abstract}

\begin{keywords}
Microbiome;  Poisson-multinomial  distribution; Nuclear norm penalty; Proximal gradient descent.
\end{keywords}

\section{Introduction}
The human microbiome is the totality of all microbes at different body sites, whose contribution to human health and disease has increasingly been recognized. Recent studies have demonstrated that the microbiome composition varies across individuals due to different health and environmental conditions \citep{Huma:fram:2012}, and may be associated with complex diseases such as obesity, atherosclerosis, and Crohn's disease \citep{Turn:Hama:Yats:Cant:Dunc:core:2009,Koet:Wang:Levi:Buff:Org:inte:2013,Lewi:Chen:Wu:Bush:infl:2015}. With the development of next-generation sequencing technologies, the human microbiome can be quantified by using direct DNA sequencing of either marker genes or the whole metagenomes. After aligning the sequence reads to the reference microbial genomes, one obtains counts of sequencing reads that can be assigned to a set of bacterial taxa observed in the samples.  Such count data provide information about the relative abundance of different bacteria in different samples. 

 In order to account for the large variability in the total number of reads obtained, the sequencing count  data are often normalized into  a relative measure of  abundance of the taxa observed. Such relative abundances provide information about the bacterial composition. However,  due to limited sequencing depth, under-sampling, and DNA dropouts,  some rare microbial taxa might not be captured  in the metagenomic sequencing, which results in zero read counts assigned to these taxa. Naive estimation  of taxon composition using count normalization  leads to excessive zeros, especially for rare taxa. Such a naive estimate can be inaccurate and leads to sub-optimal estimate of taxa diversity. It  also causes difficulty in downstream data analysis for compositional data. These zero counts are regarded as rounded zeros, which are not truly zeros, but rather represent observed values due to under-sampling or dropouts.  

Since the pioneering work of \citet{Aitc:stat:2003}, several techniques have been proposed to deal with such rounded zeros \citep{martin2011dealing} in count data. One approach is to estimate non-zero compositions through a Bayesian-multiplicative model \citep{martin2014bayesian} from the counts. Such a Bayesian method involves a Dirichlet prior distribution as the conjugate prior distribution of multinomial distribution and a multiplicative modification of the non-zero counts. In fact, the Bayesian-multiplicative method is essentially equivalent to the non-parametric imputation, where the zero replacement values were determined by the parameterizations of the prior distribution. In the compositional data analysis,  these zero replacement values are usually chosen  as half of the minimum non-zero values. Some references are \cite{Aitc:stat:2003,lin2014variable,shi2016regression,cao2017covariance,cao2017twosample}. In addition, \cite{cai2019differential} recently studied the detection of differential microbial community networks by discretizing the data into a binary Markov random field based on a prespecified abundance threshold.

This paper addresses the problem of estimating  microbial composition in positive simplex space from a high-dimensional sparse count table. The observed counts are assumed to follow a Poisson-multinomial model, where i) the total number of read counts for each individual is a Poisson random variable; ii) given the total count for each individual, the stratified read counts over different taxa follow a multinomial distribution with the underlying parameters given by a positive composition. If the compositions across different individuals are combined into a matrix, an approximately low-rank structure on this matrix is indicated by recent observations on co-occurrence pattern \citep{faust2012microbial} and various symbiotic relationships in microbial communities \citep{woyke2006symbiosis,horner2007comparison,chaffron2010global}. 

Motivated by nuclear norm minimization used in noisy matrix completion problem \citep{negahban2012restricted,klopp2015}, this paper solves the problem of  composition estimation using a nuclear norm regularized maximum likelihood approach. However, it  should be emphasized that our approach  is very different from  the matrix completion problem because the missing mechanism and data generation  models are different. The observed zero counts are the result of under-sampling or dropouts, rather than random missingness assumed in  matrix completion literature. Besides, the sparse counts are assumed to be generated from a Poisson-multinomial model, and the focus of this paper is to estimate the underlying composition, rather than to recover  the zero counts. In this framework,  the asymptotic upper and minimax lower bounds of the resulting regularized estimator are obtained. Simulations show that the estimator recovers low-rank composition matrix accurately. Although the observed composition can be seen as true composition plus noise  and the problem can be roughly framed as a version of matrix denoising, the classic methods in literature such as the singular value thresholding  \citep{candes2013unbiased,donoho2014minimax}, may not be suitable here due to heteroscedasticity of different observations in the Poisson-multinomial data.

 Our work can  be seen as a variant of low-rank Poisson matrix recovery.  \cite{salmon2014poisson} studied the non-local principal component analysis for Poisson matrix data. A two-step procedure was proposed: after achieving a warm start via regular singular value decomposition, the iterative Newton steps were applied until convergence. \cite{soni2014estimation} considered the Poisson denoising problem with sparse and structured dictionary models. A constrained maximum likelihood method was proposed and the $\ell_2$ risk upper bound  was developed using  complexity penalized maximum likelihood analyses. \cite{cao2015poisson} introduced the penalized and constrained likelihood methods for the Poisson matrix recovery and Poisson matrix completion, respectively. The theoretical guarantees were developed, including the near-matching minimax-optimal bounds for Frobenius norm loss in Poisson matrix completion. However,   these results are not directly applicable to our problem. In  microbiome 16S rRNA sequencing data analysis, our goal  is to estimate the microbial composition rather than their absolute values for each individual. In addition, the  hidden sparse dictionary structure imposed by \cite{soni2014estimation} is not likely to hold in our applications. The zero counts in our problem  are due to under-sampling, which is different from the missing entries in Poisson matrix completion \citep{cao2015poisson}. Theoretically, our proposed penalized nuclear norm minimization estimator is convex  and is proved to achieve the near-optimal rate of estimation risks in both Kullback-Leibler divergence and Frobenius norm.  

\ignore{The rest of the paper is organized as follows. Section $\ref{sec:model}$ presents details of the Poisson-multinomial model and the proposed regularized likelihood estimation when the underlying composition is approximately low-rank. A generalized accelerated proximal gradient method to implement the optimization is presented in Section \ref{sec:imp}. The theoretical properties of the estimators are analyzed in Section $\ref{sec:thm}$, where the upper bounds and minimax lower bounds  for the estimation errors measured by the average Kullback-Leibler divergence and Frobenius norm are established. Simulation results are shown in Section $\ref{sec:sim}$ to investigate the numerical performance of the proposed methods and to compare with the zero-replacement approach. A real data application to a human gut microbiome study is given in Section $\ref{sec:combo}$. Finally, a brief discussion is given in Section \ref{discuss}. The proofs of all theorems are given in Appendix. }

\section{A Poisson-multinomial model for microbiome count data}\label{sec:poisson-multinomial-model}\label{sec:model}

For any integer $n>0$, we write $[n]=\{1,\ldots,n\}$ and denote $e_i(n)$ as the canonical basis in $\mathbb{R}^n$ with $i$th entry as one and others as zero. We refer to any $u\in \mathbb{R}^p$ as a composition vector if $u \geq 0$ and $\sum_{i=1}^p u_i = 1$.
For any two composition vectors $u, v\in \mathbb{R}^p$, the Kullback-Leibler divergence is defined as $\D_{KL}(u, v) = \sum_{i=1}^p u_i \log (u_i/v_i)$. For two composition matrices $X^\ast$ and $\widehat{X}$ with each row being a composition vector, let $\D(X^\ast,\widehat{X})$ denote the sum of Kullback-Leibler divergence between rows of $X^\ast$ and $\widehat{X}$,
	\begin{align}
	\D(X^\ast,\widehat{X}) = \sum_{i=1}^n\D_{KL}(X_i^\ast,\widehat{X}_i)=\sum_{i=1}^n\sum_{j=1}^pX_{ij}^\ast\log(X_{ij}^\ast/\widehat{X}_{ij}).\label{align:kl-def}
	\end{align}

16S ribosomal RNA (rRNA) sequencing is a common amplicon-based sequencing method used to identify and compare bacteria present within a given sample. In such studies, the sequencing reads are mapped to a set of $p$  known bacterial taxa  and the resulting  data are summarized as a count matrix $W \in \mathbb{R}^{n\times p}$, where  the $(i, j)$th entry of $W$, i.e., $W_{ij}$, represents the observed read count of taxon $j$ in individual $i$. For the $i$th individual, the total count of all taxa, $N_i$, is determined by the sequencing depth and DNA materials that are modeled as a Poisson random variable as $N_i\sim \text{Pois}(\nu_i)$, where $\nu_i$ is an unknown positive parameter. Given $N_i$, it is natural to model the stratified count data over $p$ taxa as a multinomial distribution. Therefore, the proposed Poisson-multinomial model for count-compositional data can be written as
\begin{align*}
& N_i \sim \text{ Pois}(\nu_i), \quad  i\in [n],\\
& f_{X_i^\ast}(W_{i1},\cdots,W_{ip}\mid N_i) = \frac{N_i!}{\prod_{j=1}^p W_{ij}!}\prod_{j=1}^p {X_{ij}^\ast}^{W_{ij}},\quad  i\in [n].
\end{align*}
Here, $X^\ast = (X_{ij}^\ast)\in \mathbb{R}^{n\times p}$ is the unknown taxon composition matrix lying in the positive simplex space $\mathcal{S}=\{X\in\mathbb{R}^{n\times p}\mid X 1_p= 1_n,X > 0\}$, where $1_p$ is the $p$-vector of 1s.

Our goal is to estimate $X^\ast$ based on $W$. One might attempt to consider the maximum likelihood estimate $\widehat{X}^{\rm mle}$. Conditioning on fixed number of total count $N$ and ignoring the terms that do not depend on $X^\ast$, the negative log-likelihood of the observations is given as
\begin{align}\label{align:likelihood}
\mathcal{L}_N\left(X^\ast\right)=-{N}^{-1}\sum_{i = 1}^n \sum_{j=1}^{p}W_{ij}\log X_{ij}^\ast,
\end{align}
where $N = \sum_{i=1}^n N_i = \sum_{i=1}^n\sum_{j=1}^pW_{ij}$ is the total number of the observed counts, which follows $\text{Pois}(\sum_{i=1}^n \nu_i)$. Without further constraints, minimizing $\eqref{align:likelihood}$ leads to $\widehat{X}^{\rm mle}$, which is the naive count normalization:
\begin{align}\label{align:mle}
\widehat{X}_{ij}^{\rm mle}={W_{ij}}/{\sum_{k=1}^p W_{ik}}, \quad i\in[n],\quad j\in[p].
\end{align}
Due to dropouts in sample preparation or $N$ being not sufficiently large, $\widehat{X}^{\rm mle}$ often contains a large number of zeros. In microbiome studies, these zero counts are treated as rounded zeros, which means that their corresponding compositions are below the detection lower limit. However, the zero counts yield zero estimates of these compositions and cause difficulty in downstream log-ratio based compositional data analysis \citep{Aitc:stat:2003,lin2014variable,shi2016regression,cao2017twosample}.

To overcome this difficulty, replacing the zero counts by a below-detection value through either Bayesian-multiplicative model \citep{martin2014bayesian} or non-parametric imputation \citep{martin2003dealing} is commonly seen in literature. These two methods are essentially equivalent, and are widely used in compositional data analysis by replacing the zero counts by 0$\cdot$5 in the data
\citep{Aitc:stat:2003,lin2014variable,shi2016regression},
\begin{align}
\widehat{X}^{\rm zr} \in \mathbb{R}^{n\times p}, \quad \widehat{X}^{\rm zr}_{ij} = \left(W_{ij} \vee \text{0$\cdot$5}\right)/{\sum_{l=1}^p\left(W_{il} \vee \text{0$\cdot$5}\right)}.
\end{align}
However, the pseudo-count 0$\cdot$5 is chosen arbitrarily without any theoretical guarantee, while the downstream analysis might be highly sensitive to this value.

On the other hand, under Poisson-multinomial model, $W_{ij} = EW_{ij} + (W_{ij} - EW_{ij}) = \nu_i X_{ij}^\ast + (W_{ij} - EW_{ij})$, where $\{\nu_iX_{ij}^\ast\}_{i,j=1}^{n, p}$ is a low-rank matrix and $(W_{ij} - EW_{ij})$ can be regarded as the noise. Thus, estimating $W$ can be seen as a version of matrix denoising.  The singular value thresholding \citep{donoho2014minimax,gavish2014optimal,chatterjee2015matrix} provides an alternative method for composition estimation. Such an estimator $\widehat{X}^{\rm svt}$ is given as 
\begin{equation}\label{eq:SVT-2}
\widehat{X}^{\rm svt}\in \mathbb{R}^{n\times p}, \quad \widehat{X}^{\rm svt}_{ij} = \left(\widehat{W}_{ij}\vee \text{0$\cdot$5}\right)/{\sum_{l=1}^p (\widehat{W}_{il} \vee \text{0$\cdot$5})},\\
\end{equation}
where $W = \sum_k \sigma_k u_k v_k^\top$ is the singular value decomposition \textsc{(svd)} and
\begin{equation}\label{eq:SVT-1}
\begin{split}
\widehat{W} & = \sum_{k} I_{\{\sigma_k \geq \lambda\}}\cdot \sigma_k u_k v_k^\top.
\end{split}
\end{equation}
However, this singular value thresholding method may not be suitable for our Poisson-multinomial model due to the following reasons. First, the Poisson distribution is heteroscedastic according to the values of the means, but $\widehat{X}^{\rm svt}$ achieves the most efficiency for treating homoscedastic noisy data (see, e.g. \cite{donoho2014minimax}).  Second, since there is no guarantee for positivity in singular value decomposition, $\widehat{W}$ in \eqref{eq:SVT-1} may even contain a large number  of negative values. Third, singular value thresholding does not guarantee the correct normalization in the sense that the row sums of $\widehat{X}^{\rm svt}$ are typically not 1. More comparisons and discussions are given in Section \ref{sec:sim}.

\section{Regularized Estimation of the Compositional Matrix and Computational Algorithm}\label{sec:imp}

\subsection{Regularized estimation of the compositional matrix }
In order to improve the composition estimate,  the approximate low-rank structure of the compositional matrix $X^\ast$ is explored.  The co-occurrence patterns \citep{faust2012microbial}, various symbiotic relationships in microbial communities \citep{woyke2006symbiosis,horner2007comparison,chaffron2010global} and samples in similar microbial communities  are expected to lead to an approximately low-rank structure of the composition matrix in the sense that the singular values of $X^\ast$ decay to zero in a fast rate. Such a low-rank structure is further investigated in our real data analysis in Section \ref{sec:combo},  showing  the empirical evidence of low-rank compositional matrix. This motivates us to propose a  nuclear norm regularized maximum likelihood approach to estimate the composition matrix:
\begin{equation}
\begin{array}{c}\label{align:rml}
\displaystyle \widehat{X} = \argmin_{X\in\mathcal{S}(\alpha_X,\beta_X)} \mathcal{L}_N\left(X\right)+\lambda\|X\|_{\ast}, \\
\displaystyle \mathcal{S}(\alpha_X,\beta_X)=\left\{X\in\mathbb{R}^{n\times p}\mid X 1_p = 1_n, \alpha_X/p \le X_{ij}\le \beta_X/p, \text{for any } (i,j)\in[n]\times[p]\right\},
\end{array}
\end{equation}
where $\lambda>0$ and $\mathcal{S}(\alpha_X,\beta_X)$ is a bounded simplex space with tunning parameters $0<\alpha_X\le \beta_X$. The constrained element-wise lower bound, $X_{ij}\geq\alpha_X/p$, guarantees the positive sign of the estimator. The element-wise upper bound constraint, $X_{ij} \le \beta_X/p$, is only needed for the theoretical analysis. 

The proposed estimator \eqref{align:rml} is essentially a regularized nuclear norm minimization, which can be solved by either semidefinite programing via interior-point semi-definite programming solver \citep{liu2009interior,recht2010guaranteed}, or a first-order method via Templates for First-Order Conic Solvers \citep{becker2011templates}. However, the interior-point semi-definite programming solver computes the nuclear norm via a less efficient eigenvalue decomposition, which does not scale well with large $n$ and $p$. Templates for First-Order Conic Solvers on the other hand often results in the oscillations or overshoots along the trajectory of the iterations \citep{su2014differential}. To achieve a stable and efficient optimization  for \eqref{align:rml} in the high-dimensional setting,  we propose an algorithm  based on the generalized accelerated proximal gradient method and Nesterov's scheme \citep{su2014differential}. The general procedure is detailed  in Section \ref{sec.generalized_accelerated}, and a key step in the implementation, the Euclidean projection onto $\mathcal{S}(\alpha_X,\beta_X)$, is given in Section \ref{sec.projection}.

\subsection{Generalized accelerated proximal gradient method}\label{sec.generalized_accelerated}
We introduce an optimization algorithm for \eqref{align:rml} based on the generalized accelerated Nesterov's scheme, which follows the formulation of \cite{beck2009fast} and the spirit of \cite{su2014differential}. First, based on the count matrix $ W$, we initialize $X_0$ and $ Y_0$ as
$$X_0,  Y_0 \in \mathbb{R}^{n\times p}, \quad \text{where}\quad \left(X_0\right)_{ij} = \left( Y_0\right)_{ij} = W_{ij}/\sum_{l=1}^pW_{il}.$$
Then $X_0,  Y_0$ are essentially the row-wise normalization of $ W$. Next, we update $X_k$ and $ Y_k$ as
\begin{align}
X_k &= \argmin_{X\in\mathcal{S}(\alpha_X,\beta_X)} 2^{-1}L_k\|X- Y_{k-1}+L_k^{-1}\triangledown\mathcal{L}_N\left( Y_{k-1}\right)\|_{F}^2+\lambda\|X\|_{\ast},\label{align:prox}\\
 Y_k &=X_k + (k+\rho-1)^{-1}{(k-1)}\left(X_k-X_{k-1}\right),\label{align:k}
\end{align}
until convergence or a maximum number of iterations is  reached.  Here 
$\triangledown\mathcal{L}_N$ is the gradient function of $\mathcal{L}_N\left(X\right)$:
$$\triangledown\mathcal{L}_N\left(X\right) \in \mathbb{R}^{n\times p},\quad \text{with}\quad \left(\triangledown\mathcal{L}_N(X)\right)_{ij} = - {(NX_{ij})}^{-1}W_{ij},$$
where we treat possible $0/0$ as zero and $L_k$ is the reciprocal of step size in the $k$th iteration, which can be chosen by the following line search strategy. Denote
\[\mathcal{F}_{L}(X, Y)=\mathcal{L}_N\left(X\right)- \mathcal{L}_N( Y)-\langle X- Y,\triangledown \mathcal{L}_N( Y)\rangle-2^{-1}L\|X- Y\|_F^2\]
as the approximation error for the second order Taylor expansion of $\mathcal{L}_N( Y)$ with the second order coefficient as $L$ to $\mathcal{L}_N( X)$. In the $k$th iteration, we start with the integer $n_k=1$ and let $L_k=\gamma^{n_k}L_{k-1}$ for some scale parameter $\gamma>1$, then repeatedly increase $n_k=1,2,\cdots$ until $\mathcal{F}_{L_k}(X_k, Y_{k-1})\le 0$. 
${(k-1)}/{(k+\rho-1)}$ and $\rho$ are respectively referred to as the momentum term and friction parameter in optimization literature. We follow the suggestions by \cite{su2014differential} and set a high friction rate that $\rho\ge 9/2$.

Optimization $\eqref{align:prox}$ is the proximal mapping of the nuclear norm function, and it can be solved by singular value thresholding \citep{cai2010singular},
\[ X_k = \Pi_{\mathcal{S}(\alpha_X,\beta_X)}\left[\mathcal{D}_{\lambda L_k^{-1}}\left\{ Y_{k-1}-L_k^{-1}\triangledown\mathcal{L}_N\left( Y_{k-1}\right)\right\}\right].\]
Here $\Pi_{\mathcal{S}(\alpha_X,\beta_X)}\left( X\right)$ is Euclidean projection of $ X$ onto the positive simplex space $\mathcal{S}(\alpha_X,\beta_X)$ that we postpone the detailed discussions to Section \ref{sec.projection}. Provided that $ X= U\Sigma V^\top$ is the \textsc{svd}, $\mathcal{D}_{\tau}$, the soft-thresholding operator, is defined as
\[\mathcal{D}_{\tau}\left( X\right)=U\mathcal{D}_{\tau}\left(\Sigma\right)V^\top,\quad \mathcal{D}_{\tau}\left\{\Sigma\right)=\diag\left\{(\sigma_i-\tau)\wedge 0\right\}.\]

\ignore{
	The generalized accelerated proximal gradient method is summarized as Algorithm $\ref{APGalgorithm}$.
	\begin{algorithm}[H]
		\caption{Generalized accelerated proximal gradient method}
		\label{APGalgorithm}
			{\bf Input}: count matrix $ W$; statistical tunning parameters $\alpha_X$, $\beta_X$ and $\lambda$; optimization parameters $L_0, \gamma, \rho, k_{\max}, \epsilon$.\\
			Initialize: $ Y_0= X_0 \in \mathbb{R}^{n\times p}$, $( Y_0)_{ij} = ( X_0)_{ij} = W_{ij}/\sum_{j=1}^p W_{ij}$.
			\textbf{For} $k=1$ \textbf{to} $k_{\max}$\\
			Initialize: $L_k = L_{k-1}$
			$ X_k=\Pi_{\mathcal{S}(\alpha_X,\beta_X)}\left(\mathcal{D}_{\lambda L_k^{-1}}\left( Y_{k-1}-L_k^{-1}\triangledown\mathcal{L}_{N}\left( Y_{k-1}\right)\right)\right)$
			If $\mathcal{F}_{L_k}( X_k, Y_{k-1})\ge 0$,\\
			Update $L_k=\gamma L_k$, go to Step 5\\
			\textbf{End If}\\
			Update $ Y_{k}= X_k+\frac{k-1}{k+\rho-1}\left( X_k- X_{k-1}\right)$
			\textbf{If} {$\left|\mathcal{F}_{L_k}( X_k, Y_{k-1})\right|<\epsilon$} \textbf{Return} $ X_k$
			\textbf{End If}
			\textbf{End For}
	\end{algorithm}
}

\subsection{Euclidean projection onto the simplex space}\label{sec.projection}

The final step of the algorithm involves  Euclidean projection onto the simplex space $\mathcal{S}(\alpha_X,\beta_X)$,  a key step in the proposed generalized accelerated proximal gradient method. An efficient algorithm, summarized as Algorithm \ref{al:proj}, is used to perform such an Euclidean projection onto compositional space.

	\begin{algorithm}
	\caption{Euclidean projection onto compositional space with upper lower bounds}
	\label{al:proj}
		1: {\bf Input}: To-be-projected vector $x\in \mathbb{R}^{p}$; simplex constraint parameters $\alpha_X$ and $\beta_X$.\\
		2: {\bf Output}: $\widehat{ x} = \Pi_{\mathcal{S}(\alpha_X, \beta_X)}( x)$.\\
		3: Calculate $v = \{x_i - \alpha_X/p, x_i - \beta_X/p\}_{i=1}^p \in \mathbb{R}^{2p}$, and sort it as $v_{(1)} \le \cdots \le v_{(2p)}$.\\
		4: For $1\le j\le 2p$, calculate
		\begin{equation*}
		d_j = \sum_{i=1}^p \left\{(x_i - v_{(j)}) \wedge \left(\beta_X/p\right)\right\} \vee \left(\alpha_X/p\right) - 1.
		\end{equation*}
		5: Naturally $d_j$ is a decreasing sequence from non-negative values to non-positive values. Find $1\le j^\ast \le 2p-1$ such that
		$$d_{j^\ast} \geq0 \quad \text{and} \quad d_{j^\ast+1} \le 0.$$
		6: Calculate the final estimator $\widehat{x}_i$ as follows,
		\begin{equation}\label{eq:projection_situations}
		\widehat{x}_i = \left\{\begin{array}{ll}
		\beta_X/p, & x_i - v_{(j^\ast)} > \beta_X/p,\\
		\alpha_X/p, & x_i - v_{(j^\ast)} \le \alpha_X/p,\\
		x_i - v_{j^\ast} - \frac{d_j^\ast}{\left|\left\{i\mid \alpha_X/p \le x_i - v_{(j^\ast)} \le  \beta_X/p\right\}\right|}, & \alpha_X/p \le x_i - v_{(j^\ast)} \le  \beta_X/p.
		\end{array}
		\right.
		\end{equation}
		7: {\bf Return} $\widehat{ x}$
\end{algorithm}

The following Proposition \ref{prop:projection} provides theoretical guarantees for the performance of Algorithm \ref{al:proj}. The central idea of the proof to Proposition \ref{prop:projection} lies on the Karush-Kuhn-Tucker conditions for the optimization problem \eqref{eq:projection_optimization}.
\begin{proposition}\label{prop:projection}
	Recall
	\begin{equation}\label{eq:projection_optimization}
	\Pi_{\mathcal{S}(\alpha_X, \beta_X)}( x) = \argmin_{\widehat{ x}}\|\widehat{ x}- x\|_2^2 \quad \text{subject to}\quad \sum_{i=1}^p \widehat{x}_i = 1, \alpha_X/p \le \widehat{x}_i \le \beta_X/p.
	\end{equation}
	Then $\widehat{ x}$ calculated from Algorithm \ref{al:proj} exactly equals $\Pi_{\mathcal{S}(\alpha_X, \beta_X)}( x)$.
\end{proposition}

\subsection{Selection of the tuning parameters}\label{sec:tune}
 We  propose a variation of $K$-fold cross-validation to select the tunning parameters $\lambda$ and $\alpha_X$. We  set $\beta_X = p$ to  remove the element-wise upper bound constraint. 

 Let $W$ be the observed count matrix and $T_1, T_2$ be two sets of grids of positive values. We first randomly split the rows of $W$ into two groups of sizes $n_1\sim{(K-1)n}/{K}$ and $n_2\sim{n}/{K}$  for a total of  $L$ times. For the $l$th  split, denote by $I_l, I_l^c \subseteq [n]$ the row index sets of the two groups, respectively. For each $i\in I_l^c$, we further randomly select a subset $J_{i, l}\subseteq [p]$ with cardinality $p_1\sim p(K-1)/K$. For the $l$th split, the training set is defined as $\Omega_{l} = \{(i,j): (i, j)\in I_l \times [p] \text{ or } i\in I_l^c, j\in J_{i, l}\}\subseteq [n]\times [p]$,  which contains both complete and incomplete rows of $[n]\times [p]$. Denote by $W_{\Omega_l}$ the training matrix $W$ with all entries in $\Omega_l^c$ being set to zero. Next, for each $(\lambda, \alpha_X) \in T_1\times T_2$, we apply the proposed estimator to $W_{\Omega_l}$ with tuning parameters $(\lambda, \alpha_X)$ and obtain  the estimates  $\widehat{X}^{(l)}(\lambda, \alpha_X)$, $l=1,\ldots, L$. We use the Kullback-Leibler divergence defined in \eqref{align:kl-def} to evaluate the prediction error on the rows of $I_l^c$,
\[\widehat{R}(\lambda, \alpha_X)=\sum_{l=1}^I\sum_{i\in I_l^c}\D_{KL}\left\{\widehat{X}^{\rm mle}_{i\cdot},\widehat{X}^{(l)}_{i\cdot}(\lambda, \alpha_X)\right\},\]
where $\widehat{X}^{\rm mle}$ is the maximum likelihood estimator defined in Equation \eqref{align:mle}.  

We choose  $$(\lambda^\ast, \alpha_X^\ast)=\argmin\limits_{\lambda\in T_1, \alpha_X\in T_2}\widehat{R}(\lambda, \alpha_X)$$ 
as the final  tuning parameters and  obtain the final estimate $\widehat{ X}$ based on the full dataset. The performance of this tuning parameter selection procedure is verified through numerical studies on both simulated and real datasets in Sections \ref{sec:sim} and \ref{sec:combo}.


\section{Theoretical Properties}\label{sec:thm}

\subsection{Theoretical property under low-rank matrix assumption}\label{sec:theory}
We investigate the theoretical properties of $\widehat{ X}$ proposed in Section \ref{sec:imp}. Particularly, the upper bounds of estimation accuracy for the whole composition matrix are provided in Theorems \ref{thm:exact_lr} and \ref{thm:approx_lr}, and the lower bound results are given in Theorem \ref{thm:lower_bound}. These results  establish the optimal recovery rate over certain class of low-rank composition matrices. Additionally, we study the diversity index estimation and present the upper bound results in Corollary \ref{coro:diversity}.

Denote $R_i={\nu_i}/{\sum_{j=1}^n \nu_j}$ for $i\in[n]$, which quantifies the proportion of the total count for the $i$th subject. We establish the upper bound for $\widehat{ X}$ in Frobenius norm error and average Kullback-Leibler divergence. 
The following theorem gives upper bound result over a class of bounded low rank composition matrices:
\[\mathbb{B}_0(r, \alpha_X, \beta_X)=\left\{ X\in\mathcal{S}(\alpha_X,\beta_X)\mid \rank( X)\le r\right\},\]
where $\mathcal{S}(\alpha_X,\beta_X)$ is the set of bounded composition matrices defined in \eqref{align:rml}.
\begin{theorem}\label{thm:exact_lr} 
	Assume there exist constants $\alpha_R, \beta_R, \alpha_X$, and $\beta_X$ such that, for any $i\in [n]$, $\alpha_R/n\le R_i\le \beta_R/n$. Suppose $X^\ast\in\mathbb{B}_0(r, \alpha_X, \beta_X)$. Conditioning on fixed $N$, suppose that $N \ge (n+p)\log(n+p)$ and the tuning parameter is selected as
	\begin{align}\label{align:lambda}
	\lambda = \delta\left\{\frac{\beta_R}{\alpha_X^2}\frac{p(n\vee p)\log(n+p)}{nN}\right\}^{1/2}
	\end{align}
	with some constant $\delta\ge 7$. Then, there exists some constants $C_1(\alpha_X,\alpha_R,\beta_X,\beta_R)$ and $C_2(\alpha_X,\alpha_R,\beta_X,\beta_R)$ that only depend on $\alpha_X,\alpha_R,\beta_X$, and $\beta_R$, such that $\widehat{X}$ in \eqref{align:rml} satisfies
	\begin{align}
	\frac{p}{n}\left\|\widehat{ X}- X^\ast\right\|_F^2& \le C_1(\alpha_X,\alpha_R,\beta_X,\beta_R)\frac{(n+p)r\log(n+p)}{N},\label{align:exact_lr_frob}\\
    \frac{1}{n}\D( X^\ast,\widehat{ X})& \le C_2(\alpha_X,\alpha_R,\beta_X,\beta_R)\frac{(n+p)r\log(n+p)}{N} \label{align:exact_lr}
	\end{align}
	with probability at least $1 - 3(n+p)^{-1}$. In particular, $C_1(\alpha_X,\alpha_R,\beta_X,\beta_R)$ and $C_2(\alpha_X,\alpha_R,\beta_X,\beta_R)$ satisfy:
    \begin{align*}
    C_1(\alpha_X,\alpha_R,\beta_X,\beta_R) &= C\beta_X^4(\beta_X\vee\beta_R)/(\alpha_R^2\alpha_X^4),\\
    C_2(\alpha_X,\alpha_R,\beta_X,\beta_R) &= \left\{
    \begin{array}{ll}
    C\beta_X^2(\beta_X\vee\beta_R)/(\alpha_R^2\alpha_X^3) & \text{if } N < 6(n+p)^2\log(n+p)/(\alpha_X\alpha_R);\\
    C\beta_X^5\beta_R/(\alpha_R^2\alpha_X^6) & \text{if } N \ge 6(n+p)^2\log(n+p)/(\alpha_X\alpha_R),
    \end{array} \right.
    \end{align*}
    where $C>0$ is a uniform constant that does not depend on $X^\ast, r$, $N$, $p$, $n$, $\alpha_X,\alpha_R,\beta_X$, or $\beta_R$.
\end{theorem}

\begin{remark}\rm 
 In contrast to its population counterpart $\nu := \sum_{i=1}^n\nu_i$,
	 the total count $N$ is an observable value,  we therefore choose to present the results of Theorem \ref{thm:exact_lr} conditioning on the fixed number of $N$. If one replaces all $N$  by $\nu$ in the conditions and conclusions of in  this theorem, the unconditional results  similarly hold.
\end{remark}

\begin{remark}\rm
	The coefficient $p/n$ in Frobenius norm error in \eqref{align:exact_lr_frob} is used to calibrate  the rate effect from $\mathcal{S}\left(\alpha_X,\beta_X\right)$. For any $ X^\ast$ and $\widehat{ X}\in \mathcal{S}\left(\alpha_X,\beta_X\right)$, $\|\widehat{ X} -  X^\ast \|_F^2\le {n(\beta_X-\alpha_X)}/{p}.$
\end{remark}

\begin{remark}\label{rm:count-remark}\rm
	For technical purposes, we have imposed the entry-wise upper and lower bounds of $\alpha_X, \beta_X, \alpha_R, \beta_R$ in Theorem \ref{thm:exact_lr}. These conditions are mainly for regularizing the gradient of the likelihood function $\mathcal{L}_N$ and facilitate the follow-up analysis (particularly see \eqref{eq:to-see} and in the proof of Theorem \ref{thm:exact_lr}). In fact, the entry-wise upper and lower bounds widely appear in theoretical works for a wide range of Poisson inverse problems, especially for the ones with minimax-optimality. Examples include but not limited to Poisson sparse regression \citep[Assumption 2.1]{li2018minimax} and \citep[Assumption 2.1]{jiang2015minimax}, Poisson matrix completion \cite[Equation (10)]{cao2015poisson}, and Point autoregressive model \cite[Definition of $\mathcal{A}_s$ on Page 4]{hall2016inference}.
\end{remark}

	Conditioning on the fixed $N$, the count matrix $W$ follows a multinomial distribution: $(W_{ij}, {1\le i \le n, 1\le j \le p})\sim\text{Mult}\{N,(R_iX_{ij}^\ast, 1\le i \le n, 1\le j \le p)\}$, where  $R_i = {\nu_i}/{\sum_{j=1}^n \nu_j}$ represents the row probability and the composition $X_{ij}^\ast$ represents the column probability. Defining a probability matrix by Hadamard product (or entywise product) $\Pi = ( R 1_p^\top)\circ X^\ast$, we rewrite $ W = \sum_{k=1}^N E_k$, where $ E_k$ are independent and identically distributed copies of a Bernoulli random matrix $ E$ that satisfies $ {\rm pr}\left\{ E = e_i(n)e_j^\top(p)\right\} = \Pi_{ij}$ and the total count $N$ represents the number of copies. This product-type sampling distribution $\Pi$ is widely used in the matrix completion literature \citep{negahban2012restricted,Lafond2014,klopp2014noisy,klopp2015}.
 A key step in the proof of Theorem 1 is to bound the weighted Kullback-Leibler divergence $\sum_{i=1}^n\sum_{j=1}^pR_iX_{ij}^\ast\log ({X_{ij}^\ast}/{\widehat{X}_{ij}})$. We particularly apply  a peeling scheme by partitioning the set of all possible values of $\widehat{X}$, and then derive estimation loss upper bounds for each of these subsets based on concentration inequalities, including the matrix Bernstein inequality (Lemma A6) and an empirical process version of Hoeffding's inequality \citep[Theorem 14.2]{buhlmann2011statistics}. The techniques are related to recent works on matrix completion \citep{negahban2012restricted}, although our problem setup, method, and sampling procedure are all distinct from matrix completion.
		
Next Theorem on the minimax lower bounds shows that the upper bound in Theorem \ref{thm:exact_lr} is nearly rate-optimal.
\begin{theorem}\label{thm:lower_bound}
	Conditioning on fixed $N$, if $2 \le r\le p/2$, there exist constants $C_1$ and $C_2$ which only depend on $\alpha_X, \beta_X, \alpha_R, \beta_R$, such that
	\begin{align*}
    \inf_{\widehat{ X}} \sup_{\substack{ X^\ast \in \mathbb{B}_0(r, \alpha_X, \beta_X)\\ \alpha_R/n \le R_i \le \beta_R/n}} \frac{p}{n}  E\left(\left\|\widehat{ X} -  X^\ast \right\|_F^2\right) &\geq C_1\frac{(n+p)r}{N},\\
	\inf_{\widehat{ X}} \sup_{\substack{ X^\ast \in \mathbb{B}_0(r, \alpha_X, \beta_X)\\ \alpha_R/n \le R_i \le \beta_R/n}} \frac{1}{n}  E\left\{D( X^\ast,\widehat{ X})\right\} &\geq C_2\frac{(n+p)r}{N}.
	\end{align*}
\end{theorem}

\subsection{Theoretical property under approximate low-rank matrix assumption}
We  consider the following class of approximately low-rank composition matrices with  singular values of $ X^\ast$ belonging  to a $\ell_q$ ball, 
\begin{align*}
\mathbb{B}_q\left(\rho_q,\alpha_X,\beta_X\right)=\left\{ X\in \mathcal{S}(\alpha_X,\beta_X)\mid\sum_{i=1}^{n\land p}\left|\sigma_i\left( X\right)\right|^q\le \rho_q\right\},
\end{align*}
where $0\le q\le 1$. In particular, if $q=0$, the $\ell_0$ ball $\mathbb{B}_0\left(\rho_0,\alpha_X,\beta_X\right)$ corresponds to the set of bounded composition matrices with rank at most $\rho_0$. In general, we have the following upper bound result.
\begin{theorem}\label{thm:approx_lr}
	Assume there exist constants $\alpha_R$ and $\beta_R$ such that, for any $i\in [n]$, $\alpha_R/n\le R_i\le \beta_R/n$. The tuning parameter is selected by $\eqref{align:lambda}$. Conditioning on fixed $N$, if $N \ge (n+p)\log(n+p)$ and $N = O\left\{{\rho_qp^{q/2}(n+p)^{2+q/2}\log(n+p)}/{n^{q/2}}\right\}$, for any composition $ X^\ast\in\mathbb{B}_q\left(\rho_q,\alpha_X,\beta_X\right)$, the estimator $\widehat{ X}$ in $\eqref{align:rml}$ satisfies
	\begin{align}
	\frac{p}{n} E\left\{\|\widehat{ X}- X^\ast\|_F^2\right\} &\le C_1\frac{\rho_qp^{q/2}}{n^{q/2}}\left\{\frac{(n+p)\log(n+p)}{N}\right\}^{1-q/2},\label{align:approx_lr_frob}\\
	\frac{1}{n} E\left\{\D( X^\ast,\widehat{ X})\right\}&\le C_2\frac{\rho_qp^{q/2}}{n^{q/2}}\left\{\frac{(n+p)\log(n+p)}{N}\right\}^{1-q/2}.\label{align:approx_lr}
	\end{align}
\end{theorem}

\begin{remark}\rm
	The rates of convergence of \eqref{align:approx_lr} and \eqref{align:approx_lr_frob} reduce to the exact low rank case when $q=0$ and $\rho_0 = r$. 
\end{remark}

\subsection{Estimation of diversity index}\label{Est.index}
Various microbial diversity measures are often used to quantify the composition of  microbial communities (See, e.g., \cite{haegeman2013robust}). Given $X \in \mathbb{R}^{n\times p}$ that represents $p$-taxa compositions across $n$  individuals, two widely used measurements of microbial community diversity include
\begin{enumerate}
	\item Shannon's index\quad $ H_{\text{sh}}(X_i) = -\sum_{j=1}^pX_{ij}\log X_{ij}$, $1\le i \le n$,
	\item Simpson's index\quad $ H_{\text{sp}}(X_i) =  \sum_{j=1}^pX_{ij}^2$, $1\le i\le n$,
\end{enumerate}
where $\{H_{\text{sh}}(X_i)\}_{i=1}^n$ and $\{H_{\text{sp}}(X_i)\}_{i=1}^n$ are $n$-dimensional vectors, each component measuring the richness and evenness of microbial community in an individual. Higher value of Shannon's index, or lower value of Simpson's index, reflects more even distribution among different taxa.

We  estimate various diversity indices by plugging the proposed estimator $\widehat{ X}$ into the indices defined above.  The following Corollary provides the upper bounds of the mean squared errors of these  estimators when $ X^\ast\in\mathbb{B}_q\left(\rho_q,\alpha_X,\beta_X\right)$. 
\begin{corollary}\label{coro:diversity}
	Assume that the assumptions in Theorem \ref{thm:approx_lr} hold and the tuning parameter is selected by $\eqref{align:lambda}$. For any constants $0<\alpha_X < 1 <  \beta_X$ and $ X^\ast\in\mathbb{B}_q\left(\rho_q,\alpha_X,\beta_X\right)$, there exists some constants $C_1$, $C_2$ that do not depend on $n, N, p, r$, such that the estimate $\widehat{X}$ in $\eqref{align:rml}$ satisfies
	\begin{align}
	\frac{1}{n}\sum_{i=1}^n E\left\{ H_{\text{sh}}(\widehat{ X}_i)- H_{\text{sh}}( X_i^\ast)\right\}^2
	&\le C_1\frac{\rho_q(\log p)^2p^{q/2}}{n^{q/2}}\left\{\frac{(n+ p)\log(n+p)}{N}\right\}^{1-q/2},\label{align:shannon}\\
	\frac{1}{n}\sum_{i=1}^n E\left\{ H_{\text{sp}}(\widehat{ X}_i)- H_{\text{sp}}( X_i^\ast)\right\}^2
	&\le C_2\frac{\rho_q}{p^{2-q/2}n^{q/2}}\left\{\frac{(n+ p)\log(n+p)}{N}\right\}^{1-q/2}.\label{align:simpson}
	\end{align}
	In addition,  results for the exact low-rank case correspond to $q=0$ and $\rho_0 = r$. 
\end{corollary}

\begin{remark}\rm
	\cite{jiao2017maximum} considered the maximum likelihood estimation of functionals, particularly Shannon's and Simpson's indices, for discrete distributions. According to their results, if $R_i\in[\alpha_R/n,\beta_R/n]$ for any $i\in[n]$, conditioning on fixed $N$, we can derive the following rate of convergence for $\widehat{ X}^{\rm mle}$,
	\begin{align}
	\frac{1}{n} \sum_{i=1}^n  E\left\{ H_{sh}(\widehat{ X}_i^{\rm mle}) -  H_{sh}( X_i^{\ast})\right\}^2 &\asymp \frac{n^2p^2}{N^2} + \frac{n(\log p)^2}{N},\label{align:shannon_mle}\\
	\frac{1}{n}\sum_{i=1}^n E\left\{ H_{\text{sp}}(\widehat{ X}_i^{\rm mle})- H_{\text{sp}}( X_i^\ast)\right\}^2 &\asymp \frac{n}{N}. \label{align:simpson_mle}
	\end{align}
 \cite{wu2016minimax} studied the minimax-optimal estimation of Shannon's index. They showed that a best polynomial approximation estimator $\widehat{ X}^{\rm bpa}$ achieves the following sharper rate than the maximum likelihood estimator,
	\begin{equation*}
	\frac{1}{n}\sum_{i=1}^n E\left\{H_{sh}(\widehat{ X}^{\rm bpa} - \widehat{ X}^\ast)\right\}^2 \asymp \frac{n^2p^2}{N^2(\log p)^2} + \frac{n(\log p)^2}{N}.
	\end{equation*}
	Compared with the diversity estimation via $\widehat{X}^{\rm mle}$ or $\widehat{X}^{\rm bpa}$, our proposed diversity estimator achieves a sharper bound in the estimation error when the number of total counts increases in a small rate. In particular, when $n=p=d$ and $rank(X^\ast)=r$, the rate of convergence of Shannon's and Simpson's index provided by Corollary \ref{coro:diversity} is sharper than \eqref{align:shannon_mle} and \eqref{align:simpson_mle} respectively if the number of total counts increases under a certain rate:
	$N = O\left(d^2r\log d\vee {d^3} r^{-1}(\log d)^{-5}\right),$
	which is close to the required condition in Corollary \ref{coro:diversity}. 
\end{remark}

\section{Simulation studies}\label{sec:sim}
The numerical performances of the proposed estimator $\widehat{X}$ under various settings are evaluated by simulations. Since the Poisson-Multinomial model is equivalent to the multinomial model when the total count is fixed, the count matrix $W$ is generated as follows. 
Let $U\in \mathbb{R}^{n\times r}$ be the absolute values of an independent and identically distributed standard normal matrix. In order to simulate correlated compositional data arising from metagenomics,  let  $V= V_1+ V_2 \in \mathbb{R}^{p\times r}$, where
$$(V_1)_{ij} = \left\{\begin{array}{ll}
1, & i=j;\\
1, & i\neq j \text{  with probability 0$\cdot$3};\\
0, & i\neq j \text{  with probability 0$\cdot$7},
\end{array}\right. \quad (V_2)_{ij} {\sim} N(0, 10^{-3}).$$
The true composition matrix is generated as  $X_{ij}^\ast=Z_{ij}/\sum_{k=1}^pZ_{ik}$, where   $Z = UV^\top$.  Since this procedure may produce non-positive values in $X^\ast$ by a small chance, this is repeated until a positive matrix $X^\ast$ is generated. In order to account for the heterogeneity of total count across different samples, we generate $R_i = P_i/\sum_{k=1}^n P_k$ with $P_i\sim\text{Uniform}[1, 10]$ for each individual $i\in[n]$. Based on $R_i$ and $X^\ast$, the read counts $W$ are generated from the multinomial model, i.e. $W_i^\ast \sim \text{Mult}(N_i; X_i^\ast)$, where $N_i = \gamma np R_i$, $\gamma = 1,2,3,4,5$. The sample size and the number of taxa are set as  $n=100$, $p \in \{50,100, 200\}$ and  $r = 20$ (low rank model), or $r = n\wedge p$ (full rank model). These parameters are chosen to mimic the data dimensions of typical microbiome studies.

 The proposed nuclear norm regularized maximum likelihood estimator $\widehat{X}$ is applied to recover $X^\ast$. The simulations are repeated 100 times, and the tuning parameters $(\lambda,\alpha_X)$ are selected based on the data-driven procedure.  The estimation performances are evaluated by the means of average loss in squared Frobenius norm $\|\widehat{X}-X^\ast\|_F^2$, average Kullback-Leibler divergence $n^{-1}\D(X^\ast,\widehat{X})$ and the mean squared errors for the estimates of Shannon's and Simpson's indices. The results are compared with the standard  zero replacement estimator $\widehat{X}$ (in both exact low-rank and full rank models) and the singular value thresholding estimator $\widehat{X}^{\rm svt}$ in \eqref{eq:SVT-2} (only in exact low-rank model due to the difficulties in selecting $r$). 

The results are summarized in Tables \ref{tb:simulation-low-rank} and \ref{tb:simulation-full-rank} for the low rank and full rank compositional matrix, respectively. The proposed estimator $\widehat{X}$ outperforms the zero-replacement estimator $\widehat{X}^{\rm zr}$ and singular value thresholding estimator $\widehat{X}^{\rm svt}$ in almost all the settings. Particularly, the diversity index estimates based on the proposed estimator uniformly outperform other methods by a large margin. These results are consistent across  different model dimensions even when $X^\ast$ is full rank. In addition, the difference between the loss of $\widehat{X}$ and $\widehat{X}^{\rm zr}$ becomes more significant for smaller $\gamma$, i.e. when the number of total read counts is small. Therefore, our method enjoys greater improvement than the traditional methods especially when the sequencing depth is limited, which is exactly the purpose of our study.

\begin{table}[h]
	\def~{\hphantom{10}}
	\caption{Means of various performance measures for $\widehat{X}$, $\widehat{X}^{\rm zr}$ and $\widehat{X}^{\rm svt}$ in the low rank model over 100 replications.}
	\begin{center}
		{%
			\begin{tabular}{@{}lrrrrrrrrr@{}}
				& \multicolumn{3}{c}{$p=50$} & \multicolumn{3}{c}{$p=100$}& \multicolumn{3}{c}{$p=200$}\\[5pt]
				$\gamma$ & $\widehat{X}$ & $\widehat{X}^{\rm zr}$ & $\widehat{X}^{\rm svt}$ & $\widehat{X}$ & $\widehat{X}^{\rm zr}$ & $\widehat{X}^{\rm svt}$& $\widehat{X}$ & $\widehat{X}^{\rm zr}$ & $\widehat{X}^{\rm svt}$\\[5pt]
				\multicolumn{10}{c}{Squared Frobenius norm error {\upshape $\left( \times 10^{-2} \right)$}}\\
				1  & 40.70 & 95.01 & 84.33 & 26.74 & 68.65 & 55.20 & 19.00 & 48.98 & 36.24\\
				2  & 35.08 & 87.79 & 75.11 & 26.49 & 63.29 & 48.97 &	18.38 & 44.63 & 30.10\\
				3  & 35.37 &	80.91 & 67.98 & 27.77 & 56.80 &	42.22 & 18.99 & 40.62 & 25.72\\
				4  & 37.09 & 74.25 & 61.75 & 24.65 & 53.11 & 37.44 &	18.24 & 36.87 & 22.92\\
				5  & 36.47 & 67.76 & 55.81 & 25.22 & 49.20 & 34.03 &	18.22 & 34.26 & 20.54\\[5pt]
				\multicolumn{10}{c}{Average Kullback-Leibler divergence {\upshape $\left( \times 10^{-2} \right)$}}\\
				1  & 4.31 & 19.04 & 16.03 & 3.77 &	 19.68 & 14.16 &	3.82 &	20.02 & 12.73\\
				2  & 3.25 & 18.65 & 14.73 &	3.78 &	 19.15 & 12.47 &	3.53 &	18.90 &  9.87\\
				3  & 3.33 & 16.47 & 12.36 &	4.04 &	 16.56 &  9.80 &	3.82 &	16.59 &  7.31\\
				4  & 3.62 & 14.41 & 10.45 &	3.27 &	 14.59 &  7.71 &	3.51 &	14.39 &  5.85\\
				5  & 3.58 & 12.36 &	8.69 &	3.40 &	 12.77 &  6.42 &	3.54 &	12.57 &  4.69\\[5pt]
				\multicolumn{10}{c}{Shannon index mean squared errors {\upshape $\left( \times 10^{-3} \right)$}}\\
				1  & 3.92 & 19.90 & 13.50 &	3.25 &	19.82 & 8.88 & 3.00 & 18.63 & 6.35\\
				2  & 2.21 & 23.31 & 12.52 &	4.19 &	18.34 & 5.99 & 2.25 & 21.67 & 4.24\\
				3  & 2.66 & 18.96 &	8.78 &	3.56 &	16.91 & 4.07 & 2.54 & 17.84 & 2.18\\
				4  & 2.83 & 14.72 &	6.03 &	2.31 &	14.65 & 2.67 & 2.00 & 14.41 & 1.43\\
				5  & 2.60 & 11.78 &	4.58 &	2.20 &	12.46 & 1.95 & 2.12 & 11.60 & 0.89\\[5pt]
				\multicolumn{10}{c}{Simpson index mean squared errors {\upshape $\left( \times 10^{-6} \right)$}}\\
				1  & 5.93 & 55.73 & 35.62 & 1.21 & 14.27 & 5.71 & 0.28 & 3.45 & 0.94\\
				2  & 3.25 & 51.24 & 25.20 & 1.50 & 11.70 & 3.57 & 0.21 & 3.10 & 0.47\\
				3  & 3.94 & 40.86 & 17.77 & 1.35 &  9.12 & 2.01 & 0.23 & 2.40 & 0.24\\
				4  & 4.12 & 31.49 & 12.31 & 0.82 &  7.87 & 1.31 & 0.18 & 1.81 & 0.15\\
				5  & 3.68 & 23.85 &  8.96 & 0.77 &  6.36 & 0.89 & 0.19 & 1.45 & 0.10\\
		\end{tabular}}\label{tb:simulation-low-rank}
	\end{center}	
$\widehat{X}$: proposed estimator;
$\widehat{X}^{\rm zr}$: zero-replacement estimator;
$\widehat{X}^{\rm svt}$: singular value thresholding estimator.  

\end{table}

\begin{table}[h]
	\caption{Means of various performance measures for $\widehat{X}$ and $\widehat{X}^{\rm zr}$ in the full rank model over 100 replications.}
	\begin{center}
		{%
			\begin{tabular}{@{}lrrrrrr@{}}
				& \multicolumn{2}{c}{ $p=50$} & \multicolumn{2}{c}{ $p=100$}& \multicolumn{2}{c}{ $p=200$}\\[5pt]
				$\gamma$ & $\widehat{X}$ & $\widehat{X}^{\rm zr}$ & $\widehat{X}$ & $\widehat{X}^{\rm zr}$ & $\widehat{X}$ & $\widehat{X}^{\rm zr}$ \\[5pt]
				\multicolumn{7}{c}{Squared Frobenius norm error \upshape $\left( \times 10^{-2} \right)$}\\
				1   & 26.60  & 94.74  & 14.91  & 65.64  & 10.42  & 46.65\\
				2   & 25.60  & 87.48  & 14.35  & 62.72  &	9.82  & 44.13\\
				3   & 25.09  & 80.43  & 13.24  & 57.11  &	9.47  & 40.50\\
				4   & 24.12  & 73.73  & 12.64  & 52.58  &	8.79  & 37.15\\
				5   & 23.59  & 67.64  & 12.41  & 49.11  &	8.65  & 34.73\\[5pt]
				\multicolumn{7}{c}{Average Kullback-Leibler divergence \upshape $\left( \times 10^{-2} \right)$}\\
				1   &  1.82  & 18.67  &  1.14  & 18.04  &  1.11  & 18.17\\
				2   &  1.71  & 18.46  &  1.07  & 19.51  &	0.99  & 19.20\\
				3   &  1.66  & 16.58  &  0.90  & 17.03  &	0.91  & 16.71\\
				4   &  1.52  & 14.44  &  0.82  & 14.97  &	0.78  & 14.99\\
				5   &  1.43  & 12.66  &  0.78  & 13.09  &	0.76  & 13.23\\[5pt]
				\multicolumn{7}{c}{Shannon index mean squared errors \upshape $\left( \times 10^{-3} \right)$}\\
				1   &  0.66  & 29.13  &	0.15  & 31.15  &	0.12  & 31.04\\
				2   &  0.55  & 27.22  &	0.16  & 31.35  &	0.13  & 29.97\\
				3   &  0.80  & 21.03  &	0.17  & 23.89  &	0.14  & 23.66\\
				4   &  0.65  & 17.20  &	0.15  & 19.20  &	0.08  & 19.22\\
				5   &  0.47  & 14.49  &	0.13  & 15.90  &	0.08  & 16.00\\[5pt]
				\multicolumn{7}{c}{Simpson index mean squared errors \upshape $\left( \times 10^{-6} \right)$}\\
				1   &  1.02  & 71.18  &	0.06  & 17.77  &	0.01  &  4.43\\
				2   &  0.83  & 56.82  &	0.06  & 15.40  &	0.01  &  3.70\\
				3   &  1.14  & 42.53  &	0.06  & 11.31  &	0.01  &  2.88\\
				4   &  0.96  & 33.12  &	0.06  &  8.72  &	0.01  &  2.17\\
				5   &  0.58  & 25.82  &	0.05  &  7.21  &	0.01  &  1.78\\
		\end{tabular}}\label{tb:simulation-full-rank}
	\end{center}
$\widehat{X}$: proposed estimator;  
$\widehat{X}^{\rm zr}$: zero-replacement estimator.  \end{table}

To further compare the resulting estimates, Figure  \ref{fig:scatter-fr} shows the scatter plot between true composition matrix $X^\ast$ versus estimated composition matrix $\widehat{X}$ for a randomly chosen  simulated data set  in the low rank setting  with  $p=200$, and  $\gamma=1$ and $\gamma=5$, respectively.  Although $\widehat{X}$ is slightly biased due to nuclear norm penalty in the estimation,  
it still greatly  outperforms the commonly used zero-replacement estimator $\widehat{X}^{\rm zr}$. Estimates from the singular value thresholding are not compared since it can result in negative estimates.

\ignore{
\begin{figure}[h]
	\centering\small
	\begin{tabular}{@{}l@{\quad}l@{\quad}l@{}}
		\hskip.45in(a) & \hskip.32in(b) & \hskip.32in(c)\\
		\includegraphics[width=0.32\textwidth]{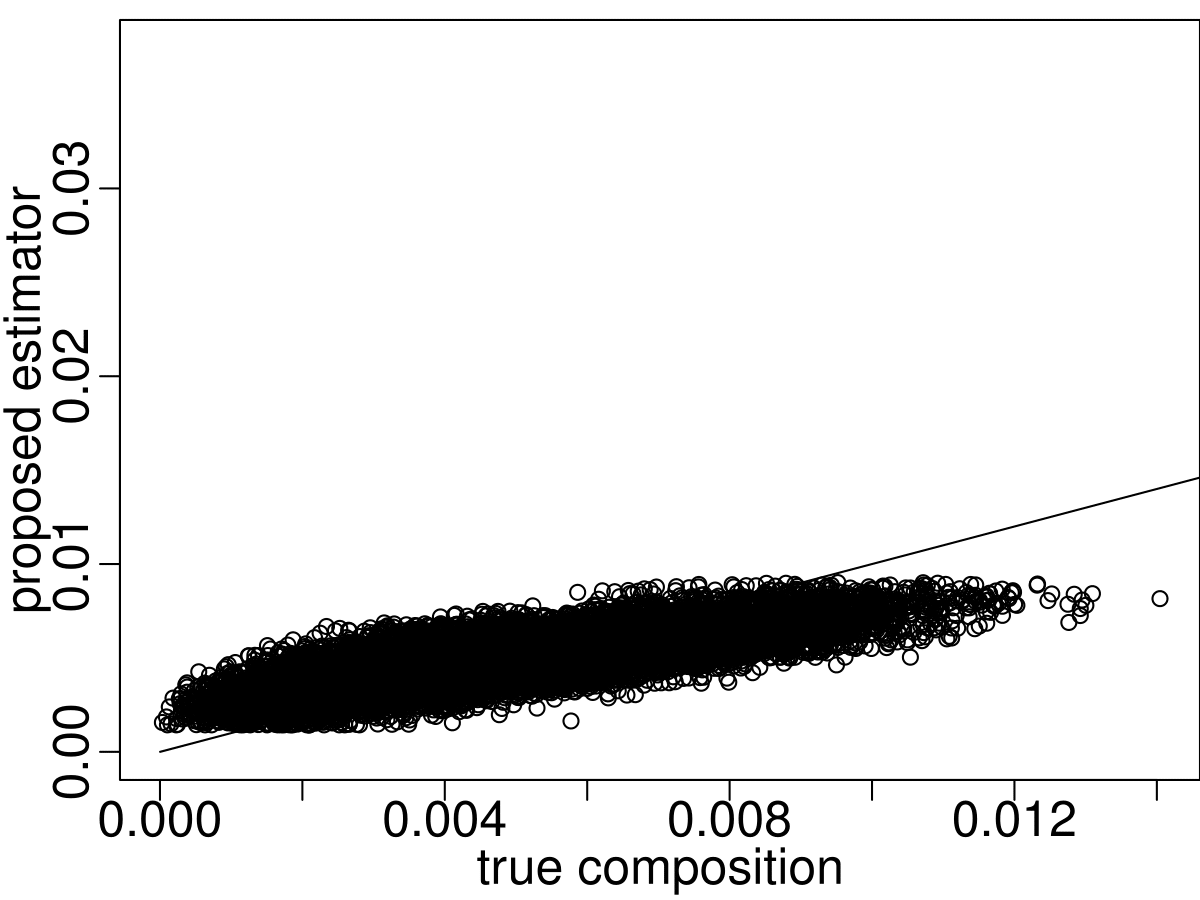} & \includegraphics[width=0.32\textwidth]{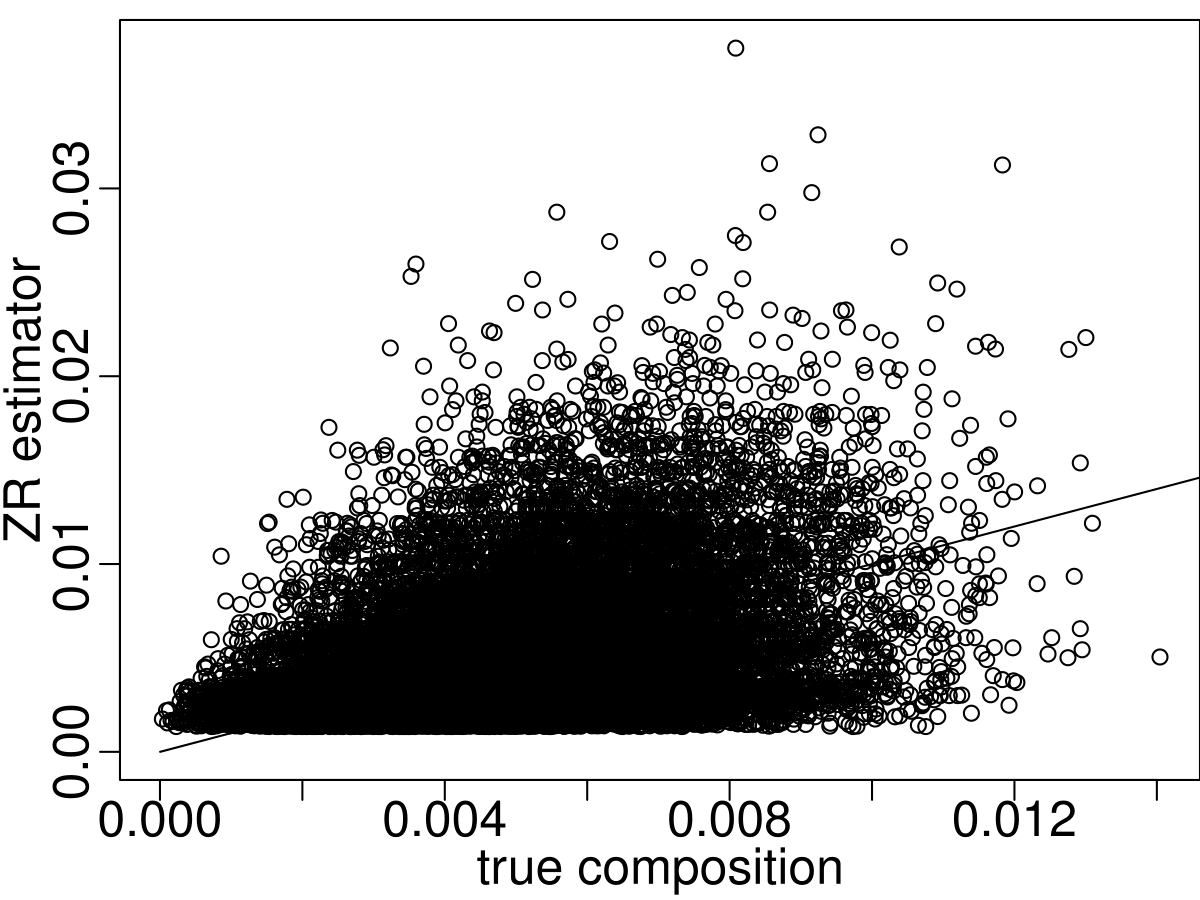} &\includegraphics[width=0.32\textwidth]{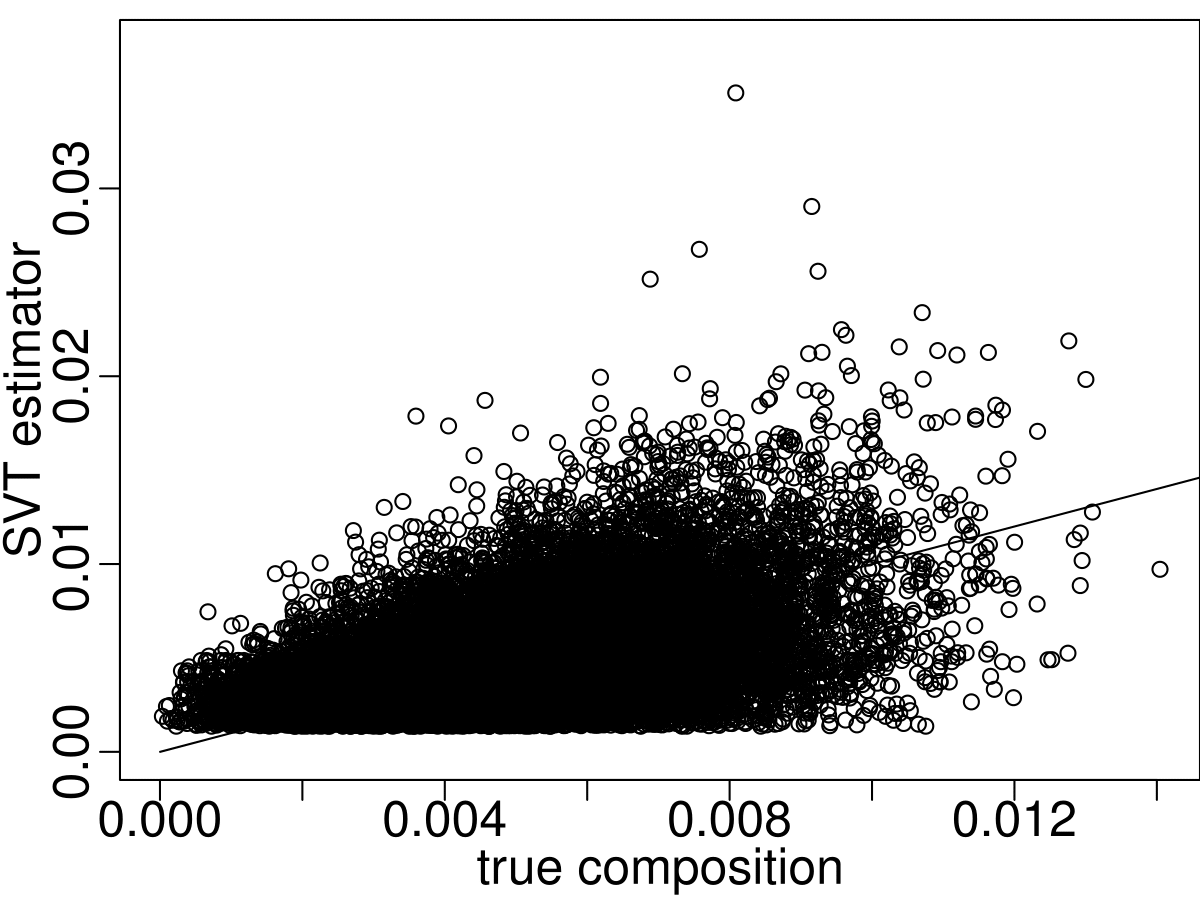} \\
		\hskip.45in(d) & \hskip.32in(e) & \hskip.32in(f)\\
		\includegraphics[width=0.32\textwidth]{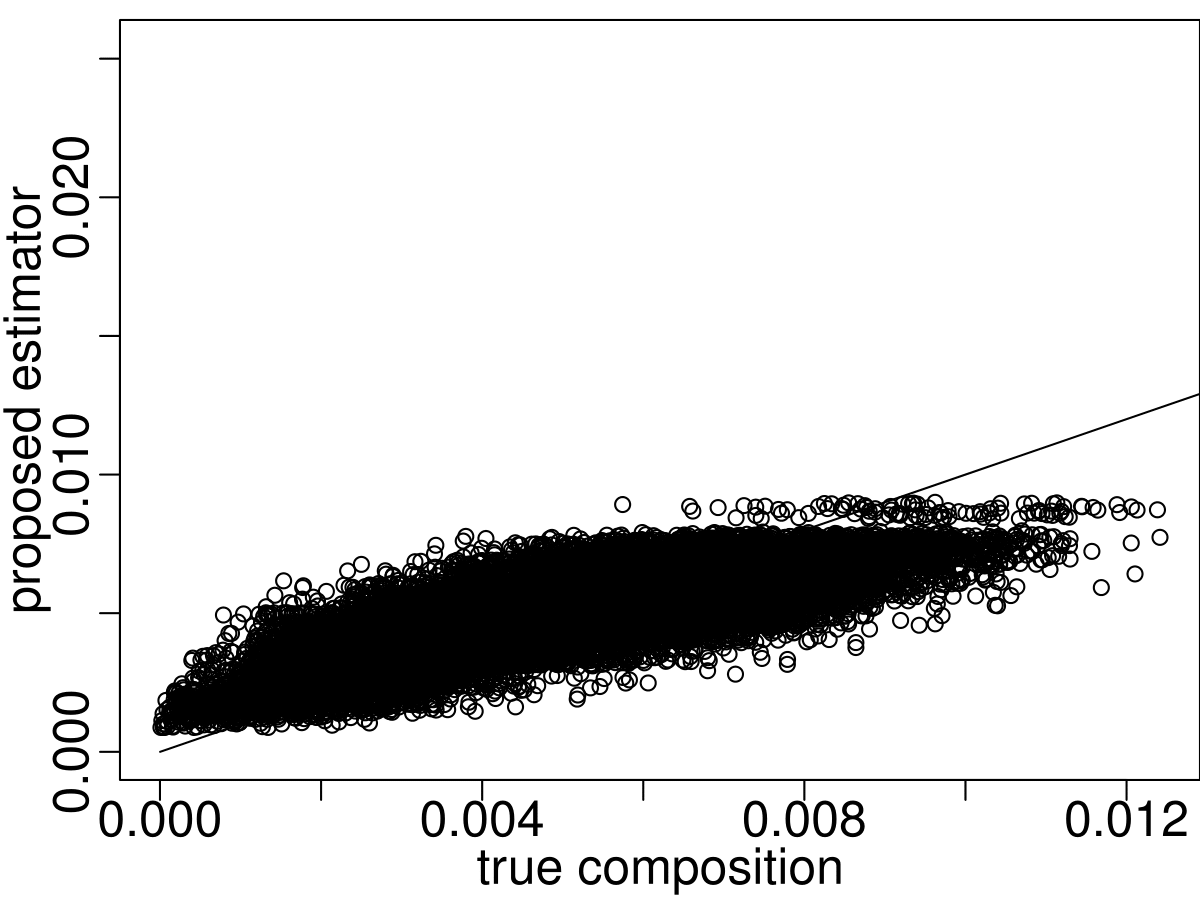} & \includegraphics[width=0.32\textwidth]{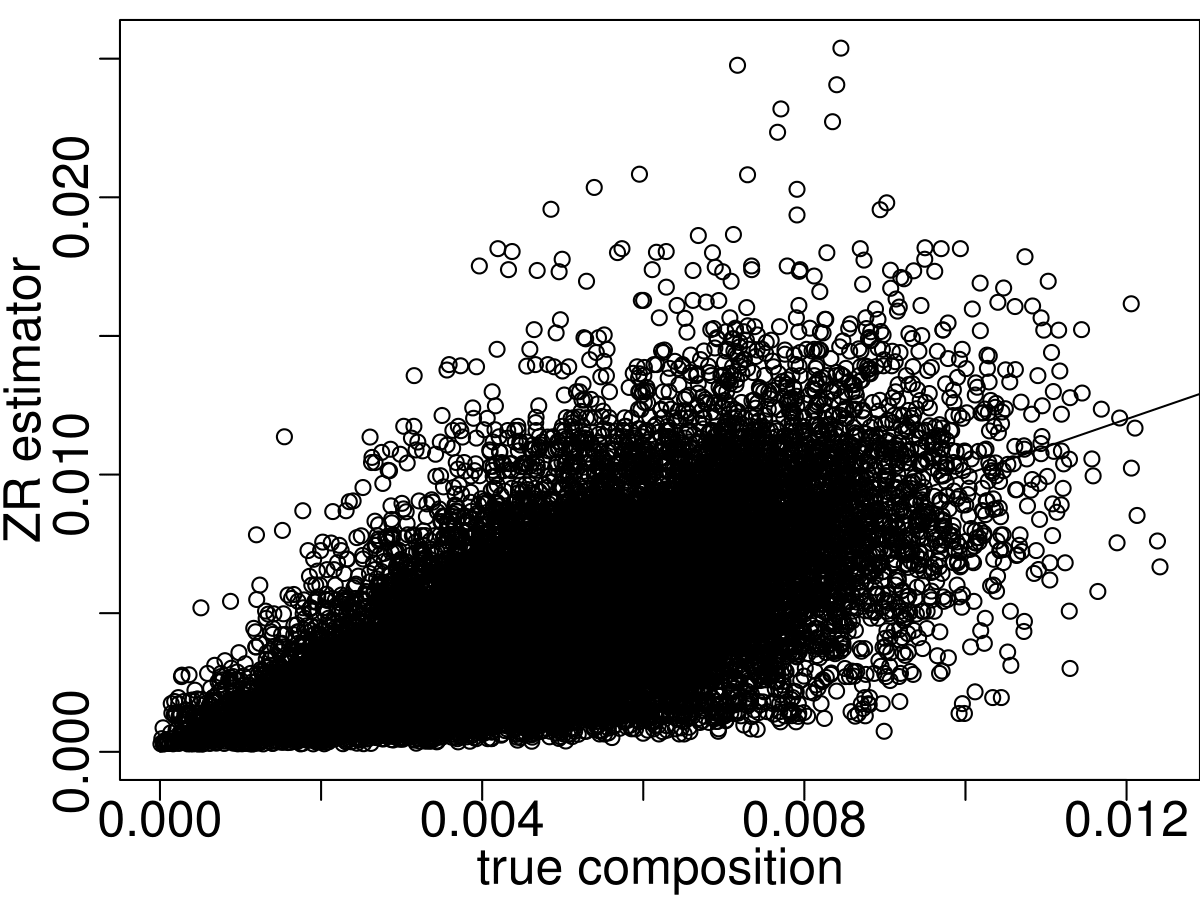} &\includegraphics[width=0.32\textwidth]{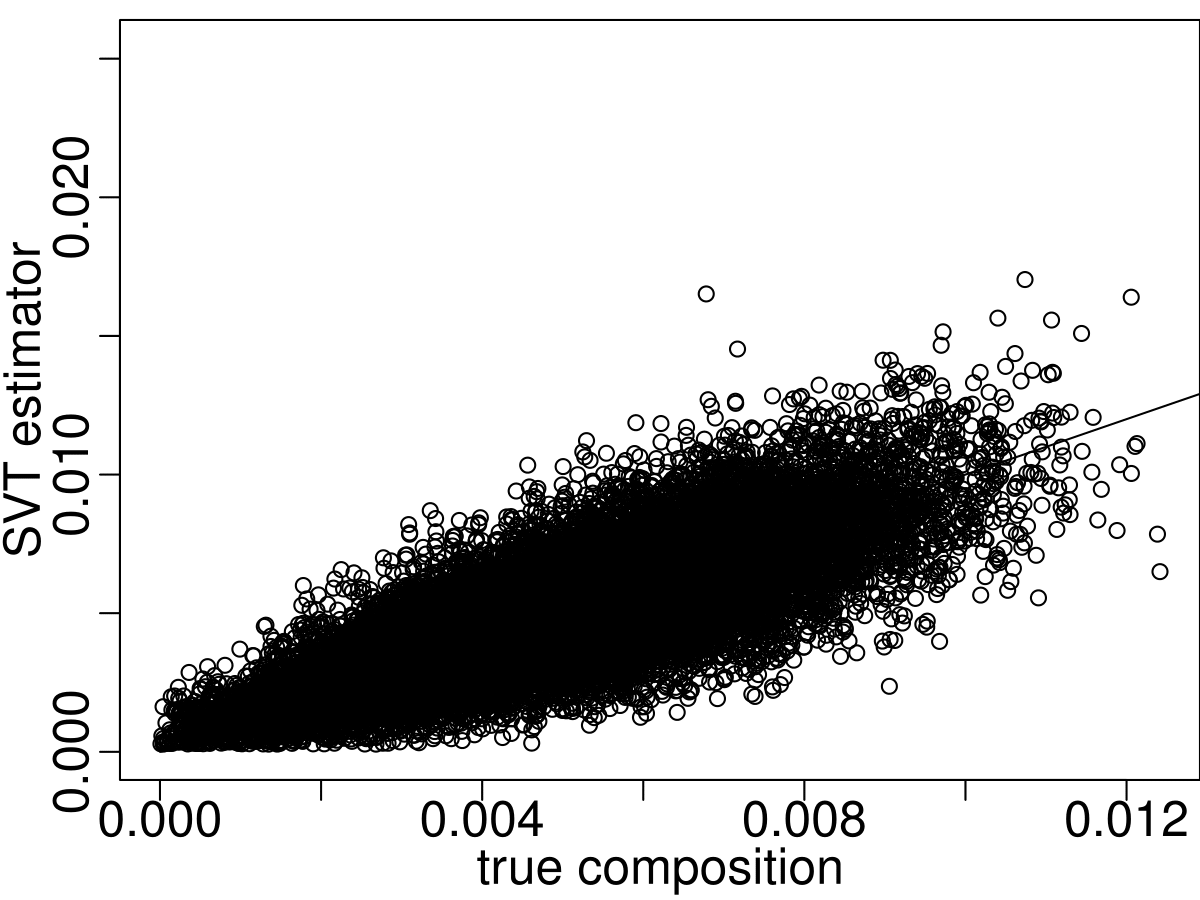}
	\end{tabular}
	\caption{Scatter plot between true composition and estimated composition in the low rank model with $p=200$. Left: the proposed estimator $\widehat{X}$; Middle: ZR estimator $\widehat{X}^{\rm zr}$; Right: singular value thresholding estimator $\widehat{X}^{\rm svt}$. Top: $\gamma = 1$; Bottom: $\gamma = 5$. Line: $y=x$.}\label{fig:scatter-lr}

}
\begin{figure}[h]	
	\begin{center}\begin{tabular}{cc}
		(a) & (b)\\
		\includegraphics[width=0.48\textwidth]{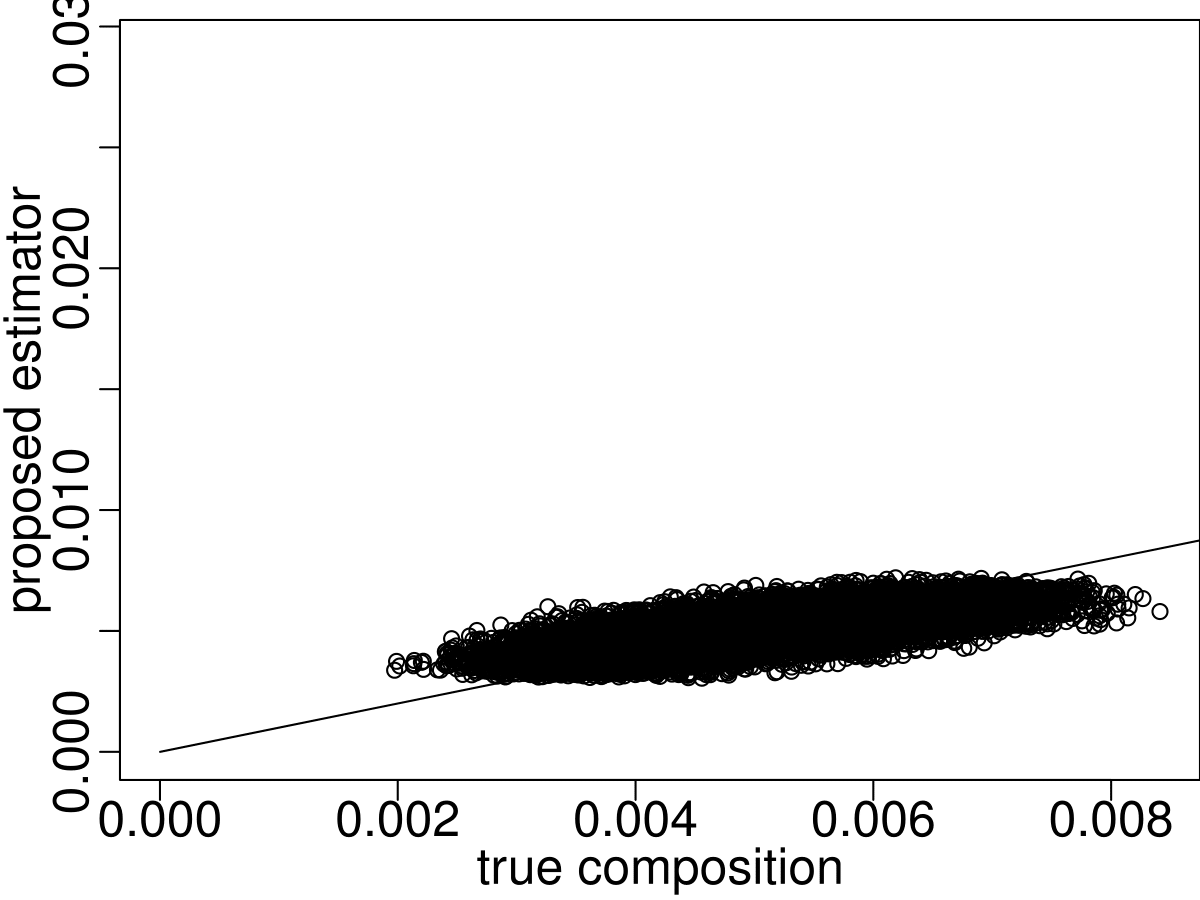} & \includegraphics[width=0.48\textwidth]{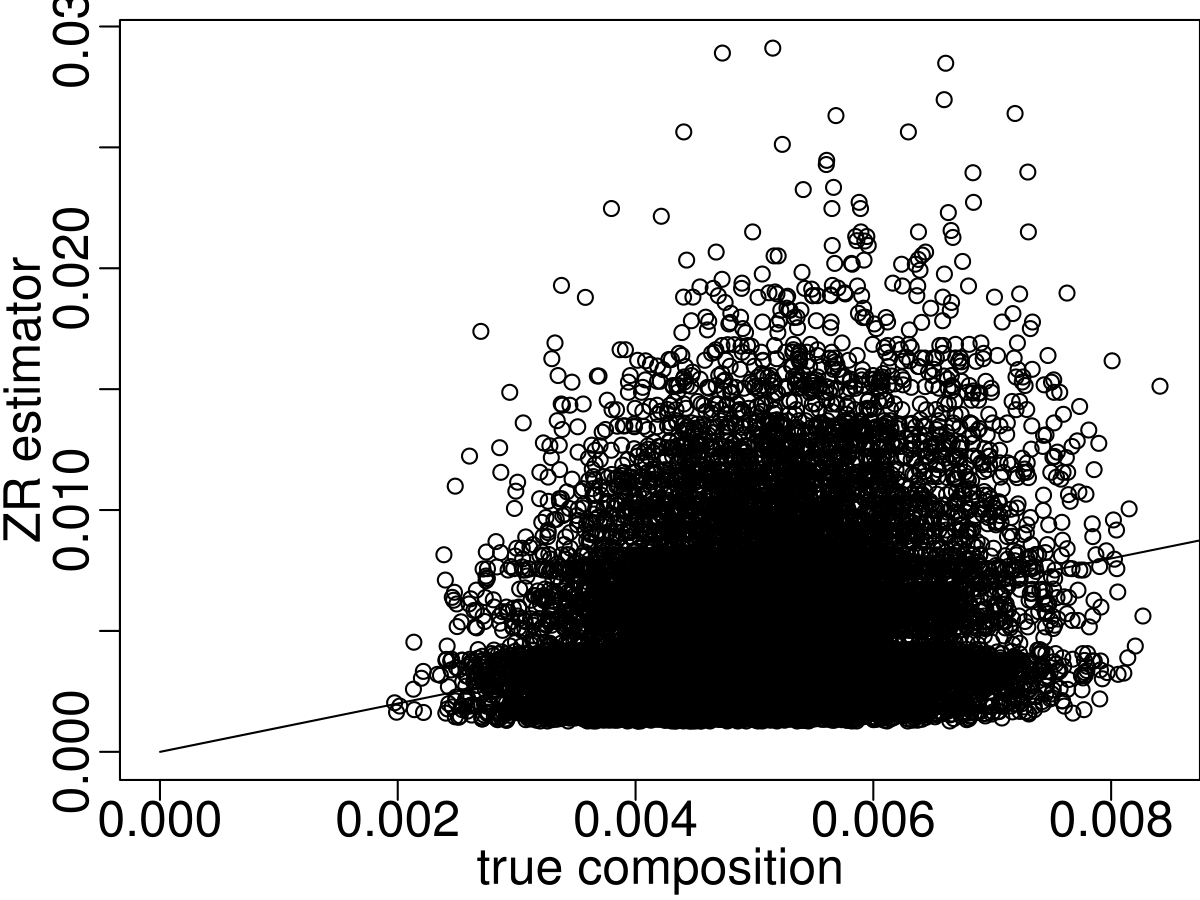} \\
		(c) &(d)\\
		\includegraphics[width=0.48\textwidth]{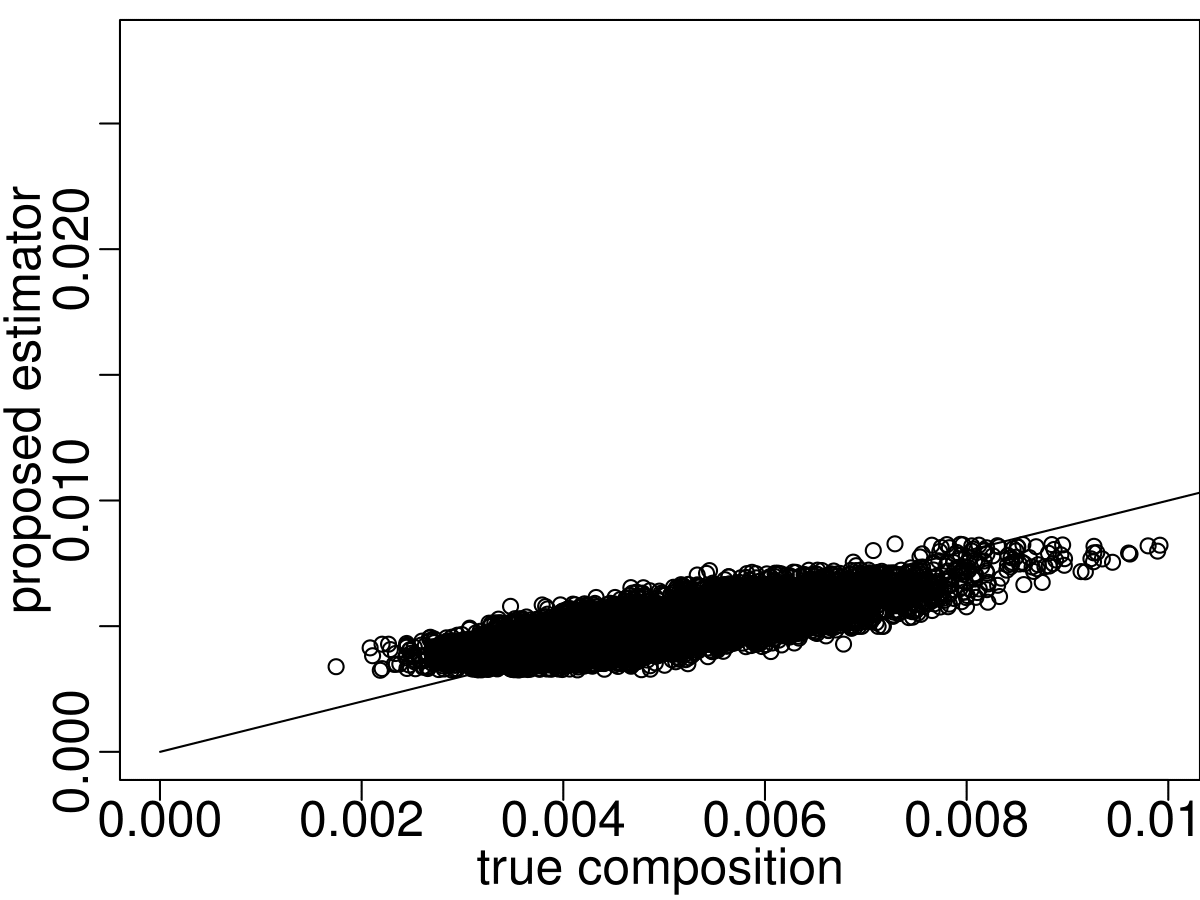} & \includegraphics[width=0.48\textwidth]{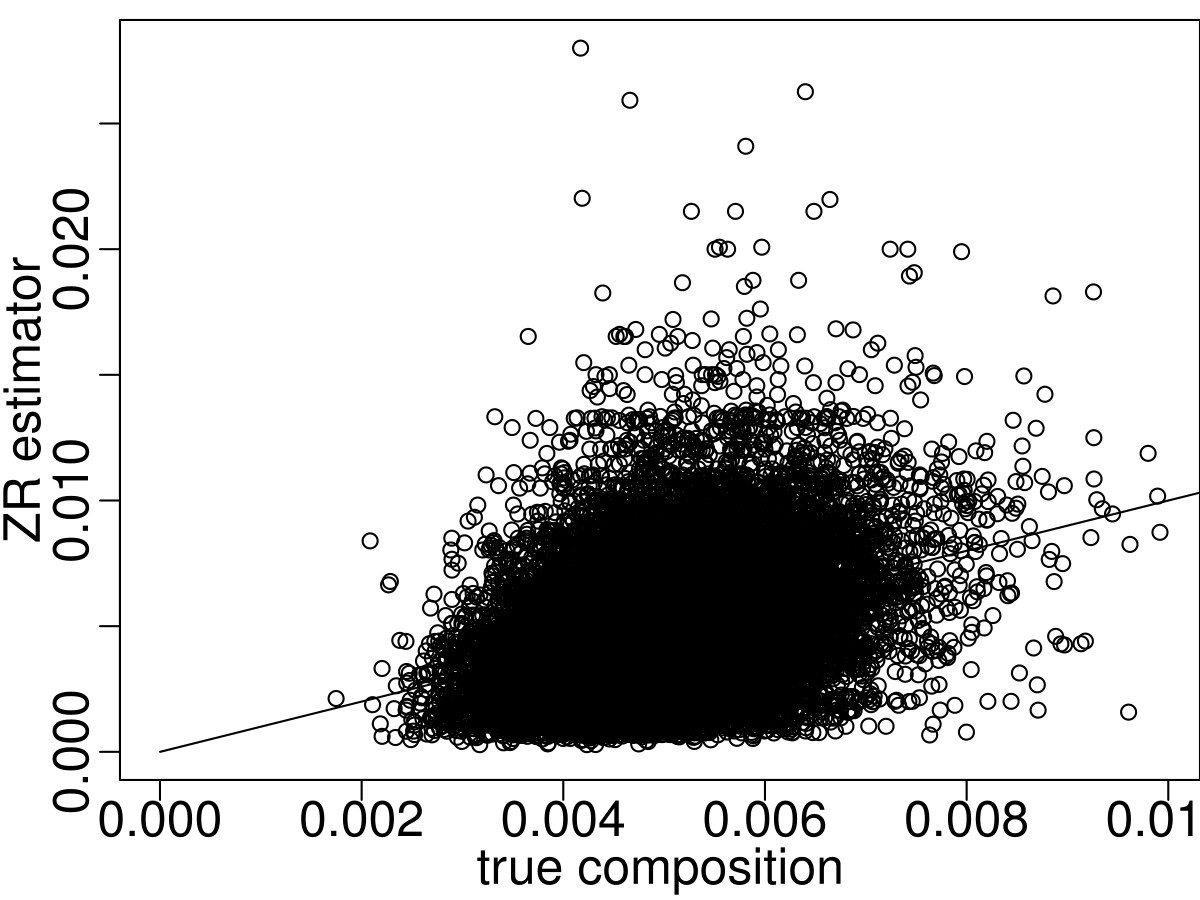}
	\end{tabular}
	\caption{Scatter plot between true composition and estimated composition in the full rank model with $p=200$ for different sequencing depths.  (a) and (c): the proposed estimator $\widehat{X}$; (b) and (d): zero-replacement  estimator. Top: $\gamma = 1$; Bottom: $\gamma = 5$. Line: $y=x$.}\label{fig:scatter-fr}
\end{center}\end{figure}

\section{Gut Microbiome Data Analysis}\label{sec:combo}
The gut microbiome plays an important role in regulating metabolic functions and  influences  human health and disease \citep{Huma:fram:2012}. We apply the proposed method to the study of \underline{C}ross-sectional study \underline{O}f diet and stool \underline{M}icro\underline{B}i\underline{O}me composition  \citep{Wu:Chen:Hoff:Bitt:Chen:Keil:link:2011}. In this study, DNAs from stool samples of 98 healthy volunteers were analyzed by 454/Roche pyrosequencing of 16S rRNA gene segments and yielded an average of 9265 reads per sample, with a standard deviation of 386, which led to identification of 3068 operational taxonomic units and 87 bacterial genera that were presented in at least one sample. Figure \ref{fig:depth-singular} (a)-(c) show the proportion of zero counts versus  total number of sequencing reads for each sample. It is clear that the samples with a smaller number of read counts often produced more zeros in the genus counts, indicating that many observed zeros are likely due to under-sampling. It is therefore reasonable to assume that the true compositions of these rare genera are positive.  Figure \ref{fig:depth-singular} (d)  shows the decay of singular values of $\widehat{X}^{\rm mle}$,  indicating an approximate low-rank composition matrix.

\begin{figure}
	\begin{center}\begin{tabular}{cc}
		(a) & (b)\\
		\includegraphics[width=0.48\textwidth]{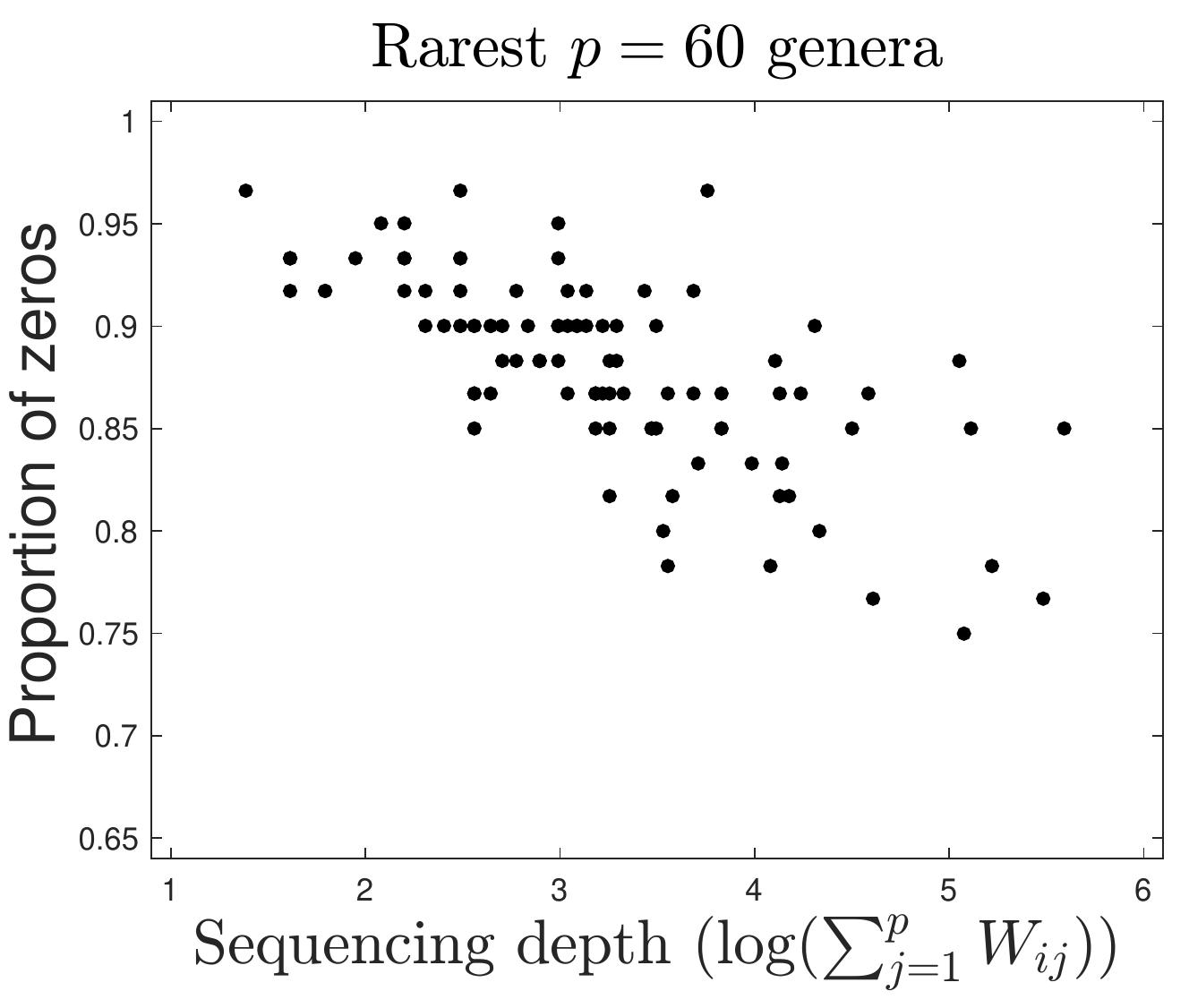} & \includegraphics[width=0.48\textwidth]{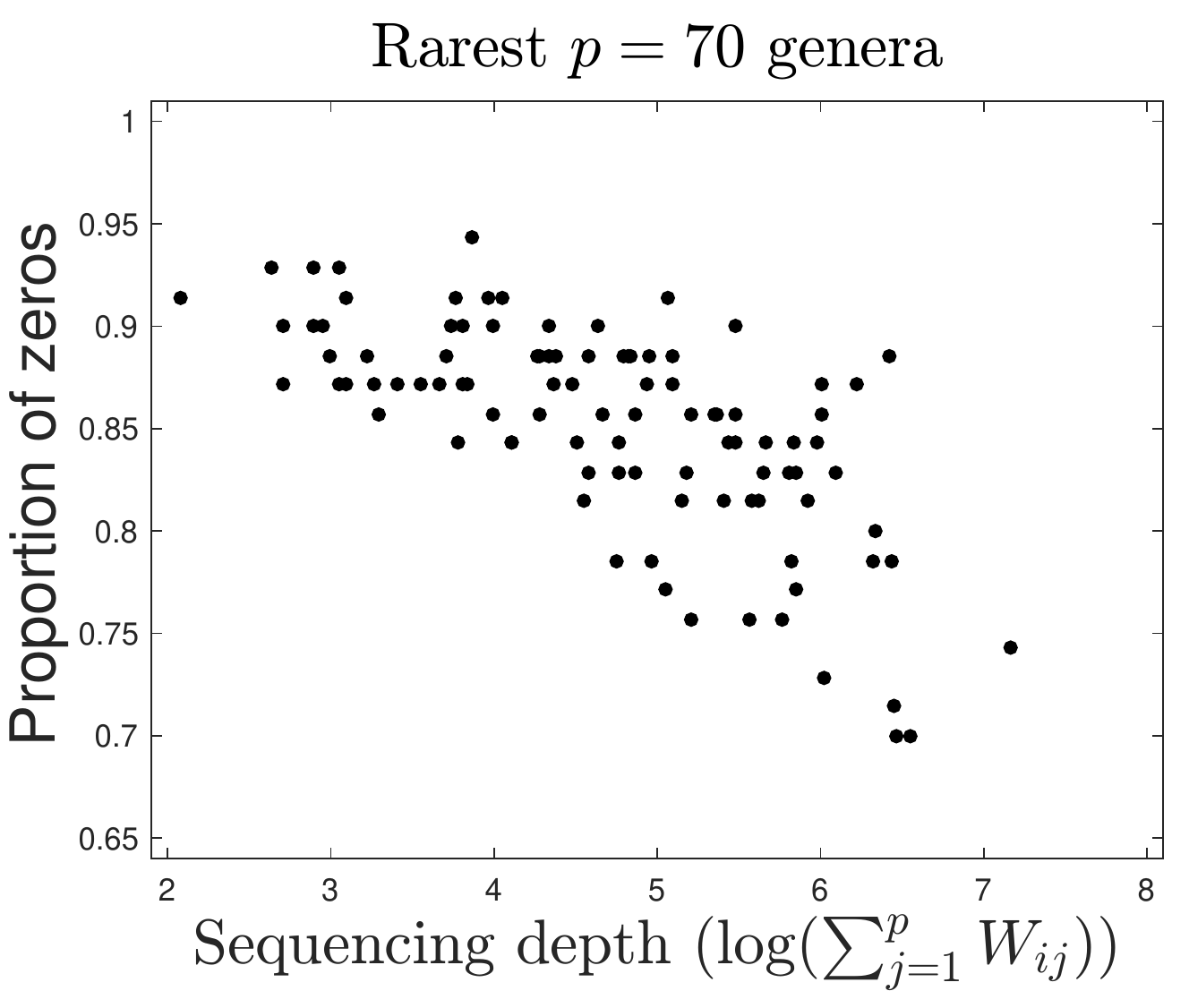}\\
		(c) & (d)\\
		\includegraphics[width=0.48\textwidth]{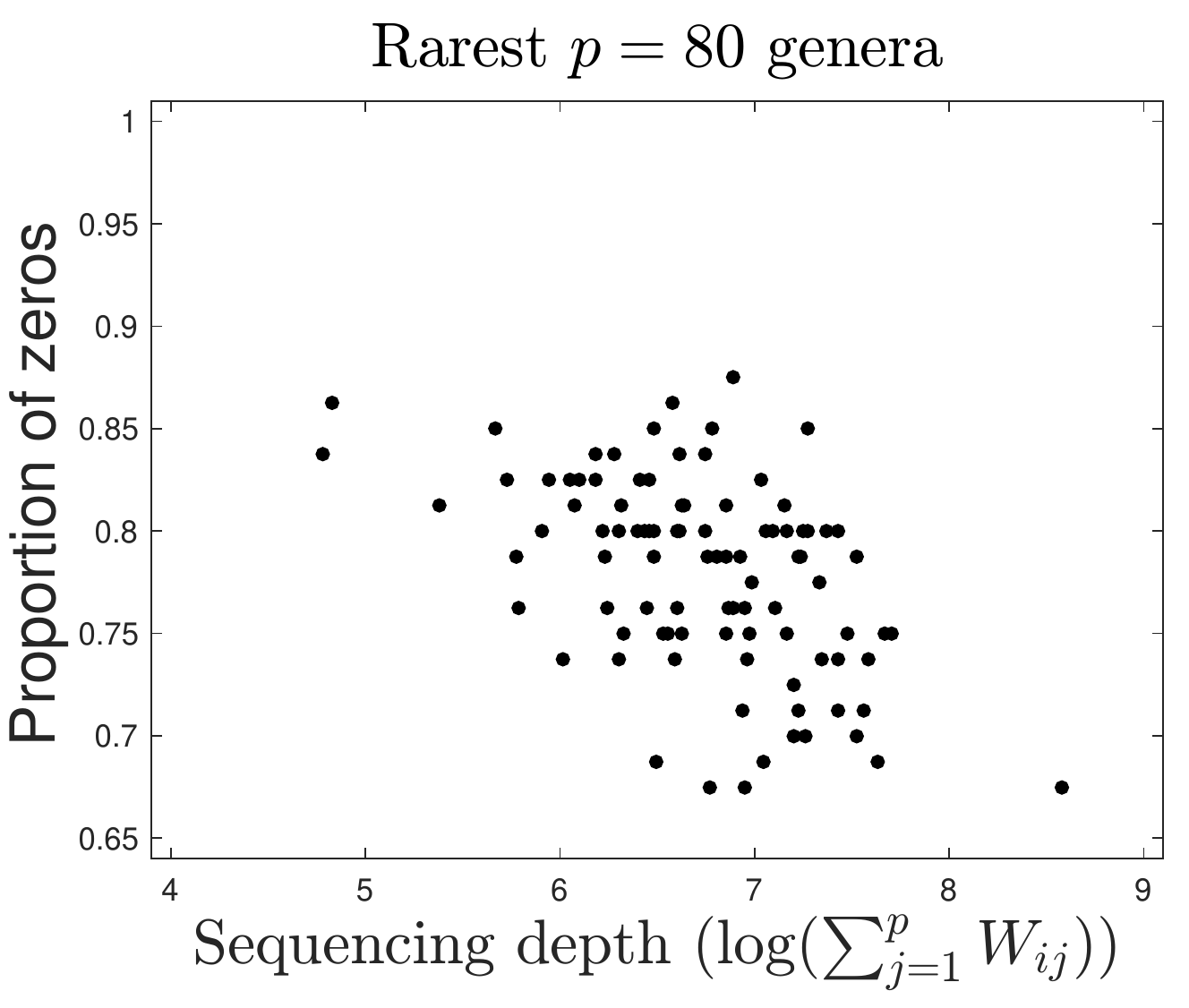} & \includegraphics[width=0.48\textwidth]{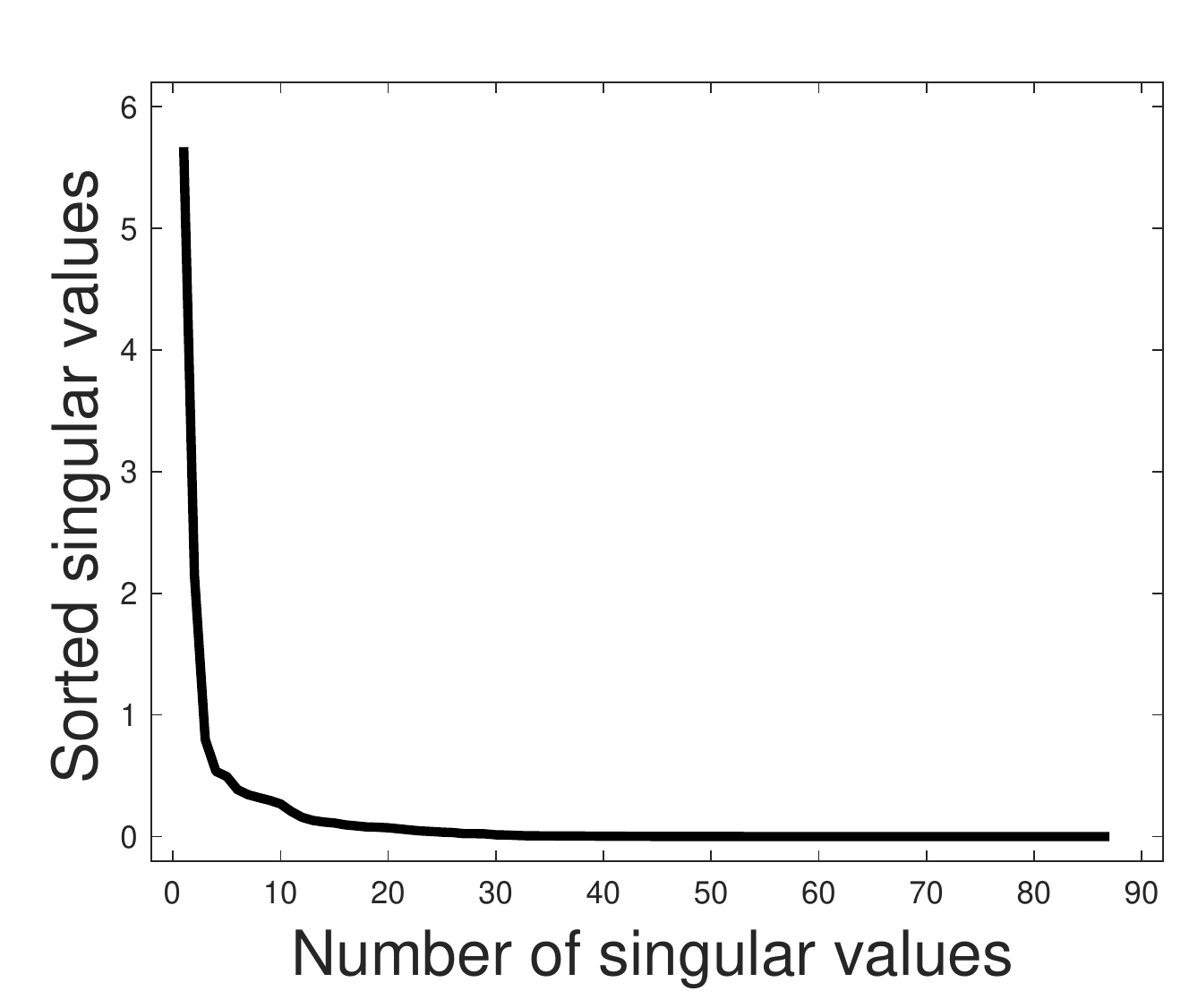}\\
	\end{tabular}
	\caption{Analysis of gut microbiome data set. (a) - (c): proportion of zero count versus the total reads count for each individual, indicating that many observed zeros are due to under-sampling. (d):  decay of singular values $d_{ii}$ based on the singular value decomposition of $\widehat{X}^{\rm mle} =  U D V^\top$, indicating the low-rank structure of the compositional matrix. }\label{fig:depth-singular}
\end{center}\end{figure}

 The proposed regularized maximum likelihood estimator $\widehat{X}$ is applied to the count matrix of these $p=87$ bacterial genera 
over $n=98$ samples. As a comparison, the traditional zero-replacement estimator $\widehat{X}^{\rm zr}$ is also calculated and compared. To compare the results,  define $\Omega=\{(i,j)\in[n]\times[p]\mid W_{ij} > 0\}$ and $\Omega^c$ as the support and the zero count indices set of $W$, respectively. The top panel of Figure \ref{fig:boxplot} shows the boxplots of the estimated compositions $\widehat{X}$ excluding three common genera Bacteroides, Blautia and Roseburia that have been observed in all individuals. For $\widehat{X}$, the observed non-zero compositions have an effect on estimating the compositions of the genera that were observed as zeros. The estimated compositions of $\widehat{X}$ in $\Omega^c$ tend to be shrank towards those in $\Omega$. In contrast, the zero replacement estimator $\widehat{X}^{\rm zr}$ (bottom panel, Figure \ref{fig:boxplot}) provides almost the same estimates for all samples/taxa in $\Omega^c$, and $\{W_{ij}\}_{(i,j)\in \Omega}$, i.e. the non-zero counts, have little effects on $\{\widehat{X}_{ij}^{\rm zr}\}_{(i,j)\in \Omega^c}$. 

\begin{figure}
	\centering
	\begin{tabular}{c}
		\includegraphics[height=0.3\textheight]{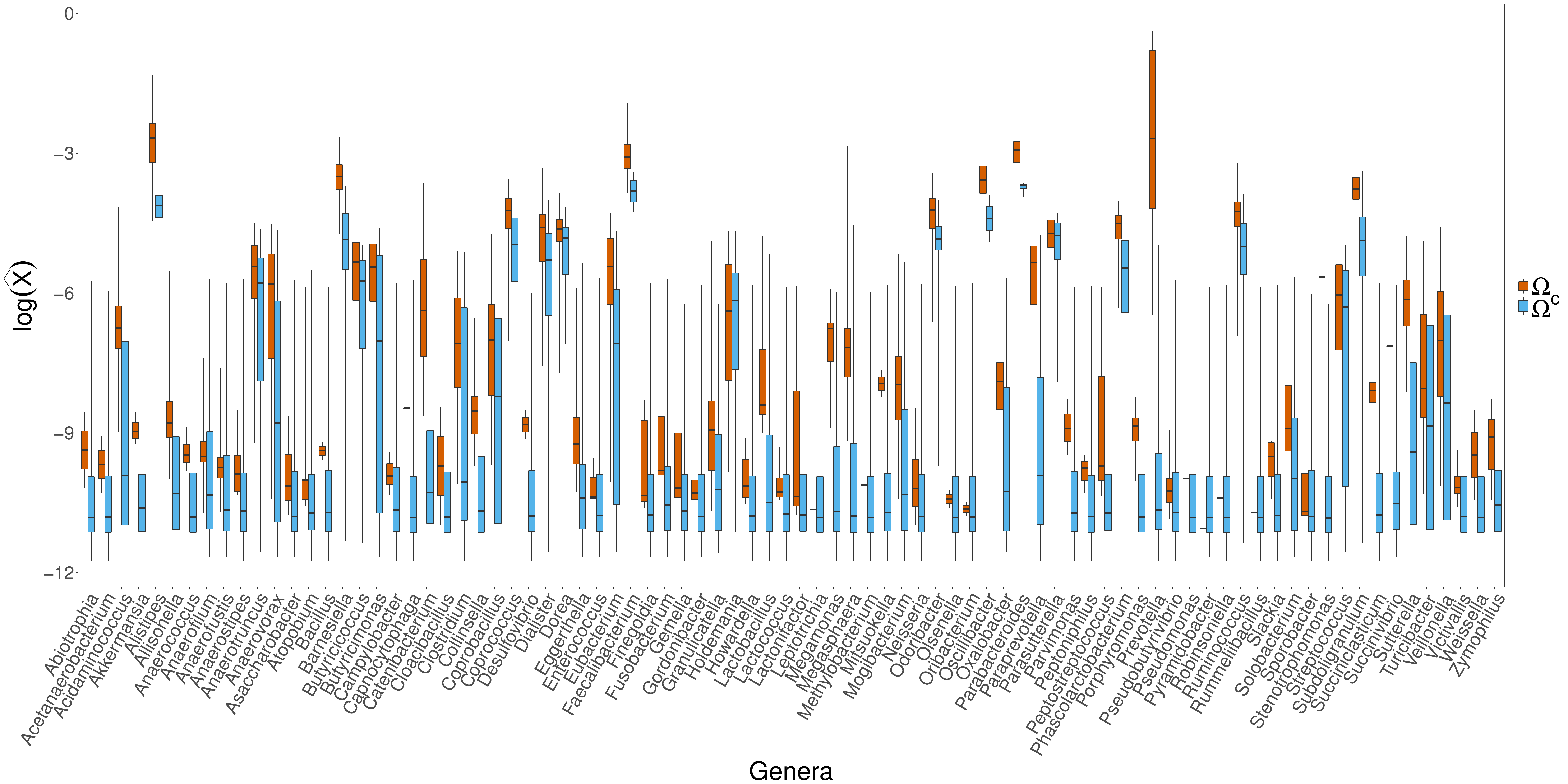}\\
		\includegraphics[height=0.3\textheight]{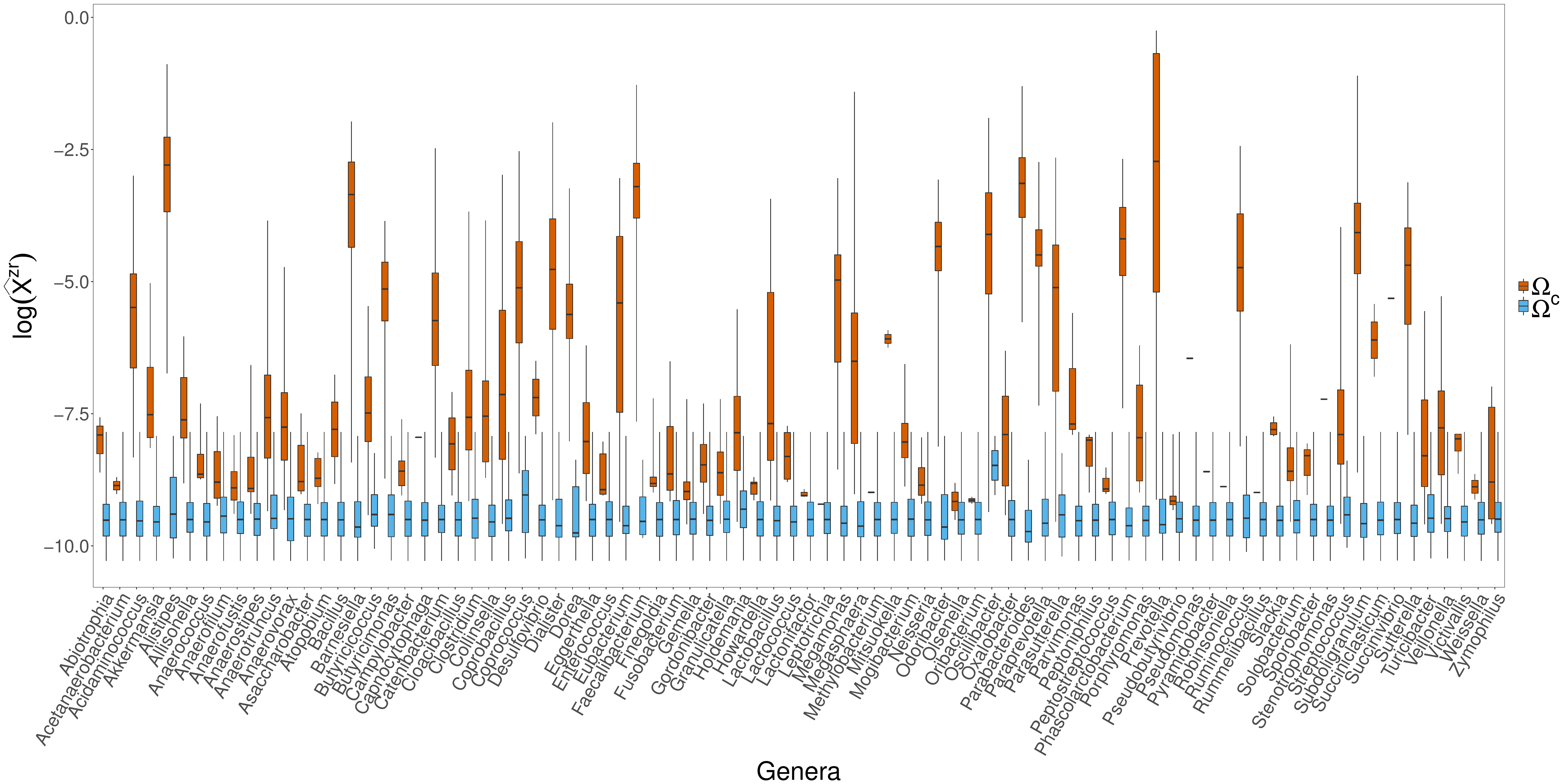}
	\end{tabular}
	\caption{Boxplots of the estimated compositions for the genera corresponding to non-zero observations ($\Omega$) and zero observations ($\Omega^c)$ in \textsc{combo} data set.  Top panel: the proposed estimator $\widehat{X}$; Bottom  panel: the zero replacement estimator $\widehat{X}^{\rm zr}$.}\label{fig:boxplot}
\end{figure}

Furthermore, as shown in Figure \ref{fig:index} (a), $\{\widehat{X}_{ij}\}_{(i,j)\in \Omega^c}$ tends to decrease as the total number of counts for each individual, i.e. $N_i$, increases. This is reasonable, as the zero counts are more likely to correspond to the very rare taxa when the sequencing gets deeper.  However, in contrast to the simple zero replacement estimates,  sequencing depth is not the only factor that determines the compositions of the taxa with zero counts. The compositional data observed in samples with non-zero counts also contribute to the final estimates. 

\begin{figure}
	\centering
	\begin{tabular}{cc}
		(a) & (b)\\
		\includegraphics[width=0.48\textwidth]{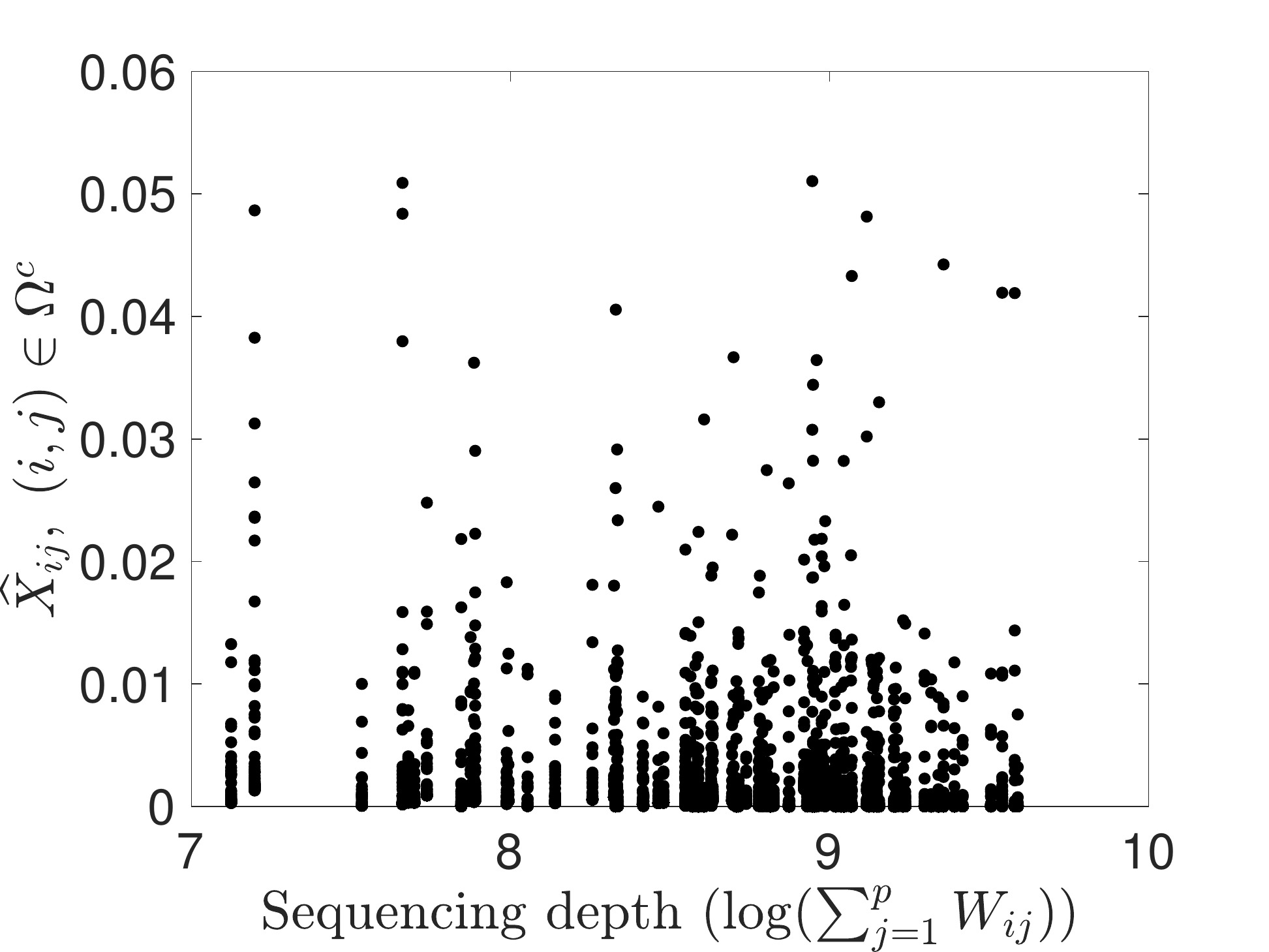} &	\includegraphics[width=0.48\textwidth]{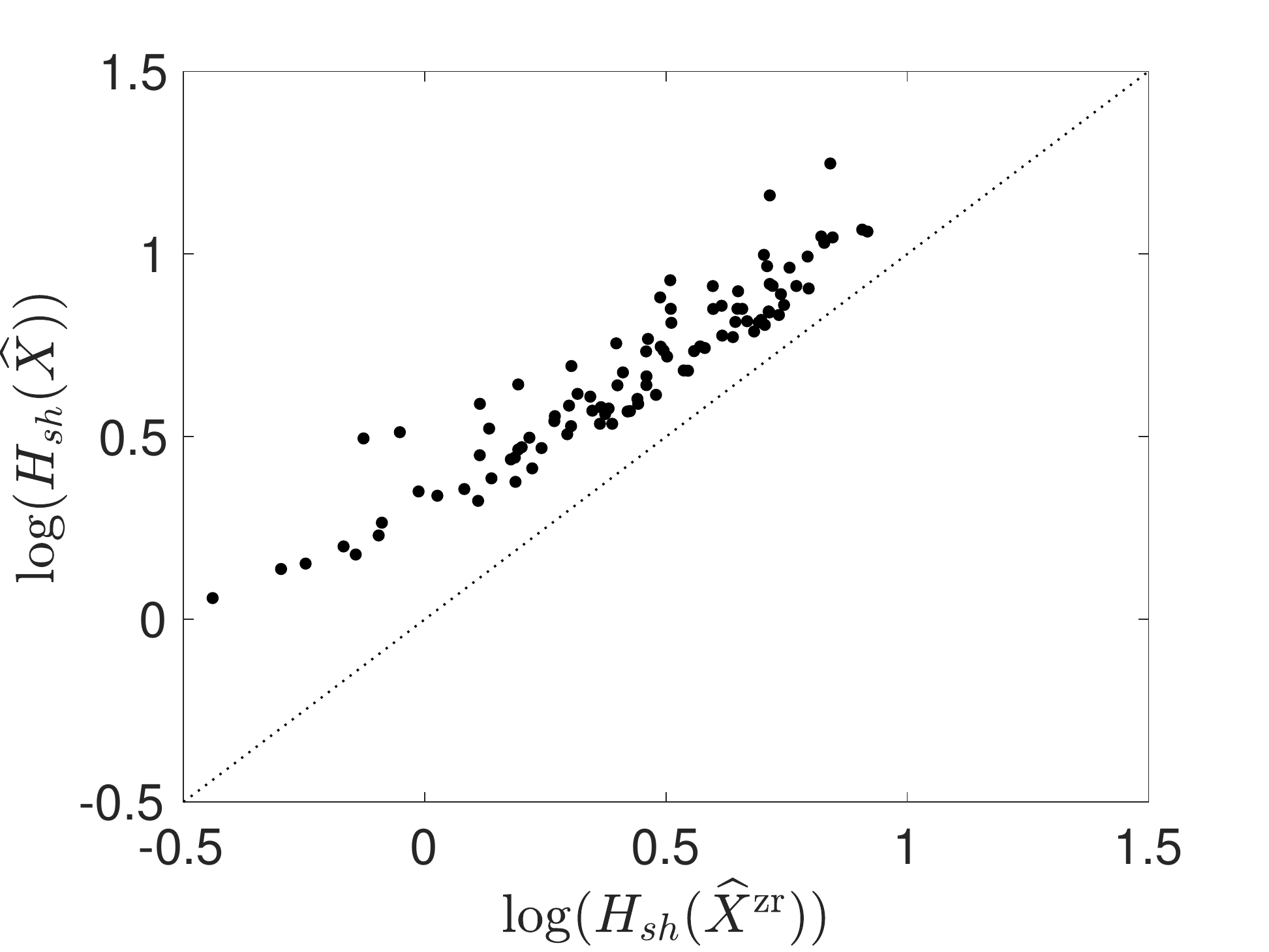} 
	\end{tabular}
	\caption{Analysis of \textsc{combo} data set. (a) The estimated compositions $\widehat{X}$ for genera corresponding to zero observations ($\Omega^c)$ versus the sequencing depth. (b) 
		Logarithm of the estimated Shannon's index  from regularized estimator $\widehat{X}$ versus zero replacement estimator $\widehat{X}^{\rm zr}$, where each dot represents one sample. 
	}\label{fig:index}

\end{figure}

Figures \ref{fig:index} (b) shows the estimates of Shannon's index for each individual using the proposed estimator $\widehat{X}$ versus the index based on  the zero-replacement estimator $\widehat{X}^{\rm zr}$, indicating that  $\widehat{X}^{\rm zr}$  produces uniformly smaller Shannon's index than $\widehat{X}$. This is mainly due to that $\widehat{X}^{\rm zr}$  replace all non-positive counts with the small value 0$\cdot$5, which yields an uneven distribution between taxa in $\Omega$ and $\Omega^c$, then results in lower diversity among all taxa and  smaller Shannon's index.

\section*{Acknowledgement}
We thank the Editor, Associate Editor, and two anonymous referees for their insightful comments.  This research was supported by grants from the National Institutes of Health and the National Science Foundation. 

\bibliographystyle{biometrika}
\bibliography{comc}

\newpage
\appendix

\setcounter{page}{1}
\setcounter{section}{0}

\begin{center}
	{\LARGE Supplement to ``Multi-sample Estimation of Bacterial}
	
	\bigskip	
	{\LARGE Composition Matrix in Metagenomics Data"	
	}
	
	\bigskip\medskip
	{Yuanpei Cao, ~ Anru Zhang, ~ and ~ Hongzhe Li}
\end{center}

\numberwithin{equation}{section}

\section*{Appendix: Proofs}
In the following analyses, we denote $\log X$ as the matrix with entry-wise logarithm of $ X$, i.e., $\left(\log  X\right)_{ij} = \log X_{ij}$.
\addtocounter{section}{1}
\setcounter{equation}{0}
\subsection{Proofs of Theorems \ref{thm:exact_lr} and \ref{thm:approx_lr}.}
Theorems \ref{thm:exact_lr} and \ref{thm:approx_lr} can be considered as two special cases of following theorem:
\begin{theorem}\label{thm:ub}
	Suppose that $N\ge (n+p)\log(n+p)$, $ X^\ast\in\mathcal{S}(\alpha_X,\beta_X)$, and $R_i\in[\alpha_R/n,\beta_R/n]$ for any $i\in[n]$. Then conditioning on fixed $N$, for any integer $1\le r\le n\wedge p$, $\widehat{ X}$ calculated from \eqref{align:rml} with the tuning parameter selected by \eqref{align:lambda} satisfies 
    
    \begin{align}
	\frac{p}{n}\|\widehat{ X} -  X^\ast\|_F^2 \notag
	\le C\Bigg[ & \frac{\beta_X^4(\beta_R\vee\beta_X)}{\alpha_R^2\alpha_X^4}\frac{(n\vee p)r\log(n+p)}{N}\\
	& +\left\{\frac{\beta_X^4(\beta_R\vee\beta_X)}{\alpha_R^2\alpha_X^4}\frac{p(n\vee p)\log(n+p)}{nN}\right\}^{1/2}\sum\limits_{i=r+1}^{n\land p}\sigma_i( X^\ast)\Bigg]. 
	\label{align:frob}
	\end{align}
    In addition, if $N<6(n+p)^2\log(n+p)/(\alpha_R\alpha_X)$, we have
	\begin{align}\label{align:kl}
	\frac{1}{n}\D( X^\ast,\widehat{ X})\notag
    \le C\Bigg[& \frac{\beta_X^2(\beta_R\vee\beta_X)}{\alpha_R^2\alpha_X^3}\frac{(n\vee p)r\log(n+p)}{N}\\
    & +\left\{\frac{\beta_R\vee\beta_X}{\alpha_R^2\alpha_X^2}\frac{p(n\vee p)\log(n+p)}{nN}\right\}^{1/2}\sum\limits_{i=r+1}^{n\land p}\sigma_i( X^\ast)\Bigg].
	\end{align}
    Otherwise, if $N\ge 6(n+p)^2\log(n+p)/(\alpha_R\alpha_X)$, we have
	\begin{align}\label{align:kl_2}
	\frac{1}{n}\D( X^\ast,\widehat{ X})\notag \le C\Bigg[&\frac{\beta_X^5\beta_R}{\alpha_R^2\alpha_X^6}\frac{(n\vee p)r\log(n+p)}{N}\\
	& +\left\{\frac{\beta_X^6\beta_R}{\alpha_R^2\alpha_X^8}\frac{p(n\vee p)\log(n+p)}{nN}\right\}^{1/2}\sum\limits_{i=r+1}^{n\land p}\sigma_i( X^\ast)\Bigg].
	\end{align}
	with probability no less than $1-3(n+p)^{-1}$ and some uniform constant $C>0$ that does not depend on $X^\ast, n, p, r, \alpha_X, \beta_X, \alpha_R,$ or $\beta_R$.
\end{theorem}
\begin{remark}\rm
	The rate of convergence provided by Theorem A$\ref{thm:ub}$ exhibits an interesting decomposition: $O\left(\frac{(n\vee p)r\log(n+p)}{N}\right)$ represents the estimation error corresponding to a rank-$r$ matrix, while $O\left(\left\{\frac{p\left(n\vee p\right)\log(n+p)}{nN}\right\}^{1/2}\sum\limits_{i=r+1}^{n\land p}\sigma_{i}( X^\ast)\right)$ accounts for the approximation error due to using $r$ as a proxy for the rank of $ X^\ast$. When $ X^\ast$ is exactly a rank-$r$ matrix, this approximation error vanishes. When $ X^\ast$ is approximately low-rank, the value of $r$ can be optimally chosen to obtain the sharpest bound.
\end{remark}
\begin{proof} First, if $\beta_X = \alpha_X$, we must have $\beta_X = \alpha_X = 1$ and all entries of $X^\ast$ as well as of any matrix in $\mathcal{S}(\alpha_X, \beta_X)$ are equal to $1/p$. The conclusion naturally holds . We thus focus on the case where $\beta_X > \alpha_X$.
	
As clarified in Remark \ref{rm:count-remark}, conditioning on the total count $N$, the count matrix $ W$ follows a multinomial distribution: $ W = \sum_{k=1}^N E_k$, where $ E_k$ are independent and identically distributed copies of a Bernoulli random matrix $ E$ that satisfies $ {\rm pr}\left( E = e_i(n)e_j^\top(p)\right) = \Pi_{ij}$ and $\Pi = ( R 1_p^\top)\circ X^\ast$. Consequently, the negative log-likelihood function \eqref{align:likelihood} can be rewritten as
	\begin{align}\label{align:likeli2}
	\mathcal{L}_N\left( X\right)=-{N}^{-1}\sum_{k=1}^N\log\langle X, E_k\rangle = -{N}^{-1}\sum_{i=1}^n \sum_{j=1}^p  W_{ij} \log  X_{ij}.
	\end{align}
	Then any solution $\widehat{X}$ to the optimization problem $\eqref{align:rml}$ satisfies
	\begin{align}\label{align:sol}
	 {N}^{-1}\sum_{k=1}^N\langle\log  X^\ast -\log \widehat{ X}, E_k \rangle=\mathcal{L}_N(\widehat{ X})-\mathcal{L}_N( X^\ast)\le \lambda\left(\| X^\ast\|_\ast-\|\widehat{ X}\|_\ast\right).
	\end{align}
	Next, we present the following Lemmas A\ref{lemma:nuc} and A\ref{lemma:kl_frob} to derive a lower bound for $N^{-1}\sum\limits_{k=1}^N\langle\log  X^\ast -\log \widehat{ X}, E_k \rangle$.
	\begin{lemma}\label{lemma:nuc}
		Given the selected tuning parameter from \eqref{align:lambda}, with probability at least $1-\left(n+p\right)^{-1}$, we have the following upper bound for $\|\widehat{ X}- X^\ast\|_\ast$:
		\begin{align*}
		\|\widehat{ X}- X^\ast\|_\ast \le 4\surd{(2r)}\|\widehat{ X}- X^\ast\|_F+4\sum_{i=r+1}^{n\land p}\sigma_{i}\left( X^\ast\right).
		\end{align*}
	\end{lemma}
\begin{lemma}\label{lemma:kl_frob}
	For any matrices $A, B\in\mathcal{S}(\alpha_X,\beta_X)$, we have
	\begin{align}\label{align:kl_frob}
	\frac{2\alpha_X^2}{\beta_Xp}\D(A, B)\le\|A - B\|_F^2\le\frac{2\beta_X^2}{\alpha_Xp}\D(A, B).
	\end{align}
\end{lemma}

Now we consider the proof of Theorem A\ref{thm:ub} in two regimes: $N<6(n+p)^2\log(n+p)/(\alpha_R\alpha_X)$ and $N\ge 6(n+p)^2\log(n+p)/(\alpha_R\alpha_X)$.

	First regime: $N<6(n+p)^2\log(n+p)/(\alpha_R\alpha_X)$. We divide the proof into three steps.
    \begin{enumerate}[leftmargin=*]
    	\item[Step 1.] For notational simplicity, denote 
    \begin{align*}
    \eta=n\log(\beta_X/\alpha_X)\left\{\frac{512\log(n+p)}{\log(4)\alpha_R^2N}\right\}^{1/2},\quad \D_{R}\left( X^\ast, \widehat{X}\right)=\sum_{i=1}^n\sum_{j=1}^pR_{i}X_{ij}^\ast\log\frac{X_{ij}^\ast}{\widehat{X}_{ij}},
    \end{align*}
    \begin{equation}\label{align:enpr}
    \begin{split} 
    E(&n,p, r)=\frac{2048\beta_X^2npr}{\alpha_R\alpha_X^3}\left[\left\{\frac{28\log(n+p)\left(\beta_R/n\vee \beta_X/p\right)}{N}\right\}^{1/2}+\frac{28\log(n+p)}{N}\right]^2 \\
    &+ \frac{16p}{\alpha_X}\left[\left\{\frac{28\log(n+p)\left(\beta_R/n\vee \beta_X/p\right)}{N}\right\}^{1/2}+\frac{28\log(n+p)}{N}\right]\sum_{i=r+1}^{n\land p}\sigma_{i}\left( X^\ast\right).
    \end{split}
    \end{equation}
	Here, $\left\{28\log(n+p)\left(\beta_R/n\vee \beta_X/p\right)N^{-1}\right\}^{1/2}$ and $28N^{-1}\log(n+p)$ in the formulation of $E(n, p, r)$ originate from the matrix concentration inequality, i.e., the upper bound of Lemmas A5 and A6. These terms are crucial to the proof of the upper bound of $E(Z_T)$ in Lemma A7. By the condition that $N>(n+p)\log(n+p)$, there exists some uniform constant $C_2>0$ that does not rely on $\alpha_X, \beta_X, \alpha_R, \beta_R, p, n, r$, such that
	\begin{align}\label{ineq:E(n,p,r)-upper}
	\begin{split}
	E(n,p,r) \le C_2\Bigg[& \frac{\beta_X^2(\beta_R\vee \beta_X)}{\alpha_R\alpha_X^3}\frac{(n\vee p)r\log(n+p)}{N}\\ 
	& +\left\{\frac{(\beta_R\vee\beta_X)}{\alpha_X^2}\frac{p(n\vee p)\log(n+p)}{nN}\right\}^{1/2}\sum_{i=r+1}^{n\land p}\sigma_{i}\left( X^\ast\right)\Bigg].
	\end{split}
	\end{align}
	We also define the following sets
	\[\mathcal{C}(\alpha_X,\beta_X)=\left\{ A\in\mathcal{S}(\alpha_X,\beta_X)\Bigg| \begin{array}{l} 
	\D( X^\ast, A)\ge \eta\\
	\|A- X^\ast\|_\ast \le 4\surd{(2r)}\|A- X^\ast\|_F+4\sum_{i=r+1}^{n\land p}\sigma_{i}\left( X^\ast\right)
	\end{array}\right\},\]
	\[\mathcal{D}(T)=\left\{ A\in\mathcal{S}(\alpha_X,\beta_X)\Bigg| \begin{array}{l}
	\D_R( X^\ast, A)\le T\\
	\|A-X^\ast\|_\ast \le 4\surd{(2r)}\|A- X^\ast\|_F+4\sum_{i=r+1}^{n\land p}\sigma_{i}\left( X^\ast\right)
	\end{array}\right\}.\]
	We separate the constraint set $\mathcal{C}(\alpha_X,\beta_X)$ into pieces and focus on a sequences of small sets $\mathcal{C}_l(\alpha_X,\beta_X)$,
	\[\mathcal{C}(\alpha_X,\beta_X) = \cup_{l=1}^\infty \mathcal{C}_l(\alpha_X,\beta_X), \quad\mathcal{C}_l(\alpha_X,\beta_X)=\left\{ A\in\mathcal{C}(\alpha_X,\beta_X)\big|2^{l-1}\eta\le \D( X^\ast, A) <  2^l\eta\right\}.\]

    \item[Step 2] Next, we use a peeling argument to prove that the probability of the following unfavorable event is small
    \[\mathcal{B}=\left\{\exists A\in\mathcal{C}(\alpha_X,\beta_X)\text{ s.t. }\left|\frac{1}{N}\sum_{k=1}^N\langle\log  X^\ast-\log A, E_k\rangle-\D_R\left(X^\ast, A\right)\right|\ge\frac{1}{2}\D_R\left( X^\ast, A\right)+E(n,p,r)\right\}.\]
    Under the assumption of $\min\limits_{1\le i\le n}R_i\ge \alpha_R/n$, we have 
    \begin{equation}\label{ineq:D_R-D}
    \D_R(X^\ast, A)\ge n^{-1}\alpha_R\D( X^\ast, A).
    \end{equation}
    It suffices to estimate the probability of the following events and then apply a union bound.
    \begin{equation*}
   	\begin{split}
   	\mathcal{B}_l= & \left\{\exists A\in\mathcal{C}_l(\alpha_X,\beta_X)\text{ s.t. }\left|\frac{1}{N}\sum_{i=1}^N\langle\log  X^\ast-\log  A, E_i\rangle-\D_R( X^\ast, A)\right|\ge \frac{1}{2}\D_R(X^\ast, A)+E(n,p,r)\right\}\\
   	\subseteq & \left\{\exists A\in\mathcal{C}_l(\alpha_X,\beta_X)\text{ s.t. }\left|\frac{1}{N}\sum_{i=1}^N\langle\log  X^\ast-\log  A, E_i\rangle-\D_R( X^\ast, A)\right|\ge \frac{\alpha_R}{2n}\D(X^\ast, A)+E(n,p,r)\right\}\\
   	\subseteq & \left\{\exists A\in\mathcal{C}_l(\alpha_X,\beta_X)\text{ s.t. }\left|\frac{1}{N}\sum_{i=1}^N\langle\log  X^\ast-\log  A, E_i\rangle-\D_R( X^\ast, A)\right|\ge \frac{2^{l}\eta\alpha_R}{4n}+E(n,p,r)\right\}\\
   	\subseteq & \left\{\exists A\in\mathcal{D}(2^l\eta)\text{ s.t. }\left|\frac{1}{N}\sum_{i=1}^N\langle\log  X^\ast-\log  A, E_i\rangle-\D_R( X^\ast, A)\right|\ge \frac{2^{l}\eta\alpha_R}{4n}+E(n,p,r)\right\}.
   	\end{split}
    \end{equation*}
    Here, we use the fact that $\D_R(X^\ast, A)\le \D(X^\ast, A)$, then $\mathcal{C}_l(\alpha_X,\beta_X)\subseteq \mathcal{D}\left(2^l\eta\right)$. Now, we can establish the upper bound of the probability of event $\mathcal{B}$ by using a union bound 
    and Lemma A$\ref{lemma:hoeffd}$,
    \begin{align*}
    & {\rm pr}\left(\mathcal{B}\right) \le \sum_{l=1}^\infty {\rm pr}(\mathcal{B}_l)\\
    \le & \sum_{l=1}^\infty {\rm pr}\left(\exists A\in\mathcal{D}(2^l\eta), \text{ s.t. }\left|\frac{1}{N}\sum_{i=1}^N\langle\log  X^\ast-\log A, E_i\rangle-\D_R( X^\ast, A)\right|\ge \frac{2^{l}\eta\alpha_R}{4n}+E(n,p,r)\right)\\
    \le & \sum_{l=1}^\infty\exp\left[-\frac{4^l\alpha_R^2N\eta^2}{512\left\{n\log(\beta_X/\alpha_X)\right\}^2}\right] \le \sum_{l=1}^\infty\exp\left[-\frac{\log(4)\alpha_R^2N\eta^2l}{512\left\{n\log(\beta_X/\alpha_X)\right\}^2}\right].
    \end{align*}
    Plugging in $\eta=n\log(\beta_X/\alpha_X)\left[{512\log(n+p)}/\left\{\log(4)\alpha_R^2N\right\}\right]^{1/2}$, we obtain 
    \begin{equation}\label{ineq:P-B}
    {\rm pr}\left(\mathcal{B}\right)\le 2(n+p)^{-1}.
    \end{equation} 
    \item[Step 3] We finalize the proof in this step. Define the following `favorable event"
    \begin{equation}
    \mathcal{G} = \left\{\|\widehat{ X} - X^\ast\|_\ast \le 4\surd{(2r)}\|\widehat{X}-X^\ast\|_F + 4\sum_{i=r+1}^{n\wedge p}\sigma_i(X^\ast)\right\}.
    \end{equation}
    By Lemma A\ref{lemma:nuc}, ${\rm pr}(\mathcal{G}) \geq 1 - (n+p)^{-1}$. Thus,
    \begin{equation}\label{ineq:Probability-G-B}
    {\rm pr}(\mathcal{G}^c \cup \mathcal{B}) \le {\rm pr}(\mathcal{G}^c) + {\rm pr}(\mathcal{B}) \le 3(n+p)^{-1}.
    \end{equation}
    Next, we develop an upper bound for estimation error when $\mathcal{G} \cap \mathcal{B}^c$, i.e., the favorable event holds while the unfavorable event does not hold. In fact, if $\mathcal{G}$ and $\mathcal{B}^c$ are both true, either of the following must hold for $\widehat{X}$:
    \begin{enumerate}[leftmargin=*]
    	\item $\widehat{X}\in \mathcal{C}(\alpha_X, \beta_X)$ and 
    	\begin{equation}\label{ineq:case-1}
    	\left|\frac{1}{N}\sum_{k=1}^N\langle\log  X^\ast-\log \widehat{X}, E_k\rangle-\D_R\left(X^\ast, \widehat{X}\right)\right| < \frac{1}{2}\D_R\left( X^\ast, \widehat{X}\right)+E(n,p,r).
    	\end{equation}
    	which also implies
    	\begin{equation*}
    	\begin{split}
    	& -\frac{1}{N}\sum_{k=1}^N\langle\log  X^\ast-\log \widehat{X}, E_k\rangle+\D_R\left(X^\ast, \widehat{X}\right) < \frac{1}{2}\D_R\left( X^\ast, \widehat{X}\right)+E(n,p,r)\\
    	\text{i.e.,}\quad & \frac{1}{2}\D_R(X^\ast, \widehat{X}) \le \frac{1}{N}\sum_{k=1}^N\langle\log  X^\ast-\log \widehat{X}, E_k\rangle + E(n, p, r).
    	\end{split}
    	\end{equation*}
   		\item $\widehat{X} \notin \mathcal{C}(\alpha_X, \beta_X)$, which also implies $\D(X^\ast, \widehat{X}) < \eta$.
    \end{enumerate}
    \begin{enumerate}[leftmargin=*]
    	\item [Under i.,]
    	By \eqref{ineq:E(n,p,r)-upper} and \eqref{ineq:D_R-D}, we have 
    \begin{equation}\label{align:upperbound2}
    \begin{split}
    &\frac{\alpha_R}{2n}\D( X^\ast,\widehat{X})\le \frac{1}{2}\sum_{i=1}^n\sum_{j=1}^pR_iX_{ij}^\ast\log\frac{X_{ij}^\ast}{\widehat{X}_{ij}} = \frac{1}{2}\D_R(X^\ast, \widehat{X})\\
    \le&\frac{1}{N}\sum_{k=1}^N\langle\log  X^\ast -\log \widehat{ X}, E_k \rangle + C_2\Bigg[\frac{\beta_X^2(\beta_R\vee \beta_X)}{\alpha_R\alpha_X^3}\frac{(n\vee p)r\log(n+p)}{N} \\
    & \qquad + \left\{\frac{(\beta_R\vee\beta_X)}{\alpha_X^2}\frac{p(n\vee p)\log(n+p)}{nN}\right\}^{1/2}\sum\limits_{i=r+1}^{n\land p}\sigma_i( X^\ast)\Bigg].
    \end{split}
    \end{equation}
    	By applying Lemma A$\ref{lemma:kl_frob}$, we obtain the upper bound of $\| X^\ast\|_\ast-\|\widehat{ X}\|_\ast$ as
    \begin{equation}\label{align:nuc_ub}
    \begin{split}
    \| X^\ast\|_\ast-\|\widehat{ X}\|_\ast & \le \| X^\ast-\widehat{ X}\|_\ast \le 4\surd{(2r)}\| X^\ast-\widehat{ X}\|_F+4\sum_{i=r+1}^{n\land p}\sigma_i( X^\ast)\\
    &\le 8\left\{{(\alpha_Xp)}^{-1}{\beta_X^2r}\D( X^\ast,\widehat{ X})\right\}^{1/2}+4\sum_{i=r+1}^{n\land p}\sigma_i( X^\ast).
    \end{split}
    \end{equation}
    Therefore, combining $\eqref{align:sol}$, $\eqref{align:upperbound2}$, and $\eqref{align:nuc_ub}$, we obtain
    \begin{align*}
    &\frac{\alpha_R}{2n}\D( X^\ast,\widehat{X})\le \lambda \left[8\left\{\frac{\beta_X^2r}{\alpha_Xp}\D( X^\ast, \widehat{X})\right\}^{1/2}+4\sum_{i=r+1}^{n\land p}\sigma_i( X^\ast)\right]\\
    &+ C_2\Bigg[\frac{\beta_X^2(\beta_R\vee \beta_X)}{\alpha_R\alpha_X^3}\frac{(n\vee p)r\log(n+p)}{N} \\
    & \qquad + \left\{\frac{(\beta_R\vee\beta_X)}{\alpha_X^2}\frac{p(n\vee p)\log(n+p)}{nN}\right\}^{1/2}\sum\limits_{i=r+1}^{n\land p}\sigma_i( X^\ast)\Bigg].
    \end{align*}
	The above formula can be treated as a quadratic inequality for $\D({ X}^\ast,  \widehat{X})$. We plug in  $\lambda = \delta\left\{{\beta_Rp(n\vee p)\log(n+p)}/{(\alpha_X^2nN)}\right\}^{1/2}$ for constant $\delta>0$, solve this quadratic inequality and obtain
    \begin{equation}
    \begin{split}
    n^{-1}\D( X^\ast,\widehat{ X}) &\le C_3\Bigg[\frac{\beta_X^2(\beta_R\vee\beta_X)}{\alpha_R^2\alpha_X^3}\frac{(n\vee p)r\log(n+p)}{N}\\
    & +\left\{\frac{(\beta_R\vee\beta_X)}{\alpha_R^2\alpha_X^2}\frac{p(n\vee p)\log(n+p)}{nN}\right\}^{1/2}\sum\limits_{i=r+1}^{n\land p}\sigma_i( X^\ast)\Bigg], \label{align:upperbound3}
    \end{split}
    \end{equation}
    where $C_3>0$ is some uniform constant that does not rely on $\alpha_X, \beta_X, \alpha_R, \beta_R, n, p, r$. 
    \\
    \item[Under ii.,] we have
    \begin{equation}\label{align:upperbound1}
    \begin{split}
    n^{-1}\D( X^\ast,\widehat{X})< & n^{-1}\eta = \log(\beta_X/\alpha_X)\left\{\frac{512\log(n+p)}{\log(4)\alpha_R^2N}\right\}^{1/2} \\
    \le & \frac{C_1\beta_X}{\alpha_R^{3/2}\alpha_X^{3/2}}\frac{(n\vee p)r\log(n+p)}{N}
    \end{split}
    \end{equation}
    for some uniform constant $C_1 >0$ that does not depend on $\alpha_X, \beta_X, \alpha_R, \beta_R, n, p$, or $r$. The last inequality is due to the regime assumption that $N \le 6(n+p)^2\log(n+p)/(\alpha_R\alpha_X)$.
	\end{enumerate}
	Under both i. or ii., by Equations \eqref{ineq:Probability-G-B}, \eqref{align:upperbound3}, and \eqref{align:upperbound1}, we have arrived at the Kullback-Leibler divergence upper bound \eqref{align:kl}; 
	By Lemma A\ref{lemma:kl_frob}, we have further reached the Frobenius upper bound \eqref{align:frob}.
    This provides the desired upper bound for the proof of Theorem A\ref{thm:ub}.

\end{enumerate}
	
	\ \par
	
	Second regime: $N>6(n+p)^2\log(n+p)/(\alpha_R\alpha_X)$. We denote $\Delta = \widehat{ X}- X^\ast$. According to \eqref{align:likeli2} and Taylor's expansion, there exists $\xi = (\xi_{ij})_{1\le i\le n, 1\le j\le p}$ such that
\begin{align*}
\mathcal{L}_N(\widehat{ X}) - \mathcal{L}_N( X^\ast) - \langle \triangledown \mathcal{L}_N( X^\ast), \Delta \rangle = \frac{1}{2N}\sum_{k=1}^N \frac{\langle \Delta,  E_k \rangle^2 }{\langle \xi,  E_k \rangle^2},\quad \xi_{ij} \text{ is between } \widehat{X}_{ij}\text{ and } X^\ast_{ij}.
\end{align*}
	Since $\widehat{X}_{ij}, X_{ij} \le \beta_X/p$, $W = \sum_{k=1}^N  E_k$, we have
	\begin{equation*}
	\mathcal{L}_N(\widehat{ X}) - \mathcal{L}_N( X^\ast) - \langle \triangledown \mathcal{L}_N( X^\ast), \Delta \rangle \geq \frac{1}{2N}\sum_{k=1}^N \frac{\langle \Delta,  E_k\rangle^2}{(\beta_X/p)^2} \geq \frac{p^2}{2N\beta_X^2}\sum_{i=1}^n\sum_{j=1}^p \Delta_{ij}^2W_{ij}.
	\end{equation*}
	On the other hand, note that $W_{ij}\sim {\rm Bin}(N, R_i X_{ij}^\ast)$, $N\geq 6(n+p)^2\log(n+p)/(\alpha_R\alpha_X)$, $R_i \geq \alpha_R/n$, and $X_{ij}^\ast \geq \alpha_X/p$. By the Chernoff bound of binomial distribution\footnote{See, e.g., \url{https://en.wikipedia.org/wiki/Chernoff_bound#Multiplicative_form_(relative_error)}} and $(n+p)^2\geq 4np$, we have
	\begin{equation*}
	\begin{split}
	{\rm pr}\left(W_{ij} \le NR_i X_{ij}^\ast/2\right) = & {\rm pr}\left(W_{ij} \le \left(1-1/2\right) EW_{ij}\right) \le \exp\left(-NR_iX_{ij}^\ast/8\right) \\
	\le & \exp\left(-\frac{6(n+p)^2\log(n+p)\alpha_R\alpha_X}{8\alpha_R\alpha_Xnp}\right)\le \left(n+p\right)^{-3}.
	\end{split}
	\end{equation*}
	By a union bound argument,
	\begin{align}\label{align:bino_chernoff}
	{\rm pr}\left(W_{ij} \geq NR_i X_{ij}^\ast/2 \text{ for any } (i,j) \right) \geq 1 - (n+p)^{-1}.
	\end{align}
	If $W_{ij} \geq NR_i X_{ij}^\ast/2$ holds for any $(i,j)$, we have the following strong convexity for $\mathcal{L}_N(\widehat{X})$,
	\begin{align}\label{align:bino_lower}
	\mathcal{L}_N(\widehat{ X}) - \mathcal{L}_N( X^\ast) - \langle \triangledown \mathcal{L}_N( X^\ast), \Delta \rangle \geq \frac{p^2}{4\beta_X^2} \sum_{i=1}^n \sum_{j=1}^p \Delta_{ij}^2 R_i X_{ij}^\ast \geq \frac{p\alpha_R\alpha_X}{4n\beta_X^2}\|\Delta\|_F^2.
	\end{align}

In addition, by using the identity $\langle  R 1_p^\top,\Delta\rangle=\langle  R,\Delta 1_p\rangle =\langle  R, 0_n\rangle = 0$ and H$\mathrm{\ddot{o}}$lder's inequality between the nuclear norm and operator norm, the upper bound of $\langle\triangledown \mathcal{L}_N( X^\ast),\Delta \rangle$ can be controlled by
\begin{equation}\label{align:grad_lambda_upper_bound}
\begin{split}
-\langle\triangledown \mathcal{L}_N( X^\ast),\Delta \rangle = & -\langle\triangledown \mathcal{L}_N( X^\ast)+ R1_p^\top,\Delta \rangle\le \|\triangledown \mathcal{L}_N( X^\ast)+ R1_p^\top\|_2\|\Delta\|_\ast.
\end{split}
\end{equation}
According to Lemma A\ref{lemma:nuc} and A\ref{lemma:lambda}, by combining $\eqref{align:sol}$, $\eqref{align:bino_chernoff}$, $\eqref{align:bino_lower}$, and $\eqref{align:grad_lambda_upper_bound}$, with probability at least $1-3(n+p)^{-1}$, we have
\begin{equation*}
\begin{split}
& \frac{p\alpha_R\alpha_X}{4n\beta_X^2}\|\Delta\|_F^2 \le \mathcal{L}_N(\widehat{X}) - \mathcal{L}_N(X^\ast) - \langle\nabla\mathcal{L}_N(X^\ast), \Delta\rangle\\
\le & \lambda(\|X^\ast\|_\ast - \|\widehat{X}\|_\ast) + \|\nabla\mathcal{L}_N(X^\ast) + R1_p^\top\|_2\cdot\|\Delta\|_\ast\\
\le & \lambda\|\widehat{X} - X^\ast\|_\ast + \|\nabla\mathcal{L}_N(X^\ast) + R1_p^\top\|_2\cdot\left(4\surd{(2r)}\|\Delta\|_F + 4\sum_{i=r+1}^{n\wedge p}\sigma_i(X^\ast)\right)\\
\le & \lambda\left(4\sqrt{2}r\|\Delta\|_F + 4\sum_{i=r+1}^{n\wedge p}\sigma_i(X^\ast)\right) \\
& + \left\{\frac{2}{3\alpha_X}+2\left(\frac{1}{9\alpha_X^2}+\frac{\beta_R}{\alpha_X}\right)^{1/2}\right\}\left\{\frac{p(n\vee p)\log(n+p)}{nN}\right\}^{1/2}\cdot \left(4\surd{(2r)}\|\Delta\|_F + 4\sum_{i=r+1}^{n\wedge p}\sigma_i(X^\ast)\right)\\
\le & C_4 \left\{\frac{\beta_R}{\alpha_X^2}\frac{p(n\vee p)\log(n+p)}{nN}\right\}^{1/2}\left(\surd{r}\|\Delta\|_F +\sum\limits_{i=r+1}^{n\land p}\sigma_i( X^\ast)\right),
\end{split}
\end{equation*}
where $C_4>0$ is some constant that does not depend on $\alpha_X,\alpha_R,\beta_X$, $\beta_R, n, p$, or $r$. Solving this quadratic inequality, we obtain
	\begin{align}
	\frac{p}{n}\|\widehat{X} -  X^\ast\|_F^2 \notag \le C_5\Bigg[&\frac{\beta_X^4\beta_R}{\alpha_R^2\alpha_X^4}\frac{(n\vee p)r\log(n+p)}{N}\\
	& +\left\{\frac{\beta_X^4\beta_R}{\alpha_R^2\alpha_X^4}\frac{p(n\vee p)\log(n+p)}{nN}\right\}^{1/2}\sum\limits_{i=r+1}^{n\land p}\sigma_i( X^\ast)\Bigg]. 
	\label{align:upperbound3_Frob}
	\end{align}
	\begin{equation*}
	\begin{split}
	\frac{1}{n}\D(X^\ast, \widehat{X}) \le \frac{\beta_Xp}{2\alpha_X^2n}\|\widehat{X}^\ast - X\|_F^2 \le C_5\Bigg[&\frac{\beta_X^5\beta_R}{\alpha_R^2\alpha_X^6}\frac{(n\vee p)r\log(n+p)}{N}\\
	& +\left\{\frac{\beta_X^6\beta_R}{\alpha_R^2\alpha_X^8}\frac{p(n\vee p)\log(n+p)}{nN}\right\}^{1/2}\sum\limits_{i=r+1}^{n\land p}\sigma_i( X^\ast)\Bigg]
	\end{split}
	\end{equation*}
	with probability at least $1 - 3(n+p)^{-1}$.

In summary, we have finished the proof of Theorem A\ref{thm:ub}.
\end{proof}

\ \par

We are ready to prove Theorems \ref{thm:exact_lr} and \ref{thm:approx_lr} with the result in Theorem A\ref{thm:ub}.

\emph{Proof of Theorem \ref{thm:exact_lr}.} Since  $\sum\limits_{i=r+1}^{n\land p}\sigma_i( X^\ast)$ vanishes when $\text{rank}( X^\ast)\le r$, \eqref{align:exact_lr_frob} and \eqref{align:exact_lr} can be obtained by applying Theorem A\ref{thm:ub}. 

Moreover, when $N = O\left((n+p)^2r\log(n+p)\right)$, we can provide the following expected risk upper bounds: there exists some constants $C_6$ and $C_7$ that does not depend on $p, n$, or $r$, such that the risks of the estimates have the following upper bounds: 
\begin{align}
    \frac{p}{n}  E\left\{\left\|\widehat{ X} -  X^\ast \right\|_F^2\right\} &\le C_6\frac{(n+p)r\log(n+p)}{N},\label{align:expect1}\\
	\frac{1}{n}  E\left\{\D( X^\ast,\widehat{ X})\right\} &\le C_7\frac{(n+p)r\log(n+p)}{N}.\label{align:expect2}
\end{align}
Since $ X^\ast$ and $\widehat{ X}$ belong to $\mathcal{S}(\alpha_X,\beta_X)$, we always have the trivial bound $\|\widehat{ X}- X^\ast\|_F^2\le np^{-1}(\beta_X-\alpha_X)^2$. Define the event
$$\mathcal{Q} = \left\{\frac{p}{n}\|\widehat{X}-X^\ast\|_F^2\le  C_1(\alpha_X,\alpha_R,\beta_X,\beta_R)\frac{(n + p)r\log(n+p)}{N}\right\}.$$
Then, ${\rm pr}(\mathcal{Q}) \ge 1 - 3(n+p)^{-1}$. Applying \eqref{align:exact_lr_frob}, we get 
\begin{align*}
\frac{p}{n} E\|\widehat{ X} -  X^\ast\|_F^2& = \frac{p}{n} E\|\widehat{ X} -  X^\ast\|_F^2\mathbb{I}_{\mathcal{Q}^c} + \frac{p}{n} E\|\widehat{ X} -  X^\ast\|_F^2\mathbb{I}_{\mathcal{Q}}\\
& \le (\beta_X-\alpha_X)^2\cdot {\rm pr}(\mathcal{Q}^c) + C_1(\alpha_X,\alpha_R,\beta_X,\beta_R)\frac{(n + p)r\log(n+p)}{N}\\
&\le C_1'(\alpha_X,\alpha_R,\beta_X,\beta_R)\frac{(n+p)r\log(n+p)}{N},
\end{align*}
where the second inequality comes from the assumption that $N=O((n+p)^2r\log(n+p))$. The proof of \eqref{align:expect2} is essentially the same by applying \eqref{align:exact_lr}.\quad $\square$

\ \par

\emph{Proof of Theorem \ref{thm:approx_lr}.} If the composition $ X^\ast\in\mathbb{B}_q\left(\rho_q,\alpha_X,\beta_X\right)$, we set $r=\max\left\{i\mid\sigma_i\left( X^\ast\right)>\tau\right\}$ for some thresholding level $\tau>0$ to be determined later. Then we obtain
\[r\tau^q\le\sum_{j=1}^r\sigma_j( X^\ast)^q\le\rho_q,\]
which implies $r\le\tau^{-q}\rho_q$. In addition, since $\sigma_{i}(X^\ast) \le \tau$ for $i\geq r+1$, we have
\[\sum_{i=r+1}^{n \wedge p}\sigma_i\left( X^\ast\right)\le \sum_{i=r+1}^{n\wedge p}\sigma_i(X^\ast)^q \tau^{1-q}\le \tau^{1-q}\rho_q.\]
By taking the above inequality in $\eqref{align:frob}$ and setting  $\tau = \left\{\frac{\beta_X^4(\beta_R\vee\beta_X)}{\alpha_R^2\alpha_X^4}\frac{(n\vee p)n\log(n+p)}{pN}\right\}^{1/2}$, we have
\begin{align*}
\frac{p}{n}\|\widehat{ X}- X^\ast\|_F^2\le  C\frac{\rho_qp^{q/2}}{n^{q/2}}\left\{\frac{\beta_X^4(\beta_R\vee\beta_X)}{\alpha_R^2\alpha_X^4}\frac{(n\vee p)\log(n+p)}{N}\right\}^{1-q/2}
\end{align*}
with probability no less than $1-3(n+p)^{-1}$. Under the assumption $N = O\left({\rho_qp^{q/2}(n+p)^{2+q/2}\log(n+p)}/{n^{q/2}}\right)$, \eqref{align:approx_lr_frob} can be proved by
\begin{align*}
\frac{1}{n} E\|\widehat{ X}- X^\ast\|_F^2&\le \frac{3}{n+p}\frac{\beta_X(\beta_X-\alpha_X)^2}{2\alpha_X^2} + C\left\{\frac{\beta_X^4(\beta_R\vee\beta_X)}{\alpha_R^2\alpha_X^4}\frac{(n\vee p)\log(n+p)}{N}\right\}^{1-q/2}\\
&\le C'\frac{\rho_qp^{q/2}}{n^{q/2}}\left\{\frac{(n\vee p)\log(n+p)}{N}\right\}^{1-q/2},
\end{align*}
for some positive constant $C'$. The proof for \eqref{align:approx_lr} is similar .\quad $\square$

\subsection{Proof of Theorem \ref{thm:lower_bound}}
We first establish the Kullback-Leibler divergence between two multinomial distribution random vectors. Suppose $x = (x_1,\ldots, x_p) \sim \text{Multi}(N, u), y = (y_1,\ldots, y_p) \sim \text{Multi}(N, v)$, where $u = (u_1,\ldots, u_p), v = (v_1,\ldots, v_p)\in \mathbb{R}^p$ satisfies $u_i, v_i \geq 0$ and $\sum_i u_i = \sum_i v_i = 1$. Then,
\begin{equation}\label{eq:KL-multinomial}
\begin{split}
\D_{KL}(x, y) = & \sum_{\substack{z_1,\ldots, z_p\in \mathbb{N}\\z_1+\cdots +z_p = N}} P\{x = (z_1,\ldots, z_p)\} \log\left[\frac{P\{x=(z_1,\ldots, z_p)\}}{P\{y=(z_1,\ldots, z_p)\}}\right]\\ 
= & \sum_{\substack{z_1,\ldots, z_p\in \mathbb{N}\\z_1+\cdots +z_p = N}} \frac{N!}{z_1!\cdots z_p!} u_1^{z_1} \cdots u_p^{z_p}\log\left(\frac{u_1^{z_1}\cdots u_p^{z_p}}{v_1^{z_1}\cdots v_p^{z_p}}\right)\\
= & \sum_{\substack{z_1,\ldots, z_p\in \mathbb{N}\\z_1+\cdots +z_p = N}}\sum_{i=1}^p \frac{N!}{z_1!\cdots z_p!} u_1^{z_1}\cdots u_p^{z_p} z_i \log(u_i/v_i)\\
= & N\sum_{i=1}^p \sum_{\substack{z_1,\ldots, z_p\in \mathbb{N}, z_i\geq 1\\z_1+\cdots +z_p = N}}\frac{(N-1)!}{\prod_{j\neq i}z_j!\cdot (z_i-1)!} \prod_{j\neq i} u_j^{z_j}\cdot u_i^{z_i-1} u_i \log(u_i/v_i)\\
= & N \sum_{i=1}^p \sum_{\substack{z_1',\ldots, z_p'\in \mathbb{N}\\z_1'+\cdots+z_p' = N-1}}\frac{(N-1)!}{\prod_{j}z_j'!} \prod_{j} u_j^{z_j'} u_i \log(u_i/v_i) \quad \text{($z_j' = z_j$ for $j\neq i$; $z_i' = z_i - 1$)}\\
= & N \sum_{i=1}^p u_i \log(u_i/v_i) = N \D_{KL}(u, v).
\end{split}
\end{equation}

We discuss the proof for Theorem \ref{thm:lower_bound} under two different scenarios.
\leftmargini=0mm
\begin{itemize}
	\item If $n\geq p$, we randomly generate $M$ copies of independent and identically distributed Rademachar random matrices: $B_1, \cdots, B_M \in \mathbb{R}^{n \times (r-1)}$.
	Since $(B_{k, ij} - B_{l, ij})^2$ has the following probability distribution
	\begin{align*}
	{\rm pr}\left((B_{k, ij} - B_{l, ij})^2 = x \right) = \left\{\begin{array}{ll}
	1/2, & x = 4,\\
	1/2, & x = 0,\\
	\end{array}\right.
	\end{align*}
	based on Bernstein's inequality,
	\begin{align*}
	&  {\rm pr}\left(\|B_k - B_l\|_F^2 \le n(r-1) \right) =  {\rm pr}\left(\sum_{i=1}^n\sum_{j=1}^{r-1}(B_{k, ij} - B_{l, ij})^2 - 2n(r-1) \le -n(r-1) \right) \\
	\le & \exp\left\{- \frac{n^2(r-1)^2/2}{4n(r-1) + 2n(r-1)/3}\right\} < \exp\left\{- n(r-1)/10\right\} .
	\end{align*}
	Therefore, whenever $M \le \exp(n(r-1)/20)$, there is a positive probability that
	\begin{equation}\label{eq:B_k - B_l}
	\min_{1\le k < l\le M}\left\{\|B_k - B_l\|_F^2\right\} \geq n(r-1),
	\end{equation}
	which means that we can find  fixed $B_1,\ldots, B_M \in \{-1, 1\}^{n\times(r-1)}$ such that \eqref{eq:B_k - B_l} holds. For the rest of proof, we assume $B_1,\ldots, B_M$ are such fixed matrices while $M = \lfloor\exp(n(r-1)/20)\rfloor$. Note that $r-1 \le p/2$, we consider the following set of random rank-$r$ matrices,
	\begin{align*}
	 X_k = \begin{bmatrix}
	p^{-1} & \cdots & p^{-1}\\
	\vdots & & \vdots\\
	p^{-1} & \cdots & p^{-1}
	\end{bmatrix}_{n\times p} + \begin{blockarray}{ccc}
	r-1 & r-1 & p - 2r+2\\
	\begin{block}{[ccc]}
	\nu B_k & -\nu B_k & 0\\
	\end{block}
	\end{blockarray},	
	\end{align*}
	where $0 < \nu < (\beta_X-1)/{p}\land (1-\alpha_X)/p$ is a to-be-determined constant. Then for $k\neq l$,
	\begin{align*}
    \left\| X_k -  X_l\right\|_F^2 &= 2\nu^2 \| B_k -  B_l\|_F^2 \geq 2\nu^2 n(r-1),\\
	\D( X_k,  X_l) &\geq c_1p \| X_k -  X_l\|_F^2 \geq 2c_1\nu^2 np(r-1),
	\end{align*}
	where we use Lemma A\ref{lemma:kl_frob} in the second inequality, and $c_1=\alpha_X/(2\beta_X^2)$. We also fix $ R = n^{-1}1_n$ as the uniform distribution for each row. Suppose $ P_k \sim {\rm Mult}(N, n^{-1} X_k)$, i.e., the multinomial distribution corresponding to composition $ X_k$ and $ R$. Based on \eqref{eq:KL-multinomial} and Lemma A\ref{lemma:kl_frob}, we have the following upper bound for the Kullback-Leibler divergence between random matrices $P_k$ and $P_l$,
	\begin{align*}
	\begin{split}
	\D_{KL}( P_k,  P_l) = & N \sum_{i=1}^n\sum_{j=1}^p n^{-1}X_{k, ij} \log(X_{k, ij}/X_{l, ij}) \\
	= & \frac{N}{n}\D( X_k, X_l) \le \frac{\beta_XNp}{2\alpha_X^2n}\| X_k- X_l\|_F^2 = \frac{c_2\nu^2Np}{4n}\| B_k -  B_l\|_F^2 \le c_2\nu^2N p(r-1),
	\end{split}
	\end{align*}
	where $c_2=2\alpha_X^{-2}\beta_X$. Using the generalized Fano's method (Lemma 3 in \cite{yu1997assouad}),
	\begin{align*}
	\inf_{\widehat{ X}} \sup_{ X \subseteq\left\{ X_1,\cdots,  X_M\right\}}  E\left\|\widehat{ X} -  X \right\|_F^2 \geq \nu^2n(r-1)\left[1 - (\log M)^{-1}\left\{c_2\nu^2 Np(r-1) + \log 2\right\}\right].
	\end{align*}
	We further set $\nu^2 = C_{\nu}n/(Np)$ for some small constant $C_{\nu}>0$ such that $\left\{C_{\nu}c_2n(r-1) + \log(2)\right\}/ \left\{n(r-1)/20\right\} < 1/2$, then the lower bound above becomes
	\begin{align*}
	\inf_{\widehat{ X}} \sup_{ X \subseteq\left\{ X_1,\cdots,  X_M\right\}}  E\left\|\widehat{ X} -  X \right\|_F^2 \geq {C_\nu n^2(r-1)}/{(2Np)},
	\end{align*}
	which implies
	\begin{align*}
	\inf_{\widehat{ X}} \sup_{ X \subseteq\left\{ X_1,\cdots,  X_M\right\}} \frac{p}{n}  E\left\|\widehat{ X} -  X \right\|_F^2 \geq \frac{C_\nu n(r-1)}{2N} = \frac{C_\nu(r-1)(n\vee p)}{2N}.
	\end{align*}	
	We can similarly derive that, for some constant $C'$,
	\begin{align*}
	\inf_{\widehat{ X}} \sup_{ X \subseteq\left\{ X_1,\cdots,  X_M\right\}} n^{-1} \D ( X,\widehat{ X}) \geq C'{(r-1)(n\vee p)}/{N}.
	\end{align*}	
	If $r\geq 2$, $r-1\geq r/2$ and the desired lower bound is obtained. 
	\item If $n < p$, the proof is essentially the same as the case of $n\geq p$. Here we construct $M$ copies of $i.i.d.$ Rademachar random matrices: $ B_1, \ldots,  B_M\in \mathbb{R}^{(r-1) \times \lfloor p/2\rfloor}$, and the following set of random rank-$r$ matrices,
	\begin{align*}
	 X_k = \begin{bmatrix}
	p^{-1} & \cdots & p^{-1}\\
	\vdots & & \vdots\\
	p^{-1} & \cdots & p^{-1}
	\end{bmatrix}_{n\times p} + \begin{blockarray}{cccc}
	& \lfloor p/2\rfloor & \lfloor p/2\rfloor & p - 2\lfloor p/2\rfloor\\
	\begin{block}{c[ccc]}
	r-1 & \nu B_k & -\nu B_k & 0\\
	n-r+1 & 0 & 0 & 0\\
	\end{block}
	\end{blockarray}.
	\end{align*}	
	We omit the rest of the proof as it is essentially the same as the part for $n\geq p$.
\end{itemize}

\subsection{Proof of Corollary \ref{coro:diversity}}
First, by the mean-value theorem for the function $f(x)=x\log(x)$, there exists $\xi_{ij}$ between $X_{ij}$ and $\widehat{X}_{ij}$ such that
$$X_{ij}^\ast \log X_{ij}^\ast =  \widehat{X}_{ij} \log \widehat{X}_{ij} + \log\xi_{ij}\left(X_{ij}^\ast - \widehat{X}_{ij}\right) + (X_{ij}^\ast - \widehat{X}_{ij}). $$
Thus,
\begin{align*}
& \frac{1}{n}\sum_{i=1}^n( H_{\text{sh}}(\widehat{ X}_i)- H_{\text{sh}}( X_i^\ast))^2= \frac{1}{n}\sum_{i=1}^n\left(\sum_{j=1}^p X_{ij}^\ast\log X_{ij}^\ast-\widehat{X}_{ij}\log \widehat{X}_{ij}\right)^2\\
= & \frac{1}{n}\sum_{i=1}^n\left\{\sum_{j=1}^p \log \xi_{ij}(X_{ij}^\ast-\widehat{X}_{ij})+\sum_{j=1}^p (X_{ij}^\ast-\widehat{X}_{ij})\right\}^2 =\frac{1}{n}\sum_{i=1}^n\left\{\sum_{j=1}^p \log \xi_{ij} (X_{ij}^\ast-\widehat{X}_{ij})\right\}^2\\
\le & \log^2 (p/\alpha_X) \frac{p}{n}\|\widehat{ X}- X^\ast\|_F^2.
\end{align*}
The third equality uses $\sum_{j=1}^p X_{ij}^\ast=\sum_{j=1}^p\widehat{X}_{ij}=1$ and the last inequality uses Cauchy-Schwarz inequality and $0< \alpha_X/p\le \xi_{ij}\le \beta_X/p\le 1$. The above inequality, together with \eqref{align:approx_lr_frob}, implies \eqref{align:shannon}. 

We can similarly prove \eqref{align:simpson} using mean-value theorem and Cauchy-Schwarz inequality:
\begin{align*}
&\frac{1}{n}\sum_{i=1}^n( H_{\text{sp}}(\widehat{ X}_i)- H_{\text{sp}}( X_i^\ast))^2 =\frac{1}{n}\sum_{i=1}^n \left(\sum_{j=1}^p\widehat{X}_{ij}^2-{X_{ij}^\ast}^2\right)^2\\
=& \frac{1}{n}\sum_{i=1}^n \left\{\sum_{j=1}^p2\xi_{ij}(\widehat{X}_{ij}-X_{ij}^\ast)\right\}^2 \le \frac{4\beta_X^2}{p^2}\frac{p}{n}\|\widehat{ X}- X^\ast\|_F^2.
\end{align*}

\subsection{Proof of Proposition \ref{prop:projection}}
First, the object function with Lagrange multipliers of the optimization problem   \eqref{eq:projection_optimization} is
$$2^{-1}\|\widehat{ x} -  x\|_2^2 + \mu\left(\sum_{i=1}^p \widehat{x}_i - 1\right) + \sum_{i=1}^p \lambda_i (\widehat{x}_i - \beta_X/p) + \sum_{i=1}^p \gamma_i (\alpha_X/p - \widehat{x}_i), \quad \lambda_i, \gamma_i \geq0. $$
Then the Karush-Kuhn-Tucker condition for optimization problem \eqref{eq:projection_optimization} is:
\begin{equation}\label{eq:to-check-1}
\widehat{ x}-  x + \mu 1_p + (\lambda_1, \ldots, \lambda_p)^\top - (\gamma_1,\ldots, \gamma_p)^\top = 0,
\end{equation}
\begin{equation}\label{eq:to-check-3}
\sum_{i=1}^n \widehat{x}_i= 1, \quad \alpha_X/p\le \widehat{x}_i \le \beta_X/p,
\end{equation}
\begin{equation}\label{eq:to-check-4}
\lambda_i, \quad \gamma_i \geq 0,
\quad \lambda_i(\widehat{x}_i - \beta_X/p) = 0, \quad \gamma_i (\alpha_X/p - \widehat{x}_i) = 0,
\end{equation}
Since the optimization problem \eqref{eq:projection_optimization} has an optimum, there exist dual variables $\lambda_i, \gamma_i, \mu$ such that \eqref{eq:to-check-1}--\eqref{eq:to-check-4} all hold. Now we prove the following fact:
\begin{equation}\label{eq:important_fact}
\widehat{x}_i = \left\{\left(x_i + \mu \right) \wedge \left(\beta_X/p\right)\right\}\vee \left(\alpha_X/p\right)
\end{equation}
for all $i \in [p]$. In fact, by \eqref{eq:to-check-1},
\begin{equation}\label{eq:fact}
\widehat{x}_i = x_i - \mu  - \lambda_i + \gamma_i , \quad \lambda_i \geq 0, \gamma_i \geq 0
\end{equation}
for any $i$. Then we prove \eqref{eq:important_fact} in five different situations:
\begin{itemize}
	\item when $x_i - \mu >\beta_X/p$, since $\widehat{x}_i \le \beta_X/p$, by \eqref{eq:fact} we have $\lambda_i >0$, which means $\widehat{x}_i = \beta_X/p$ by \eqref{eq:to-check-4};
	\item when $x_i - \mu < \alpha_X/p$, we can similarly show that $\widehat{x}_i = \alpha_X/p$ by symmetry;
	\item when $\alpha_X/p < x_i-\mu < \beta_X/p$, let us consider the situation if $\widehat{x}_i = \beta_X/p$. In such a case $\widehat{x}_i \neq \alpha_X/p$ thus $\gamma_i = 0$ by \eqref{eq:to-check-4}. Also by \eqref{eq:fact}, $\widehat{x}_i = x_i-\mu - \lambda_i \le x_i-\mu < \beta_X/p$, which is a contradiction, therefore t $\widehat{x}_i \neq \beta_X/p$. Similarly $\widehat{x}_i \neq \alpha_X/p$. Thus we must have $\lambda_i = \gamma_i=0$ and $\widehat{x}_i = x_i-\mu = \{(x_i - \mu)\wedge (\beta_X/p)\}\vee (\alpha_X/p)$ in this case;
	\item when $x_i - \mu = \beta_X/p$, we first have $\widehat{x}_i \le \beta_X/p$ by \eqref{eq:to-check-3}. Let us consider the situation if $\widehat{x}_i < \beta_X/p$. In such case, $\lambda_i = 0$ by \eqref{eq:to-check-4}, then $\widehat{x}_i = x_i-\mu -0 + \gamma_i \geq \beta_X/p$, which is a contradiction. Therefore, we must have $\widehat{x}_i = \beta_X/p$;
	\item when $x_i - \mu = \alpha_X/p$, by symmetry we can still prove $\widehat{x}_i = \alpha_X/p$.
\end{itemize}
To sum up, in all situations we have $\widehat{x}$ satisfies \eqref{eq:important_fact}. Next, in order to solve $\widehat{ x}$, we only need to find out $\mu$. Let
$$F(\mu) = \sum_{i=1}^p \{(x_i - \mu) \wedge \left(\beta_X/p\right)\}\vee \left(\alpha_X/p\right). $$
Then $F(\mu)$ is a continuous, non-increasing and piece-wise linear function of $\mu$ with all possible end points of segments in $\{x_i - \alpha_X/p, x_i - \beta_X/p\}_{i=1}^p$. Also, by $\sum_{i=1}^p x_i =1$, we must have $F(\mu) = 1$. In order to search for the solution for $F(\mu) = 1$, we let $v = \{x_i - \alpha_X/p, x_i - \beta_X/p\}_{i=1}^{2p}$, sort $v$ to $v_{(1)},\ldots, v_{(2p)}$, calculate $d_j = F(v_{(j)}) - 1$, and search for $j^\ast$ such that $d_{j^\ast} \geq 0, d_{j^\ast+1} \le 0$. In such a case, the desired $\mu$ must be between $v_{(j^\ast)}$ and $v_{(j^\ast+1)}$. This is exactly what the Proposition \ref{prop:projection} proposed. Therefore we have finished the proof for Proposition \ref{prop:projection}.

\subsection{Proof of Lemma A\ref{lemma:nuc}}
For notational simplicity, we denote $\Delta = \widehat{ X}- X^\ast$. According to \eqref{align:likeli2} and Taylor's expansion, 
\begin{equation}\label{eq:to-see}
\mathcal{L}_N(\widehat{ X}) - \mathcal{L}_N( X^\ast) - \langle \triangledown \mathcal{L}_N( X^\ast), \Delta \rangle = \frac{1}{2N}\sum_{k=1}^N \frac{\langle \Delta,  E_k \rangle^2 }{\langle \xi,  E_k \rangle^2},\quad \xi_{ij} \text{ is between } \widehat{ X}_{ij}\text{ and }  X^\ast_{ij}.
\end{equation}
Adding $\langle\triangledown \mathcal{L}_N( X^\ast),\Delta \rangle$ on both sides, we further obtain
\begin{align}
\frac{1}{2N}\sum_{k=1}^N \frac{\langle\Delta,  E_k\rangle^2}{\langle \xi,  E_k\rangle^2}&=\mathcal{L}_N(\widehat{ X})-\mathcal{L}_N( X^\ast)-\langle\triangledown \mathcal{L}_N( X^\ast),\Delta \rangle\notag\\
&\le-\langle\triangledown \mathcal{L}_N( X^\ast),\Delta \rangle+\lambda\left(\| X^\ast\|_\ast-\|\widehat{ X}\|_\ast\right)\notag\\
&= -\langle\triangledown \mathcal{L}_N( X^\ast)+ R1_p^\top,\Delta \rangle+\lambda\left(\| X^\ast\|_\ast-\|\widehat{ X}\|_\ast\right)\notag\\
&\le \|\triangledown \mathcal{L}_N( X^\ast)+ R1_p^\top\|_2\|\Delta\|_\ast+\lambda\left(\| X^\ast\|_\ast-\|\widehat{ X}\|_\ast\right).\label{align:grad_lambda}
\end{align}
The third line comes from the identity $\langle  R1_p^\top,\Delta\rangle=\langle  R,\Delta 1_p\rangle =\langle  R, 0_n\rangle = 0$ and the forth inequality is the H$\mathrm{\ddot{o}}$lder's inequality between the nuclear norm and operator norm. To further upper bound the nuclear norm $\|\Delta\|_\ast$, we state two useful technical results.
\begin{lemma}\label{lemma:lambda}
	Assume there exist constants $\alpha_R,\beta_R, \alpha_X$, and $\beta_X$ such that, for any $i\in [n]$ and $j\in [p]$, $\alpha_R/n\le R_i\le \beta_R/n$, $\alpha_X/p \le X_{ij}^\ast \le \beta_X/p$. Conditioning on fixed $N$, where $N>(n\vee p)\log(n+p)$, with probability at least $1-\left(n+p\right)^{-1}$, we have
	\[\| \triangledown\mathcal{L}_N\left( X^\ast\right)+ R 1_p^\top\|_2\le \left\{\frac{2}{3\alpha_X}+2\left(\frac{1}{9\alpha_X^2}+\frac{\beta_R}{\alpha_X}\right)^{1/2}\right\}\left\{\frac{p(n\vee p)\log(n+p)}{nN}\right\}^{1/2}.\]
\end{lemma}
Based on Lemma A$\ref{lemma:lambda}$, with probability proceeding $1-\left(n+p\right)^{-1}$, the selected tuning parameter $\lambda$ satisfies $\lambda\ge 2\|\triangledown\mathcal{L}_N( X^\ast)+ R 1_p^\top \|_2$. We  complete the proof by using the following lemma.
\begin{lemma}\label{lemma:nuc_frob}
	If $\eqref{align:grad_lambda}$ holds and $\lambda\ge 2\|\triangledown\mathcal{L}_N( X^\ast)+ R 1_p^\top \|_2$, we have the following upper bound for the nuclear norm of $\Delta = \widehat{X} - X^\ast$:
	\begin{align*}
	\|\Delta\|_\ast \le 4\surd{(2r)}\|\Delta\|_F+4\sum_{i=r+1}^{n\land p}\sigma_{i}\left( X^\ast\right).
	\end{align*}
\end{lemma}

\subsection{Proof of Lemma A\ref{lemma:kl_frob}}
By using Taylor's expansion of function $\log(B_{ij})$ at $A_{ij}$, we can rewrite the Kullback-Leibler divergence $\D(A, B)$ as
\[\D(A, B)=\sum_{i,j}-A_{ij}\log\frac{B_{ij}}{A_{ij}}=\sum_{i,j}\left\{-(B_{ij}-A_{ij})+\frac{A_{ij}}{2\xi_{ij}^2}(B_{ij}-A_{ij})^2\right\}=\sum_{i,j}\frac{A_{ij}}{2\xi_{ij}^2}(B_{ij}-A_{ij})^2,\]
where $\xi_{ij}$ is a quantity between $B_{ij}$ and $A_{ij}$. Here, we used the fact that $\sum_{ij}A_{ij}=\sum_{ij}{B}_{ij}=n$ in the third equality. Since $A_{ij}$ and $B_{ij}\in [\alpha_X/p,\beta_X/p]$, we complete the proof by
\begin{align*}
\D(A, B) &=\sum_{i,j}\frac{A_{ij}}{2\xi_{ij}^2}(B_{ij}-A_{ij})^2\le \frac{\beta_X/p}{2(\alpha_X/p)^2}\sum_{i,j}(B_{ij}-A_{ij})^2=\frac{\beta_Xp}{2\alpha_X^2}\|A-B\|_F^2,\\
\D(A, B)&=\sum_{i,j}\frac{A_{ij}}{2\xi_{ij}^2}(B_{ij}-a_{ij})^2\ge \frac{\alpha_X/p}{2(\beta_X/p)^2}\sum_{i,j}(B_{ij}-A_{ij})^2=\frac{\alpha_Xp}{2\beta_X^2}\|A- B\|_F^2.
\end{align*}

\subsection{Proof of Lemma A\ref{lemma:lambda}}
We rewrite $\triangledown\mathcal{L}_N\left( X^\ast\right)+ R1_p^\top=N^{-1}\sum_{k=1}^N Y_k$, where $ Y_k=-\langle X^\ast, E_k\rangle^{-1} E_k+ R1_p^\top$ with $E (Y_k)= 0$. Note that, under the assumption that  $\alpha_R/n\le R_i\le \beta_R/n$, using Weyl's inequality, for any $ X^\ast\in\mathcal{S}(\alpha_X,\beta_X)$, we have
\[\left\| Y_k\right\|_2=\left\|-\langle E_k, X^\ast\rangle^{-1} E_k+ R1_p^\top\right\|_2\le \max_{i,j}\left\| {X_{ij}^{\ast}}^{-1}e_{i}e_{j}^\top\right\|_2+\left\| R1_p^\top\right\|_2\le \alpha_X^{-1}{p}+\beta_R\surd{(n^{-1}p)}.\]
We also observe that
\[ E Y_k^\top Y_k=\sum_{i,j}{{X_{ij}^\ast}}^{-1}{R_i}e_{j}e_{j}^\top-\| R\|_2^21_p1_p^\top\quad \text{and} \quad  E Y_k Y_k^\top=\sum_{i,j}{{X_{ij}^\ast}}^{-1}{R_i}e_{i}e_{i}^\top-p R R^\top.\]
Hence, we apply Weyl's inequality to $ E Y_k^\top Y_k$ and $ E Y_k Y_k^\top$ and obtain
\begin{equation*}
\begin{split}
\left\| E Y_k^\top Y_k\right\|_2 \le & \left\|\sum_{i,j}\frac{R_i}{X_{ij}^\ast} e_{j}e_{j}^\top\right\|_2+\| R\|_2^2\left\|1_p1_p^\top\right\|_2=\max\limits_{1\le j\le p}\sum_{i=1}^n\frac{R_i}{X_{ij}^\ast}+p\sum_{i=1}^nR_i^2\\
\le & \frac{p}{\alpha_X}\sum_{i=1}^nR_i + p \max_i R_i \cdot \sum_{i=1}^n R_i \le \frac{p}{\alpha_X} + \frac{\beta_R p}{n},
\end{split}
\end{equation*}
\begin{equation*}
\begin{split}
\left\| E Y_k Y_k^\top\right\|_2&\le \left\|\sum_{i,j}\frac{R_i}{X_{ij}^\ast}e_{i}e_{i}^\top\right\|_2+\left\| p R R^\top\right\|_2=\max\limits_{1\le i\le n}\sum_{j=1}^p\frac{R_i}{X_{ij}^\ast}+p\sum_{i=1}^nR_i^2\le \frac{\beta_Rp^2}{\alpha_Xn}+\frac{\beta_Rp}{n}.
\end{split}
\end{equation*}
Write $M=\alpha_X^{-1}p+\beta_R\surd{(n^{-1}p)}$ and $\sigma^2 = n^{-1}{Np}\left\{\beta_R+\alpha_X^{-1}(n\vee \beta_Rp)\right\}$, and set
\begin{align*}
t_0&=\frac{2\log(n+p)M}{3N}+\left\{\frac{4\log^2(n+p)M^2}{9N^2}+\frac{4\sigma^2\log(n+p)}{N^2}\right\}^{1/2}.
\end{align*} 
Then applying Lemma A$\ref{lemma:aw}$, with any $t>t_0$, we have
\begin{align*}
P&\left(\left\|\triangledown\mathcal{L}_N\left( X^\ast\right)+ R1_p^\top\right\|_2\ge t\right)\le\left(n+p\right)\exp\left(-\frac{N^2t_0^2/2}{\sigma^2 + MNt_0/3}\right)\leq (n+p)^{-1}.
\end{align*}
Since $M \le (\alpha_X^{-1}+o(1))p$ and $\sigma^2 \le (\alpha_X^{-1}\beta_R+o(1))Np(n\vee p)/n$, by the condition that $N>(n+p)\log(n+p)$, we have
\begin{align*}
t_0\le\left\{\frac{2}{3\alpha_X}+2\left(\frac{1}{9\alpha_X^2}+\frac{\beta_R}{\alpha_X}\right)^{1/2}\right\}\left\{\frac{p(n\vee p)\log(n+p)}{nN}\right\}^{1/2}.
\end{align*}
and have completed the proof.

\subsection{Proof of Lemma A\ref{lemma:nuc_frob}}
We observe that $\eqref{align:grad_lambda}$ is essentially equivalent to  Lemma 1 (B.2) in \cite{negahban2011estimation}. Therefore, following their results, under the assumption $\lambda\ge 2\|\triangledown\mathcal{L}_N( X^\ast)+ R 1_p^\top \|_2$, for each $r\le n\land p$, there exists an orthogonal decomposition $\Delta=\Delta^{'}+\Delta^{''}$, where the rank of $\Delta^{'}$ is no more than $2r$ and $\Delta^{''}$ satisfies
\begin{align*}
\|\Delta^{''}\|_\ast\le3\|\Delta^{'}\|_\ast+4\sum_{i=r+1}^{n\land p}\sigma_{i}\left( X^\star\right)\text{ and } \|\Delta\|_F^2 = \|\Delta^{'}\|_F^2+\|\Delta^{''}\|_F^2.
\end{align*}
Using the triangle inequality and $\|\Delta^{'}\|_\ast\le \surd{(2r)}\|\Delta^{'}\|_F\le\surd{(2r)}\|\Delta\|_F$, we obtain
\[\|\Delta\|_\ast\le \|\Delta^{'}\|_\ast+\|\Delta^{''}\|_\ast\le 4 \|\Delta^{'}\|_\ast+4\sum_{i=r+1}^{n\land p}\sigma_{i}\left( X^\ast\right)\le 4\surd{(2r)}\|\Delta\|_F+4\sum_{i=r+1}^{n\land p}\sigma_{i}\left( X^\ast\right),\]
which completes the proof.

\subsection{Concentration Inequalities}
\begin{lemma}\label{lemma:aw}
	Let $\{ Y_k\}_{k=1}^N$ be independent $n\times p$ zero-mean random matrices such that $\| Y_k\|_2\le M$ and define $\sigma^2 =\max\left\{\sum_{k=1}^N\left\| E Y_k^\top Y_k\right\|_2,\sum_{k=1}^N\left\| E Y_k Y_k^\top\right\|_2\right\}$. For all $t>0$, we have
	\begin{align}\label{align:tropp}
	P\left[\left\|\frac{1}{N}\sum_{k=1}^N  Y_k\right\|_2\ge t\right]\le\left(n+p\right)\exp\left(-\frac{N^2t^2/2}{\sigma^2+MNt/3}\right).
	\end{align}
	In addition, the expected spectral norm satisfies
	\begin{align}\label{align:expectsum}
	\left( E\left\|\sum_{k=1}^N Y_k\right\|_2^2\right)^{1/2}\le \left\{28\log(n+p)\sigma^2\right\}^{1/2}+28\log(n+p)M.
	\end{align}
\end{lemma}
\begin{proof}
	The proof of concentration inequality $\eqref{align:tropp}$ follows, for example, Theorem 1.6 of \cite{tropp2011user}. The proof of inequality $\eqref{align:expectsum}$ follows, for example, Theorem 1 of \cite{tropp2015expected}.
\end{proof}

\begin{lemma}\label{lemma:uprdm}
	Let $n\times p$ random matrices $\{ E_k\}_{k=1}^N$ be independent and identically distributed with distribution $\Pi=( R1_p^\top)\circ X^\ast$ on $\{e_i(n)e_j^\top(p), (i,j)\in[n]\times[p]\}$ and $\{\varepsilon_k\}_{k=1}^N$ is an i.i.d Rademacher sequence. Assume that $\alpha_R/n\le R_i \le \beta_R/n$, for any $ X^\ast\in\mathcal{S}(\alpha_X,\beta_X)$, we have the upper bound
	\begin{align}\label{align:uprdm}
	E\left\|\frac{1}{N}\sum_{k=1}^N\varepsilon_k E_k\right\|_2\le\left\{\frac{28\log(n+p)\left({\beta_R}/{n}\vee{\beta_X}/{p}\right)}{N}\right\}^{1/2}+\frac{28\log(n+p)}{N}.
	\end{align}
\end{lemma}
\begin{proof}
	We establish this bound by applying Lemma A$\ref{lemma:aw}$. Let $ Y_k=\varepsilon_k E_k$, we first calculate the terms $M$ and $\sigma^2$ involved in Lemma A$\ref{lemma:aw}$. According to the definition of $\varepsilon_k$ and $ E_k$,
	\[M=\max_{1\le k\le N}\left\| Y_k\right\|_2=\max_{1\le k\le N}\left\|\varepsilon_k E_k\right\|_2=1.\]
	Also note that,
	\[ E\left[ Y_k^\top Y_k\right]= E\left[\varepsilon_k^2 E_k^\top E_k\right]=\sum_{i,j}R_iX_{ij}^\ast e_{j}(p) e_{j}^\top(p)\text{ and }  E\left[ Y_i^\top Y_i\right]= E\left[\varepsilon_k^2 E_k E_k^\top\right]=\sum_{i,j}R_iX_{ij}^\ast e_{i}(n)e_{i}^\top(n).\]
	We observe that, under the assumption that $\alpha_R/n\le R_i\le \beta_R/n$, for any $ X^\ast\in\mathcal{S}(\alpha_X,\beta_X)$, we have
	\begin{align*}
	\left\|\sum_{i,j}R_i X_{ij}^\ast e_j(p) e_j^\top(p)\right\|_2=\max_j\sum_{i=1}^n R_iX_{ij}^\ast\le\sum_{i=1}^n R_i\cdot\max_{i,j}X_{ij}^\ast\le\beta_X/p,\\
	\left\|\sum_{i,j}R_i X_{ij}^\ast e_i(n) e_i^\top(n)\right\|_2=\max_i\sum_{j=1}^p R_iX_{ij}^\ast\le\sum_{j=1}^p X_{ij}^\ast\cdot\max_{i}R_{i}\le\beta_R/n.
	\end{align*}
	As a result, $\sigma^2\le N\left({\beta_X}/{p}\vee {\beta_R}/{n}\right)$.
	By applying Jensen's inequality and $\eqref{align:expectsum}$, we obtain
	\[ E\left\|\frac{1}{N}\sum_{k=1}^N\epsilon_k E_k\right\|_2\le\left( E\left\|\frac{1}{N}\sum_{k=1}^N\epsilon_k E_k\right\|_2^2\right)^{1/2}\le \left\{\frac{28\log(n+p)\left({\beta_R}/{n}\vee {\beta_X}/{p}\right)}{N}\right\}^{1/2}+\frac{28\log(n+p)}{N},\]
	which has finished the proof of this lemma.
\end{proof}

\begin{lemma}\label{lemma:hoeffd}
	Suppose $X^\ast \in \mathcal{S}(\alpha_X, \beta_X)$. Let $n\times p$ random matrices $\{ E_k\}_{k=1}^N$ be independent and identically distributed with distribution $\Pi=( R1_p^\top)\circ X^\ast$ on $\{e_i(n)e_j^\top(p), (i,j)\in[n]\times[p]\}$. We define a constraint set $\mathcal{D}\left(T\right)$ for $T>0$,
	\begin{equation}\label{align:ct}
	\mathcal{D}\left(T\right)=\left\{ A\in\mathcal{S}(\alpha_X,\beta_X)\Big|
	\begin{array}{l}
	\D( X^\ast, A)\le T, \\
	\|X^\ast - A\|_\ast \le 4\surd{(2r)}\|X^\ast - A\|_F + 4\sum_{i=r+1}^{n\wedge p} \sigma_i(X^\ast)
	\end{array}\right\},
	\end{equation}
	where $\beta_X > \alpha_X$. We also define
	\[Z_T=\sup\limits_{ A\in\mathcal{D}\left(T\right)}\left|{N}^{-1}\sum_{k=1}^N\langle\log X^\ast-\log A, E_k\rangle-\sum_{i,j}R_iX_{ij}^\ast\log\left({X_{ij}^\ast}/A_{ij}\right)\right|.\]
	Under the assumption that $\alpha_R/n\le R_i\le \beta_R/n$, we have
	\[ P\left\{Z_T\ge \frac{\alpha_R T}{4n} + E(n,p,r)\right\}\le\exp\left[-\frac{\alpha_R^2NT^2}{512\left\{n\log(\beta_X/\alpha_X)\right\}^2}\right],\]
	where $E(n,p,r)$ is defined in \eqref{align:enpr} as
	\begin{equation*}
	\begin{split} 
	E(&n,p, r)=\frac{2048\beta_X^2npr}{\alpha_X^3\alpha_R}\left[\left\{\frac{28\log(n+p)\left(\beta_R/n\vee \beta_X/p\right)}{N}\right\}^{1/2}+\frac{28\log(n+p)}{N}\right]^2 \\
	&+ \frac{16p}{\alpha_X}\left[\left\{\frac{28\log(n+p)\left(\beta_R/n\vee \beta_X/p\right)}{N}\right\}^{1/2}+\frac{28\log(n+p)}{N}\right]\sum_{i=r+1}^{n\land p}\sigma_{i}\left( X^\ast\right).
	\end{split}
	\end{equation*}
\end{lemma}
\begin{proof}
	By the entrywise upper and lower bounds of $X^\ast$ and all matrices in $\mathcal{S}(\alpha_X, \beta_X)$, 
	$$\sup\limits_{ A,  X^\ast\in\mathcal{S}(\alpha_X,\beta_X)}\max\limits_{i,j}|\log X^\ast_{ij}-\log A_{ij}|\le \log(\beta_X/\alpha_X),$$ 
	we obtain the following concentration inequality by a version of empirical process Hoeffding's inequality (\cite{massart2000constants}; also see Theorem 14.2 of \cite{buhlmann2011statistics}),
	\begin{align}\label{align:hoeffd}
	P\left\{Z_T- EZ_T\ge \alpha_RT/(8n)\right\}\le\exp\left[-\frac{\alpha_R^2NT^2}{512\left\{n\log(\beta_X/\alpha_X)\right\}^2}\right].
	\end{align}
	It remains to upper bound the quantity $ EZ_T$. By a symmetrization argument due (\cite{wellner1996weak}; also see Theorem 14.3 in \cite{buhlmann2011statistics}), we obtain
	\begin{align*}
	E\left(Z_T\right)&= E\sup\limits_{ A\in\mathcal{D}\left(T\right)}\left|N^{-1}\sum_{k=1}^N\langle\log X^\ast-\log A, E_k\rangle-\sum_{i,j}R_iX_{ij}^\ast\log\left(X_{ij}^\ast/A_{ij}\right)\right|\\
	& = E\sup\limits_{ A\in\mathcal{D}\left(T\right)}\left|N^{-1}\sum_{k=1}^N\langle\log X^\ast-\log A, E_k\rangle-E \left(N^{-1}\sum_{k=1}^N\langle\log X^\ast-\log A, E_k\rangle\right)\right|\\
	&\le 2 E\left(\sup\limits_{ A\in\mathcal{D}\left(T\right)}\left|N^{-1}\sum_{k=1}^N\varepsilon_k\langle\log X^\ast-\log A, E_k\rangle\right|\right)\\
	&=2 E\left(\sup\limits_{ A\in\mathcal{D}\left(T\right)}\left|N^{-1}\sum_{k=1}^N\varepsilon_k\sum_{i,j}\mathbb{I}_{( E_k=e_i(n)e_j^\top(p))}\log\left(X_{ij}^\ast/A_{ij}\right)\right|\right),
	\end{align*}
	where $\{\varepsilon_k\}_{k=1}^N$ is an independent and identically distributed Rademacher sequence. For any number $k\in[N]$, there exists an $(i_k, j_k)$ pair such that $e_{i_k}(n)e_{j_k}^\top(p) = E_k$. We accordingly define
	\[\phi_{k}(t)=\left\{\begin{array}{ll}\alpha_Xp^{-1} \left\{\log {X_{i_kj_k}^\ast}-\log (X_{i_kj_k}^\ast + t)\right\}, & \text{ if } t \geq \alpha_X/p - X_{i_kj_k}^\ast;\\
	\alpha_Xp^{-1}\left\{\log {X_{i_kj_k}^\ast}-\log (\alpha_X/p)\right\}, & \text{ if } t < \alpha_X/p - X_{i_kj_k}^\ast.
	\end{array}\right.\]
	Since $|\phi_k'(t)|\le 1$ for $t\geq \alpha_X/p-X_{i_kj_k}^\ast$; $\phi_k'(t)=0$ for $t < \alpha_X/p-X_{i_kj_k}^\ast$, $\phi_k(t)$ is a contraction map such that
	$$|\phi_k(x) - \phi_k(y)| \le |x-y|,\quad \forall x, y\in \mathbb{R}; \quad \phi_k(0)=0.$$ 
	Then the contraction principle (Theorem 14.4 in \cite{buhlmann2011statistics}; \cite{ledoux2013probability}), together with H$\mathrm{\ddot{o}}$lder's inequality between nuclear and operator norm, yields
	\begin{equation}\label{align:contract}
	\begin{split}
	E(Z_T)& \le 2 E\left\{\sup\limits_{ A\in\mathcal{D}\left(T\right)}\left|N^{-1}\sum_{k=1}^N\varepsilon_k\sum_{i,j}\mathbb{I}_{( E_k=e_i(n)e_j^\top(p))}\log\left(X_{i_kj_k}^\ast/A_{i_kj_k}\right)\right|\right\},\\
	& = 2 E\left\{\sup\limits_{ A\in\mathcal{D}\left(T\right)}\left|N^{-1}\sum_{k=1}^N\varepsilon_k \cdot \alpha_X^{-1}p \phi_k\left(A_{i_kj_k} - X_{i_kj_k}^\ast \right)\right|\right\}\\
	& = 2 E\left[E\left\{\sup\limits_{ A\in\mathcal{D}\left(T\right)}\left|N^{-1}\sum_{k=1}^N\varepsilon_k \cdot \alpha_X^{-1}p \phi_k\left(A_{i_kj_k} - X_{i_kj_k}^\ast \right)\right|~~\Bigg| E_1,\ldots, E_N\right\}\right]\\
	& \leq 2E\left[2E\left\{\sup\limits_{ A\in\mathcal{D}\left(T\right)}\left|N^{-1}\sum_{k=1}^N\varepsilon_k \cdot \alpha_X^{-1}p \left(A_{i_kj_k} - X_{i_kj_k}^\ast \right)\right|~~\Bigg| E_1,\ldots, E_N\right\}\right]\\
	&= 4\alpha_X^{-1}p E\left\{\sup\limits_{ A\in\mathcal{D}(T)}\left|N^{-1}\sum_{k=1}^N\langle X^\ast- A,\varepsilon_k E_k\rangle\right|\right\}\\
	&\le 4\alpha_X^{-1}p\sup\limits_{ A\in\mathcal{D}(T)}\left(\left\| X^\ast- A\right\|_\ast E\left\|N^{-1}\sum_{k=1}^N\varepsilon_k E_k\right\|_2\right).
	\end{split}
	\end{equation}
	We bound $ E\left\|N^{-1}\sum_{k=1}^N\varepsilon_k E_k\right\|_2$ by applying Lemma A$\ref{lemma:uprdm}$,
	\begin{align}\label{align:spectral}
	E\left\|N^{-1}\sum_{k=1}^N\varepsilon_k E_k\right\|_2\le \left\{\frac{28\log(n+p)\left(\beta_R/n\vee \beta_X/p\right)}{N}\right\}^{1/2}+\frac{28\log(n+p)}{N}.
	\end{align}
	Under the assumption that $A\in \mathcal{D}_T$ and applying Lemma A$\ref{lemma:kl_frob}$, we have 
	\begin{align}\label{align:KL}
	\sup\limits_{A\in\mathcal{D}(T)}\| X^\ast-A\|_\ast &\le 4\surd{(2r)}\sup\limits_{A\in\mathcal{D}(T)}\| X^\ast- A\|_F+4\sum_{i=r+1}^{n\land p}\sigma_{i}\left( X^\ast\right)\notag\\
	&\le \sup\limits_{A\in\mathcal{D}(T)} 8\left\{{\beta_X^2r}/{(\alpha_Xp)}\D( X^\ast, A)\right\}^{1/2}+4\sum_{i=r+1}^{n\land p}\sigma_{i}\left( X^\ast\right)\notag\\
	&\le 8\left\{{\beta_X^2rT}/{(\alpha_Xp)}\right\}^{1/2}+4\sum_{i=r+1}^{n\land p}\sigma_{i}\left( X^\ast\right).
	\end{align}
	Combining inequalities $\eqref{align:contract}$, $\eqref{align:spectral}$, $\eqref{align:KL}$, and the arithmetic-geometry inequality yields,
	\begin{align*}
	E(Z_T) \le&\frac{16 p}{\alpha_X}\left[\left\{\frac{28\log(n+p)\left(\beta_R/n \vee \beta_X/p\right)}{N}\right\}^{1/2}+\frac{28\log(n+p)}{N}\right]\left\{2\left({\frac{\beta_X^2rT}{\alpha_Xp}}\right)^{1/2}+\sum_{i=r+1}^{n\land p}\sigma_{i}\left( X^\ast\right)\right\}\\
	\le & \frac{2048\beta_X^2npr}{\alpha_X^3\alpha_R}\left[\left\{\frac{28\log(n+p)\left(\beta_R/n\vee \beta_X/p\right)}{N}\right\}^{1/2}+\frac{28\log(n+p)}{N}\right]^2   + \alpha_RT/(8n) \\
	& + \frac{16p}{\alpha_X}\left[\left\{\frac{28\log(n+p)\left(\beta_R/n\vee \beta_X/p\right)}{N}\right\}^{1/2}+\frac{28\log(n+p)}{N}\right]\sum_{i=r+1}^{n\land p}\sigma_{i}\left( X^\ast\right)\\
	= & E(n,p,r) + {\alpha_RT}/{(8n)}.
	\end{align*}
	Finally, plugging the upper bound of $ E(Z_T)$ into concentration inequality $\eqref{align:hoeffd}$, we obtain
	\[ {\rm pr}\left(Z_T\ge {\alpha_R T}/{(4n)}+E(n,p,r)\right)\le\exp\left[-\frac{\alpha_R^2NT^2}{512\left\{n\log(\beta_X/\alpha_X)\right\}^2}\right]\]
	and have finished the proof of this lemma.
\end{proof}

\end{document}